\documentclass[12pt,a4paper]{article}
\pdfoutput=1
\usepackage{jheppub}
\usepackage[utf8]{inputenc}
\usepackage{xcolor}
\usepackage{epstopdf}
\usepackage{graphicx}
\usepackage{epsfig}
\usepackage{dcolumn}  
\usepackage{bm}    
 \usepackage{caption}
 \usepackage{subcaption}
\usepackage{amssymb} 
\usepackage{amsmath,bm}
\usepackage{amsfonts}  
\usepackage{amsmath}  
\usepackage{slashed}  
\usepackage{enumitem}
\usepackage[mathscr]{euscript}
\usepackage{tabu}
\usepackage{epsfig}
\makeatletter
\g@addto@macro\bfseries{\boldmath}
\makeatother

\def\sltwor{\mathfrak{sl}(2,\mathbb{R})}
\def\btau{{\bar{\tau}}}
\def\bq{{\bar{q}}}
\def\Vs#1{V^{(s)}_{#1}}
\def\bVs#1{\bar{V}^{(s)}_{#1}}

\def\btau{\bar{\tau}}

\def\vev#1{\langle #1 \rangle}
\def\cA{\mathcal{A}}

\def\nn{\nonumber}
\def\pd{\partial}
\def\Re{R\'{e}nyi }
\def\l1{{{1-loop}}}

\def\by{{\bar{y}}}

\def\n1{\Bigg|_{n=1}}

\def\n{{(n)}}
\def\tr{{Tr}}

\def\I{\ \mathbf{\mathcal{I}}}

\def\tr{\text{Tr}}

\def\cZ{\mathcal{Z}}

\def\bL{\bar{L}}

\usepackage{bookmark}
  \title{\textbf{\textsf{R\'{e}nyi divergences from Euclidean quenches}} }
  \author{Barsha G. Chowdhury${}^{a}$, Shouvik Datta${}^b$, Justin R. David${}^a$}
\affiliation{\vspace{.1cm} ${}^a$Centre for High Energy Physics, \\ ${}^{~\,}$Indian Institute of Science,\\
${}^{~\,}$C. V. Raman Avenue, Bangalore 560012, India.\vspace{.2cm}\\
${}^b$Mani L.\,Bhaumik Institute for Theoretical Physics, \\
${}^{~}$Department of Physics  \& Astronomy,  University of California, \\
${}^{~}$Los Angeles, CA 90095, USA.
}
\emailAdd{barsha@iisc.ac.in, shouvik@ucla.edu, justin@iisc.ac.in}
\abstract{We study  the  generalisation of 
relative entropy, the  R\'{e}nyi divergence  $D_{\alpha} ( \rho||\rho_\beta) $
in 2$d$ CFTs
 between an excited state density matrix $\rho$, created by deforming the Hamiltonian, and the thermal density matrix $\rho_\beta$.
 Using the  path integral  representation of this quantity as a Euclidean quench,
 we  obtain the 
leading contribution to the R\'{e}nyi divergence for deformations by scalar primaries
and  
by conserved holomorphic currents  in  conformal perturbation theory.
Furthermore, we calculate the leading  contribution to the  R\'{e}nyi divergence  when the
conserved current perturbations 
 have inhomogeneous  spatial profiles which are versions of the sine-square deformation (SSD). The dependence on the \Re parameter ($\alpha$) of the leading 
contribution  have a universal form for these inhomogeneous deformations and it is identical to that seen in the R\'{e}nyi divergence
of the simple harmonic oscillator perturbed by a linear  potential.  
Our study of these  R\'{e}nyi  divergences shows that 
the family of second laws of thermodynamics, which are equivalent to the 
monotonicity of R\'{e}nyi divergences, do indeed provide stronger 
constraints for  allowed transitions  compared to  the traditional second law. 
}
 
\begin{document}

\maketitle

\section{Introduction}
\label{sec:intro}

The application of ideas  from information theory  has led to  important insights in 
quantum field theory, holography  and black hole physics. 
The most well studied measure is that of entanglement entropy which is defined as the  von-Neumann entropy of the 
reduced density matrix on a spatial region. Entanglement entropy and its holographic  realization 
in terms of the minimal surface \cite{Ryu:2006bv} 
has been the key ingredient for the  recent developments 
in AdS/CFT and black holes. 
Another concept which has recently received attention is 
the relative entropy of two density matrices which is defined as 
\begin{equation}
S(\rho_1||\rho_2) = {\rm Tr} ( \rho_1 \log \rho_1) - {\rm Tr} ( \rho_1 \log \rho_2)~. 
\end{equation}
Relative entropy assigns a positive number given two density matrices and therefore can 
be used as a measure of distance in the space of density matrices. 
It was introduced in the holographic context in \cite{Blanco:2013joa} and has subsequently found  several applications \cite{Bousso:2014sda,Jafferis:2015del}. 
For an introduction to  information theoretic measures and their applications in 
quantum field theory see \cite{Witten:2018lha}.

Just as the R\'{e}nyi entropies are a 
one parameter generalisation of the  von-Neumann entropy,  relative entropy also 
admits one parameter generalisations. 
The focus of this paper is the generalization known  
as R\'{e}nyi divergence,  also known as the Petz entropy \cite{PETZ198657}. 
This is defined as
\begin{eqnarray}
D_{\alpha}( \rho_1|| \rho_2) = \frac{{\rm sgn} (\alpha) }{\alpha - 1} \log {\rm Tr [\rho^\alpha_1\rho_2^{1-\alpha}] }~,
\end{eqnarray}
where $\rho_1, \rho_2$ are normalized density matrices. 
This quantity  forms an important distance measure in information theory. 
In \cite{Bernamonti:2018vmw}  
R\'{e}nyi divergence between a state $\rho$ deformed from the thermal state 
$\rho_\beta$  at the temperature $\beta$ 
with respect to a given Hamiltonian $H$  was used to study additional 
second law like constraints which can govern the out of equilibrium 
state $\rho$ as  it evolves to the thermal state $\rho_\beta$ by the Hamiltonian $H$. 
The excited state $\rho$  considered in \cite{Bernamonti:2018vmw} 
is  also a thermal state but  under  a deformed 
 Hamiltonian  $H_{\rm def} = H + \mu {\cal O}  $ where ${\cal O }$ is the deforming operator. 
For this class of excited states, \cite{Bernamonti:2018vmw} showed that one can evaluate the 
R\'{e}nyi divergences using its path integral as a Euclidean quench. 
The Euclidean quench  was also used to develop a
method to evaluate R\'{e}nyi divergences holographically.
In the cases where $H$ describes a $2$-dimensional conformal field theory
and ${\cal O}$ a conformal primary of dimensions $\Delta <1$, 
 it was shown that 
there are indeed situations where constraints are stronger than the conventional 
second law.  These additional constraints  resulted from the monotonicity of 
R\'{e}nyi divergences.

In this paper  we  study more properties of R\'{e}nyi divergences in 
$2$-dimensional conformal field theory. 
The formulation of \Re divergence as a Euclidean quench presents us with the interesting problem of  calculating a new class of generalized partition functions
\begin{align} \label{genpart}
\cZ_\alpha (\tau,\mu)= \tr \left[q^{\alpha H_{\rm def}(\mu)} q^{(1-\alpha)H_{\rm CFT}}\right]. 
\end{align}
Here, $H_{\rm def}(\mu)=H_{\rm CFT}+\mu \mathcal{O}$. 
After a review of the formulation of \Re divergence as a Euclidean quench,
we re-visit the evaluation of  the leading contribution to 
R\'{e}nyi divergences for excited states created by deforming 
the Hamiltonian by a scalar primary of weight $\Delta$. 
We obtain an analytical expression for the R\'{e}nyi divergence in terms of an infinite  series. 
The representation in terms of infinite series is obtained for all values of $\Delta$ 
 provided the integral 
resulting from conformal perturbation theory is regulated. 
For the case $\Delta =1$ we obtain a closed form expression in terms of known functions.

When $\mathcal{O}$ commutes with the Hamiltonian, as is the case with conserved current deformations, the above quantity can be calculated from the knowledge of the deformed partition function $\tr \left[q^{H_{\rm def}(\alpha\mu)}\right]$.  We consider the cases of $U(1)$, the stress tensor and the spin-3 current deformations. 
We obtain the   generalised partition function (\ref{genpart}) 
partition function  for these cases  in the closed form 
by considering   the perturbative expansion  in the  coupling $\mu$.  
For carrying out the conformal perturbation theory we used the methods developed earlier 
in \cite{Datta:2014ska,Datta:2014uxa,Datta:2014zpa}.    The prescription adopted for carrying 
out the integrals that arise in conformal perturbation theory ensure that each integral is 
finite despite the fact  the conformal dimension of the current  is  such that 
 $\Delta \geq 1 $.   These
  methods enable us to evaluate the R\'{e}nyi divergences, $D_\alpha( \rho||\rho_\beta)$,  
where $\rho$ is the excited state obtained by adding conserved currents to the Hamiltonian.

When $\mathcal{O}$ does not commute with $H_{\rm CFT}$, calculating $\cZ_\alpha$ is not straightforward. The effective Hamiltonian $H_{\rm eff}$, defined as $q^{H_{\rm eff}}=q^{\alpha H_{\rm def}(\mu)}q^{(1-\alpha) H_{\rm CFT}}$, then involves an appropriate re-summation of the Baker-Campbell-Hausdorff series. We note that ${H_{\rm eff}}$ is also the Floquet Hamiltonian governing the dynamics if the system is evolved  alternatively using $H_{\rm CFT}$ and $H_{\rm def}$. Some progress in finding closed form expressions for $H_{\rm eff}$ has been made in \cite{floquet:replica}. 
To render the problem tractable we work with the situation  that the deforming operator 
$\mathcal{O}$ is still constructed from conserved currents, but $\mu$ acquires  a spatial profile. 
Taking the spatial directions to be compact and choosing periodic functions in space one 
can obtain deforming operators  $\mathcal{O}$ which do not commute with Hamiltonian 
$H_{\rm{CFT}}$. 
For the case of the stress tensor, 
a special class of  deformations  was considered earlier  in \cite{ishibashi:ssd,wen:floquet,wen:ssd,vishwanath:ssd,chitra:ssd} called 
sine-square deformation (SSD).   This name is derived from the fact the spatial dependence
of the  envelope function is a sine-squared profile. 
We generalise this envelope function introducing a   parameter  $\mu$ which controls the 
amplitude of the deformation.  The deforming operator is then proportional to  combination of 
Virasoro generators, 
$L_{1} + L_{-1}$. 
We also generalise  to the cases when the deforming operator is constructed from 
higher spin currents. We call these the  higher spin SSD deformations.
Analogous to the Virasoro case,  the deforming
 operator is given by $\Vs{-k}+\Vs{k}$, where $k$ is the mode number of the 
the  spin-$s$ current and $V^{(s)}$ is the generator of higher spin symmetries.

We set up Hamiltonian perturbation theory in $\mu$ and evaluate the 
generalised partition function (\ref{genpart})  and the corresponding  R\'{e}nyi divergences. 
For the SSD and its higher spin generalization we find that the $\alpha$ dependence of the leading contribution to the \Re divergence takes the following universal form
\begin{equation} \label{exactrend}
D_\alpha(\rho_\mu||\rho_\beta) =
 \mu^2   \frac{  \sinh ( \frac{\pi  (\alpha -1) k \beta}{L} ) 
 \sinh ( \frac{\pi  \alpha  k \beta }{L}) }{ (\alpha-1)} f(\beta, L)  
 + O(\mu^4) 
~ .
\end{equation}
Here $k$ is the mode number of the deforming operator, $\Vs{-k}+\Vs{k}$ and $L$ is the size 
of the spatial circle. 
We also show that this dependence on $\alpha$ is remarkably identical to the \Re divergence of the simple harmonic oscillator deformed by a linear potential.  For the case of the simple harmonic oscillator, the 
\Re divergence in (\ref{exactrend}) is exact to all orders in $\mu$ and $\frac{\beta}{L}$ is replaced by 
$\omega \beta$ where $\omega$ is the frequency of the oscillator. 
We show the universality in \Re divergence is the 
  consequence of the same nature of commutation relations of the deforming operator with the undeformed Hamiltonian.   These operators  satisfy the Heisenberg algebra.

Armed with these results for R\'{e}nyi divergences, 
 we examine the  constraints for non-equilibrium transitions arising from 
 generalized second laws  put forward in \cite{Bernamonti:2018vmw}. 
 We see that for open systems,  indeed  monotonicity of 
 R\'{e}nyi divergences  place more constraints on allowed transitions than the conventional 
 second law.   These observations generalise those found in \cite{Bernamonti:2018vmw}. 
 Due to the analytical nature of our result we are able to translate the  constraints 
 from the generalised second laws  to  domains  in the deforming parameter $\mu$ for which 
 non-equilibrium transitions are allowed. These domains turn out to be more 
 restrictive than those allowed by the traditional second law. 
 
 The organization of the paper is as follows. 
 In section \ref{sec:rd}, we review the formulation of R\'{e}nyi divergences in terms of 
 the Euclidean quench and also discuss some of its properties. 
 In section \ref{analytdelta}, we evaluate R\'{e}nyi divergences for deformations due to
 scalar primaries of dimension $\Delta$. 
 In section  \ref{unifsour}, we consider situations with  $\mu$  constant and 
 the deforming operators are the $U(1)$ currents, stress tensor,  spin-3 
 currents and  evaluate the corresponding  
 R\'{e}nyi divergences  in closed form using conformal perturbation theory. 
 In section  \ref{inhodef},  we first consider the case of the simple Harmonic 
 oscillator deformed by the linear potential and show using the path integral
 that the R\'{e}nyi divergence is exactly given by (\ref{exactrend}). 
 We then evaluate the R\'{e}nyi divergences of  the SSD deformation and 
 its higher spin analogues using Hamiltonian perturbation theory and how that 
 to the leading order  the $\alpha$ dependence of the  R\'{e}nyi divergence is 
 given by the universal form in (\ref{exactrend}). 
In section \ref{gslt}, we use our results and show that generalized 
second laws for non-equlibrium transitions based on R\'{e}nyi divergences do 
indeed place additional constrains compared to the conventional second law. 
Section \ref{conclusion} contains our conclusions. 
Appendix \ref{appendA}  contains details of 
 conformal perturbation theory for spin-2 and spin-3 deformation. 
 Appendix  \ref{appendB}  and Appendix \ref{ham-alg}  contains evaluation of partition functions 
 and the details of the Hamiltonian perturbation theory for the SSD deformation and its
 higher spin generalizations.

\section{R\'{e}nyi divergence as  Euclidean quench}
\label{sec:rd}
In this section we review the evaluation of R\'{e}nyi divergences using its 
path integral representation as an Euclidean quench put forward in \cite{Bernamonti:2018vmw}. 
The R\'{e}nyi divergence between two density matrices $\rho$ and  $\rho_\beta$ is defined as
\begin{eqnarray}\label{defrenyd}
D_{\alpha}( \rho || \rho_\beta) = \frac{1}{\alpha - 1} \log \frac{\rm Tr [\rho^\alpha\rho_\beta^{1-\alpha}] }
{   \rm Tr {[\rho]}^\alpha ~\rm Tr{[\rho_\beta]} ^{1-\alpha} } ~. 
\end{eqnarray}
Here, $\rho$ is an excited state obtained by deforming the Hamiltonian corresponding to the 
thermal density matrix $\rho_\beta$. 
Therefore,    the path integral representation of  $\rho_\beta$  is the path integral of the 
conformal field theory with action $S_{\rm CFT}$ over  a cylinder of  circumference $\beta$. 
We write this formally 
\begin{equation}\label{path1}
\rho_\beta = \int [d\phi] \exp( - S_{\rm CFT} ( \phi) ) ~,
\end{equation}
\begin{figure}[!t]
	\centering
	\includegraphics[width=.6\textwidth]{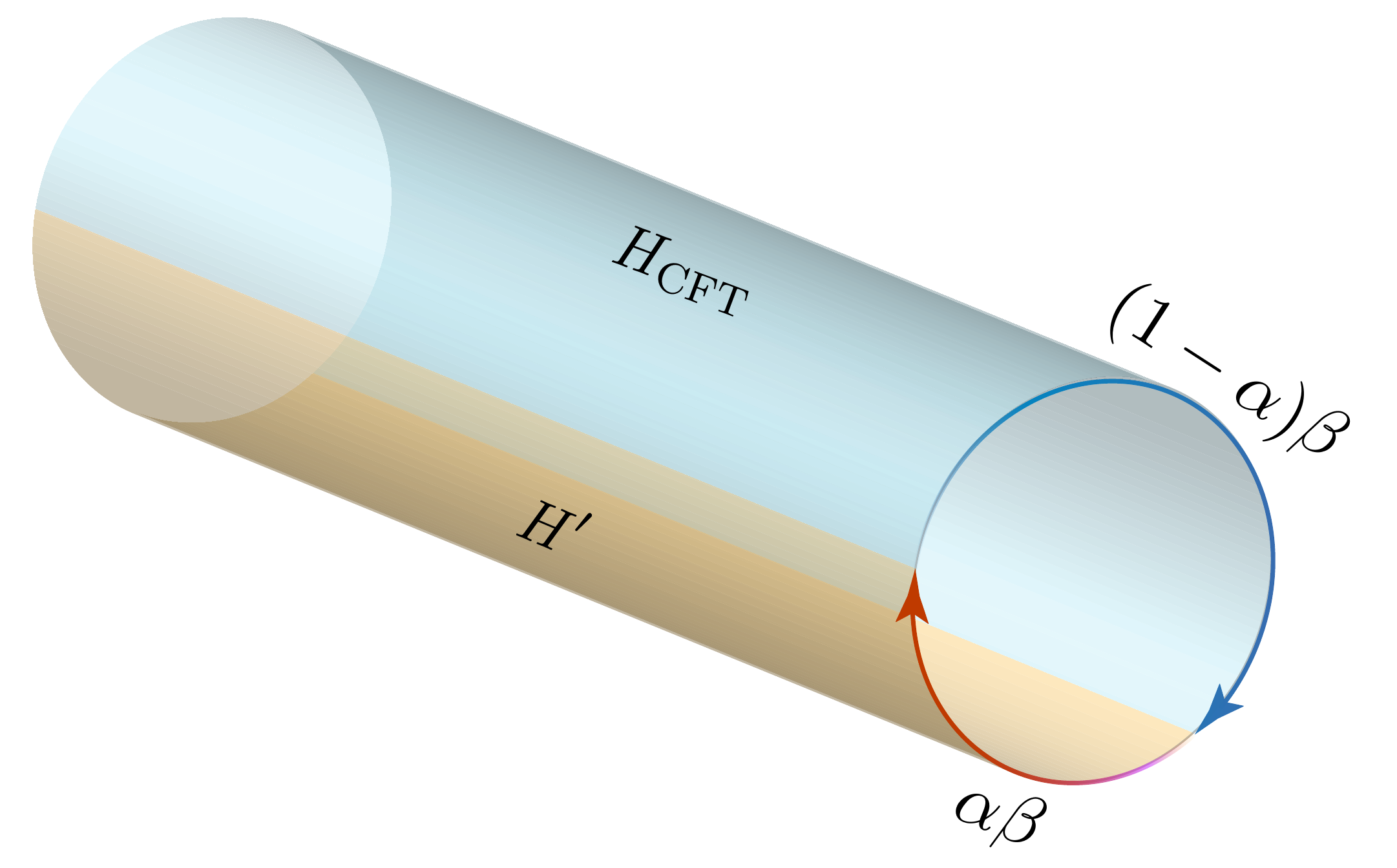}
	\caption{The Euclidean quench setup for \Re divergence.}
	\label{cyl}
\end{figure}
In the operator language the thermal density matrix can be written as
\begin{equation}
\rho_\beta =  e^{ - \beta H _{\rm CFT} }, 
\end{equation}
where  $H_{\rm CFT}$ is the Hamiltonian which generates translations along the thermal 
circle. 
The deformed density matrix $\rho$ is then defined by the path integral 
\begin{equation}\label{path2}
\rho = \int[d\phi] \exp \left[ - S_{\rm CFT} (\phi)  - 
\int dz d \bar z \, \mu \,{\cal O }_{\Delta}  ( z, \bar z)  \right],
\end{equation}
where ${\cal O }_{\Delta} ( z, \bar z ) $ is a conformal primary of weight $(h, \bar h) $ with 
$\Delta = h + \bar h$.  In the Hamiltonian form,  the excited state $\rho$  can be 
thought of a thermal state defined with respect to a new Hamiltonian 
given by 
\begin{equation}
H' = H_{\rm CFT} + \mu {\cal O }_{\Delta}~, 
\end{equation}
and therefore $\rho= e^{-\beta H'}$. 

Now, the path integral representation of the trace appearing in the numerator of 
the definition of the R\'{e}nyi divergence in (\ref{defrenyd}) can be summarised 
in  Fig.\,\ref{cyl}.
Essentially we sew the regions of the path integral given in (\ref{path1}) and (\ref{path2}). 
Thus the entire path integral can be thought of as performing the path integral   on the 
thermal cylinder with the 
Hamiltonian $H_{\rm CFT}$  deformed by the operator ${\cal O}$ coupled to a time dependent 
source given by 
\begin{equation}
\mu(\tau) = ( \theta(\tau) - \theta( \tau - \alpha\beta) ) \mu~.
\end{equation}
Here $\theta(\tau)$ refers to the Heaviside step function defined by 
\begin{equation}
\theta(\tau) =  \left\{ \begin{array}{ll} 
1 & \mbox{ for } \;  \tau \geq 0 ,\\
0 & \mbox { for } \; \tau <1.
\end{array} 
\right. 
\end{equation}
Since the deformation is turned on at $\tau= 0$ and turned off at $\tau= \alpha\beta$, this 
formulation of the R\'{e}nyi divergence is also called the Euclidean quench. 
The factors in the denominator in (\ref{defrenyd}) can be evaluated using the path integral 
(\ref{path1}) and (\ref{path2}).

It is clear  from this formulation for evaluating  \Re divergences  that we 
are restricted to the specific class of  excited states 
which are obtained by deforming the original Hamiltonian $H_{\rm CFT}$. 
As in \cite{Bernamonti:2018vmw} we wish to think of these as excited states within the theory 
in which the evolution is determined by $H_{\rm CFT}$. 
Also the path integral formulation allows the evaluation of the \Re divergence for 
the range $ 0 \leq \alpha <1$. 

The work   \cite{Bernamonti:2018vmw}  focused on deformation by 
relevant scalar primaries, 
in this paper we  first  revisit these deformations. We then focus  on
 deformation by conserved currents and also evaluate 
R\'{e}nyi divergences when  the perturbations in  $\mu$  have inhomogenous spatial profiles 
which specifically correspond to the sine-squared deformation (SSD).

Before we go ahead let us also recall a few properties R\'{e}nyi  divergences
\cite{van_Erven_2014}. 
One simple check   we perform in the next section is that 
we verify these properties are true for all  the cases 
for which we have  evaluated the R\'{e}nyi divergences. 
\begin{enumerate}[leftmargin=*]  \label{properties}
\item Positivity:  $D_\alpha \geq 0$. 
\item Monotonicity in $\alpha$:   $D_{\alpha_1} \geq D_{\alpha_2}$ for 
$\alpha_1 > \alpha_2$. 
\item Continuity in $\alpha$. 
\item Concavity:  $( 1-\alpha) D_\alpha$ is concave in $ \alpha$. 
\item Relation to relative entropy in the  $\alpha \rightarrow 1$ limit:  
\begin{equation}
D_1(\rho||\rho_\beta)  =  {\rm Tr} ( \rho \log \rho) - {\rm Tr} ( \rho  \log \rho_\beta) .
\end{equation}
Furthermore it can be shown that  relative entropy can be written in terms of the 
differences in free energies  of the density matrices $\rho$ and $\rho_\beta$
\begin{equation}\label{relalpha1}
D_1(\rho||\rho_\beta)  =  \beta ( F(\rho) - F(\rho_\beta) ) .
\end{equation}
It is important to realise that in this equation, the free energy of the excited state 
is given by 
\begin{equation} \label{defFrho}
F( \rho) = {\rm Tr} [\hat  \rho H] +  {\rm Tr}  [\hat \rho \log \hat \rho] .
\end{equation}
Note that   the expectation value of the energy is that of the undeformed  Hamiltonian 
in the excited state $\rho$ and $\hat \rho$ is the normalized density matrix given by 
\begin{equation}
\hat \rho =  \frac{\rho}{{\rm Tr}[ \rho]} 
\end{equation}
The definition of 
$F(\rho_\beta) $ is the same as that of (\ref{defFrho}) with $ \rho$ replaced by 
$ \rho_\beta$. 
\end{enumerate}

It is useful to obtain an expression for the free energy of excited state in terms of 
the partition function the excited state.
Consider the un-normalized density matrix of the excited state which is given by 
 \begin{equation} \label{exitedstate}
   \rho = e^{ - \beta ( H - \mu {\cal O}) }.
   \end{equation}
   where ${\cal O}'$ is the  operator which corresponds to the deformation in
   the Hamiltonian picture. 
   From this  definition (\ref{exitedstate})  it is easy to see that 
   \begin{equation} \label{expH}
   {\rm Tr}  [H \hat \rho] = \frac{\mu}{\beta} \frac{\partial}{\partial\mu} \log ( {\rm Tr} [\rho]) 
   - \frac{\partial}{\partial\beta} \log ( {\rm Tr} [\rho]) , 
   \end{equation}
   Note that  on the LHS  $\hat \rho$ refers to the  normalized 
   density matrix and $H$ refers to the undeformed Hamiltonian. 
    We also have the equation
   \begin{equation} \label{entropy} 
   \frac{1}{\beta} {\rm Tr} ( \hat \rho \log \hat \rho) =
     -T \frac{\partial}{\partial T} ( T \log {\rm Tr} [\rho] ) .
   \end{equation}
   Both the equations (\ref{expH}) and (\ref{entropy}) are functions of
   the deformed partition function $\cZ[\mu] = {\rm Tr } [\rho]$. 
   The free energy of the 
    unperturbed CFT is given by 
    \begin{equation}\label{freunpert}
    F(\rho_\beta) = -\frac{1}{\beta} \log {\rm Tr}  [\rho_\beta] = - \frac{\pi c L}{6 \beta^2} . 
    \end{equation}
    Now using (\ref{expH}), (\ref{entropy})  and (\ref{freunpert}), it can be seen that the
    difference in free energy (\ref{defFrho}) can be expressed
    entirely in terms of the  partition function of the excited state  
    $\cZ[\mu]= \tr[\rho]$. Therefore, the relative entropy is  
    \begin{eqnarray} \label{D1free}
    D_1(\rho||\rho_\beta) &=&  \beta ( F(\rho) - F(\rho_\beta) ), \\ \nonumber
    &=& \mu  \frac{\partial}{\partial\mu} \log  {\rm Tr} [\rho]
  - \log {\rm Tr} [\rho]
    + \frac{\pi c L}{6\beta}. 
    \end{eqnarray}

Additionally, using $\rho_\beta=e^{-\beta H}$ and $\rho=e^{-\beta (H+\mu \mathcal{O})}$ and the Golden-Thompson inequality, $\tr[e^{A}e^{B}] \geq \tr[e^{A+B}]$, we obtain that the \Re divergence \eqref{defrenyd}  is bounded from above by the following combination of the deformed and undeformed partition functions 
\begin{eqnarray}\label{bound}
\hspace{-1cm}
D_{\alpha}( \rho || \rho_\beta) 
\leq  \frac{1}{\alpha-1} \log \frac{\cZ[\alpha\mu]}{\cZ[\mu]^\alpha \cZ[0]^{1-\alpha}},\quad \text{where } \cZ[\mu] \equiv \tr [{e^{-\beta (H+\mu\mathcal{O})}}]. 
\end{eqnarray}
for $0 \leq \alpha < 1$. This inequality is saturated if and only if the deforming operator $\mathcal{O}$ commutes with the Hamiltonian.

\section{Deformations by  scalar primaries} \label{analytdelta}

In this section we revisit the evaluation of the R\'{e}nyi divergences for 
which the excited state is that obtained by 
a scalar  primary  ${\cal O }$ of weight $\Delta$ to the quadratic order in the amplitude. 
This  problem was addressed in 
 \cite{Bernamonti:2018vmw} where the integral resulting from conformal perturbation theory
 was evaluated numerically  and for $\Delta <\frac{1}{2}$, the integral was written in terms of a
 series. 
 In this section we show that the R\'{e}nyi divergences can be written as a series for 
  all $\Delta$.  Further more for the case of $\Delta =1$, we obtain the 
  R\'{e}nyi divergences in closed form.
   
Let us consider the Euclidean quench by a scalar primary, the action is deformed by 
\begin{equation}
S_{\rm CFT} \mapsto S_{\rm CFT} + 
\mu \int d^2 w [ \theta ( \tau) - \theta( \tau - \alpha \beta) ] {\cal O} ( w, \bar w)  ~.
\end{equation}
The  leading contribution to the quenched partition function
begins at the quadratic order in $\mu$ and  is given by 
\begin{equation}
\log {\rm Tr} ( \rho^\alpha \rho^{(1-\alpha)} ) = \frac{\pi cL}{6 \beta} + 
 \mu^2 \int_0^{\alpha\beta} \int_0^{\alpha\beta}  d\tau_1   d\tau_2  
\int_{-\infty}^{\infty} \int_{-\infty}^\infty 
d\sigma_1 d\sigma_2 \langle  {\cal O} (w_1, \bar w_1)  {\cal O} ( w_2, \bar w_2) \rangle
+ \cdots 
\end{equation}
The normalized  two point function of the primary on the cylinder is given by 
\begin{equation}
 \langle  {\cal O} (w_1, \bar w_1)  {\cal O} ( w_2, \bar w_2 ) \rangle =  \left(  \frac{\pi}{\beta} \right)^{2\Delta} 
\left(  \frac{1}{ \sinh \frac{\pi }{ \beta} ( w_1 - w_2)\sinh \frac{\pi}{\beta} ( \bar w_1 - \bar w_2) }
\right)^{ \Delta}. 
\end{equation}
After a change of variable and  simple manipulations we obtain 
\begin{equation}
\log {\rm Tr} ( \rho^\alpha \rho^{(1-\alpha)} ) = \frac{\pi cL}{6 \beta} 
+ \frac{\pi L }{\beta} \left( \frac{\pi}{\beta} \right)^{ (2 \Delta -4) }   I ( \Delta, \alpha)  + \cdots ,
\end{equation}
where
\begin{equation}\label{fundaint}
I (  \alpha, \Delta) =  2^{\Delta - 2}  \int_0^{2\pi \alpha }  dp  \int_0^\infty d \sigma 
\frac{  2 (2\pi \alpha - p ) }{(  \cosh \sigma - \sqrt{1-\epsilon^2}  \cos p )^{\Delta} }. 
\end{equation}
We  have cut off the second  spatial integral to a size of length $L$. After a change of variables
we have performed the integrals over one of the temporal directions. 
The integral is regulated  following \cite{Bernamonti:2018vmw}. 
Using (\ref{defrenyd}), the R\'{e}nyi divergence is then given by 
\begin{equation}\label{arbdeltarenyi}
D_\alpha( \rho||\rho_\beta)  = \mu^2 \frac{\pi L}{\beta}  \left(\frac{\pi}{\beta} \right)^{ (2 \Delta -4) } 
\frac{ I ( \alpha, \Delta) - \alpha  ( 1, \Delta) }{ \alpha - 1} + O(\mu^3) . 
\end{equation}

We now evaluate the integral  in (\ref{fundaint}) by expanding the denominator 
using the binomial expansion 
\begin{eqnarray}
I (  \alpha, \Delta) =   2^{\Delta - 2} 
\sum_{n=0}^\infty \frac{\Gamma( \Delta + n) }{\Gamma( \Delta) \Gamma( n+1) }
( 1- \epsilon^2)^{\frac{n}{2} }  \int_0^{2\pi \alpha} dp
 \int_0^\infty  d\sigma  \frac{ 2( 2\pi \alpha - p)  (\cos p)^n }{ (  \cosh   \sigma)^{n+\Delta} }.
 \nonumber \\
\end{eqnarray}
For a finite $\epsilon$, this expansion is uniformly convergent and we have interchanged the 
sum and the integral. 
Now the functions that occur in the integral are elementary and we can integrate term by term. 
We obtain
\begin{eqnarray} \label{sumformint}
I(\alpha,\Delta)&=&I(\alpha,\Delta)_{\text{even}}+I(\alpha,\Delta)_{\text{odd}}, \\
\nonumber
I(\alpha,\Delta)_{\text{even}}&=&  2^{2\Delta - 3}  \sum_{m=0}^{\infty}\Gamma_m(\Delta)\, (1-\epsilon^2)^m\left[\sum_{k=0}^{m-1}\binom{2m}{k}s(m-k)+\binom{2m}{m}2\pi^2\alpha^2\right], \\ \nonumber
I(\alpha,\Delta)_{\text{odd}}&=&   2^{2\Delta - 3}  \sum_{m=1}^{\infty}  \Gamma_{m-\frac{1}{2}}(\Delta)\, (1-\epsilon^2)^{\frac{2m-1}{2}}\sum_{k=0}^{m-1}4\binom{2m-1}{k}s(m-k-\tfrac{1}{2}), 
\end{eqnarray}
where we have defined the quantities $s(x)=\sin^2[2\pi \alpha x]/x^2$ and $\Gamma_m(\Delta)=\Gamma(m+\Delta/2)^2/(\Gamma(2m+1)\Gamma(\Delta))$. 

\begin{figure}[h]
  \centering
  \begin{subfigure}[b]{0.49\linewidth}
    \includegraphics[height=.8\textwidth, width=1\textwidth]{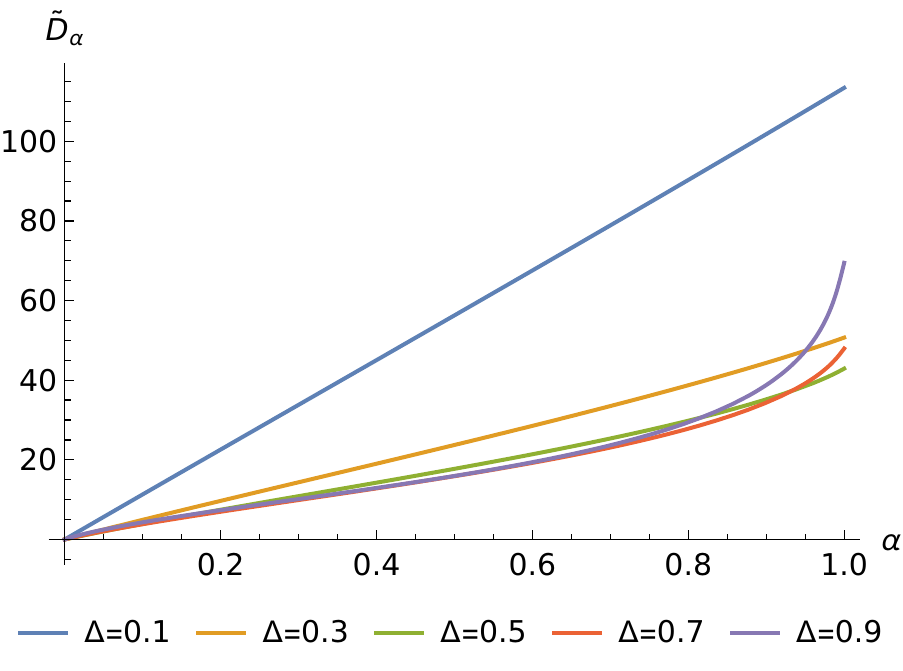}
    \caption{} \label{vardel1}
  \end{subfigure}
  \hspace{1pt}
  \begin{subfigure}[b]{0.49\linewidth}
    \includegraphics[height=.8\textwidth, width=1\textwidth]{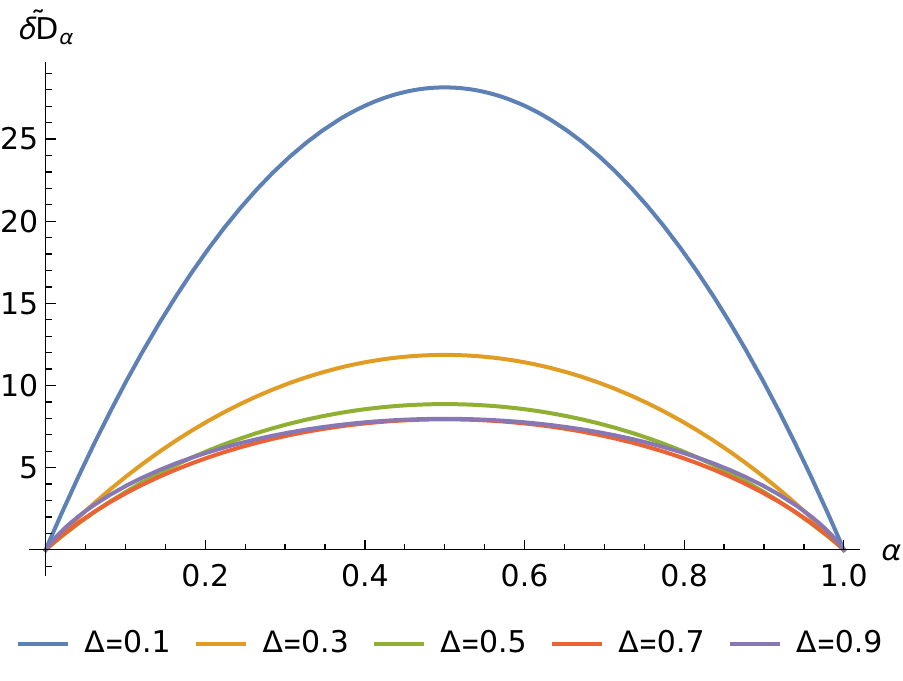}
    \caption{} \label{vardel2}
  \end{subfigure}
  \caption{(Left) Plot of $\tilde{D}_{\alpha}=[\mu^2 \frac{\pi L}{\beta}(\frac{\pi}{\beta} )^{ (2 \Delta -4) }]^{-1} D_{\alpha}(\rho_1||\rho_2)$ vs $\alpha$. (Right) Plot of $(1-\alpha)  \tilde{D}_{\alpha}(\rho_1||\rho_2)$ vs $\alpha$ .  In both the plots the values of $\Delta$ for the curves decrease from $0.1$ to $0.9$ from top to bottom.  We have taken 1000 terms in the series
given in (\ref{sumformint}). }
\end{figure}

The integral is convergent for $0<\Delta <1$, therefore we can set $\epsilon =0$ and 
we can plot the R\'{e}nyi divergences given by (\ref{arbdeltarenyi}) and (\ref{sumformint})
In figures \ref{vardel1} and \ref{vardel2}, we have plotted these  R\'{e}nyi divergences
for various values of $\Delta$ for 1000 terms in the series. 
It can be seen that the properties discussed in 
section \ref{sec:rd} hold. 
We checked the convergence of the series by evaluating the R\'{e}nyi divergences by taking 
different 
the number of terms in the series.  For instance changing the number of terms from 
800 to 1000 results in changes of less than equal to  $0.04$ to the R\'{e}nyi divergences. 
The changes  kept decreasing as we increased the number of terms. 
We have also seen by  using the ratio test to a series which bounds the one given in
 (\ref{sumformint}),  that the sum is convergent for any $\Delta$ given a finite $\epsilon$. 
 In \cite[Appendix A, equation A.19]{Bernamonti:2018vmw}  a series form of the integral 
 was obtained for strictly $\epsilon =0$ and was valid for $0<\Delta <1/2$, the series 
 we have incorporates the cut off and is convergent for any $\Delta$ given a finite $\epsilon$.

\begin{figure}[h!]
  \centering
  \begin{subfigure}[b]{0.45\linewidth}
    \includegraphics[width=\linewidth]{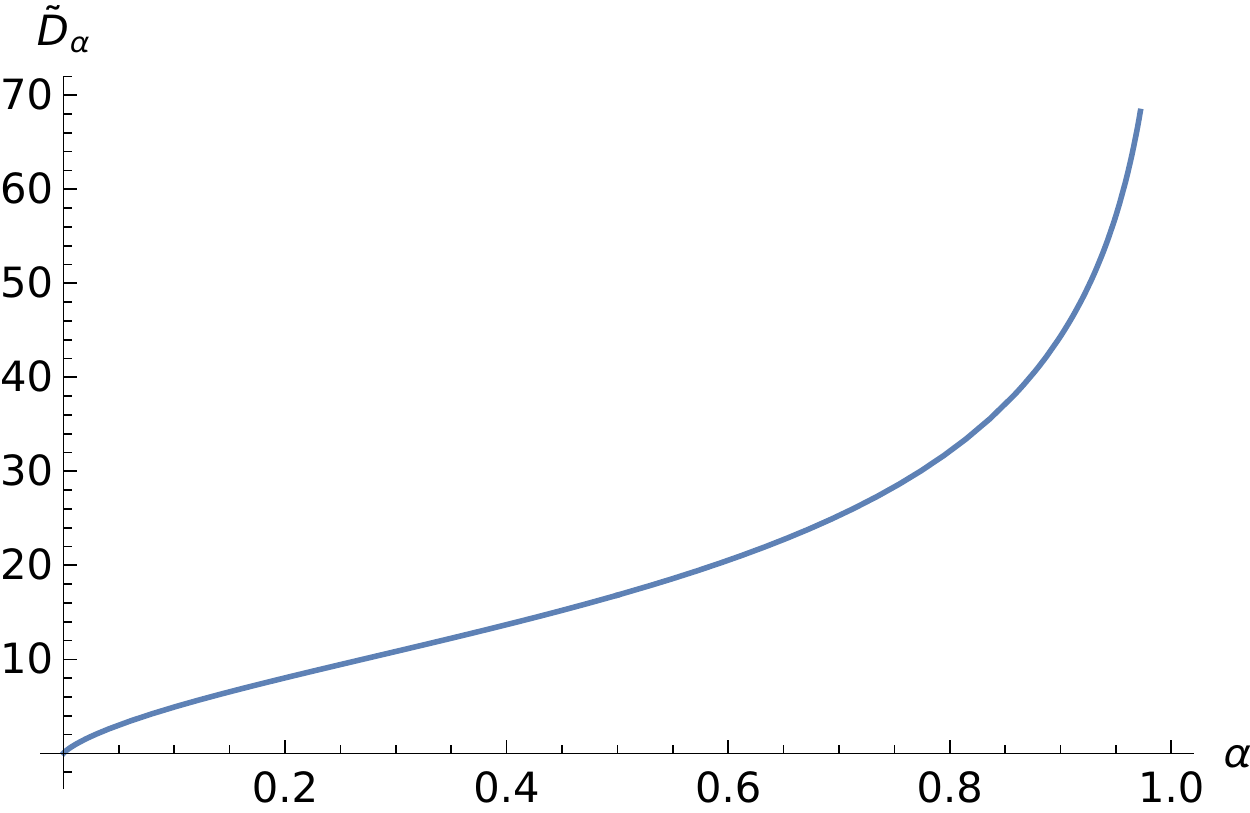}
    \caption{} \label{closed1}
  \end{subfigure}
  \hspace{2pt}
  \begin{subfigure}[b]{0.45\linewidth}
    \includegraphics[width=\linewidth]{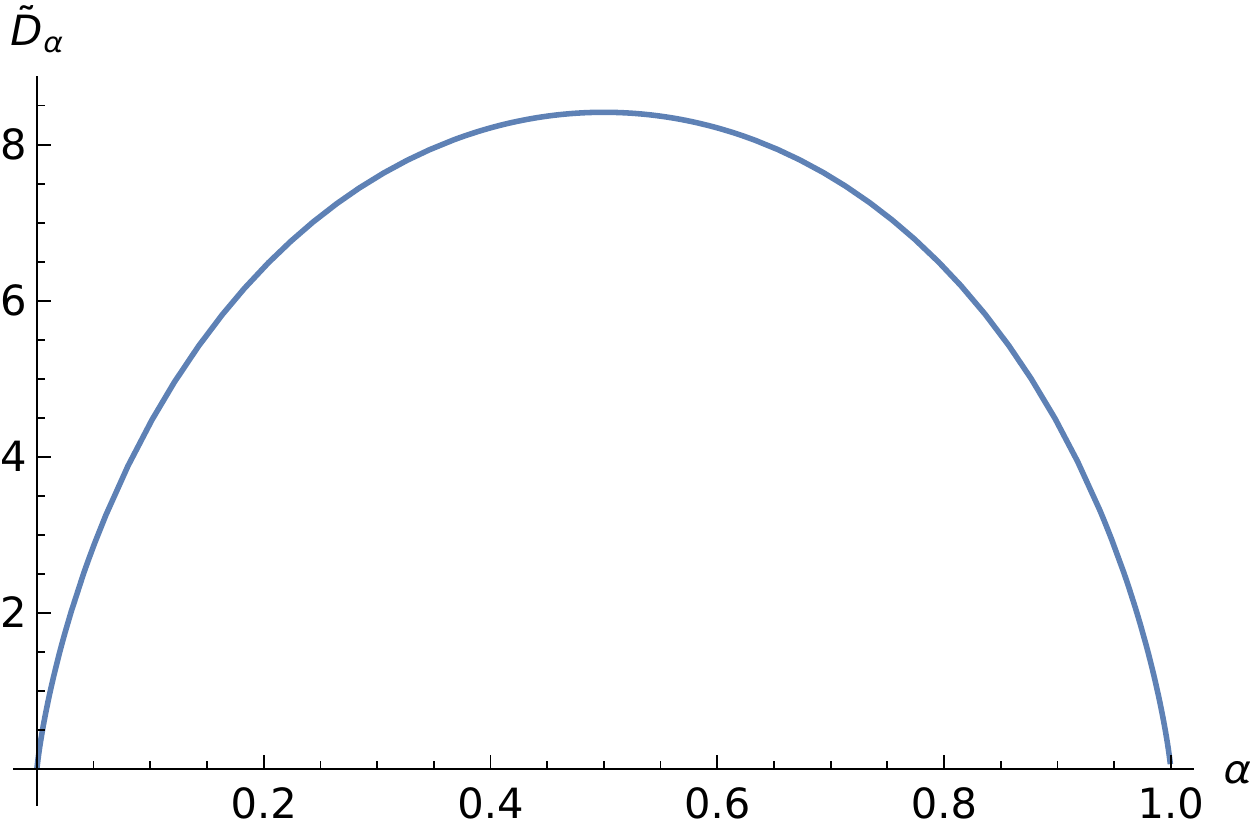}
    \caption{} \label{closed2}
  \end{subfigure}
  \caption{(Left) Plot of $  \tilde{D}_{\alpha}=\frac{1}{\mu^2(\beta L/\pi)}D_{\alpha}(\rho_1||\rho_2)$ vs $\alpha$. (Right) Plot of $(1-\alpha)  \tilde{D}_{\alpha}(\rho_1||\rho_2)$ vs $\alpha$ }
\end{figure}

As a check of our  method of integration, we consider the case of $\Delta =1$, for which 
the integral can be obtained in terms of known functions for $\epsilon =0$. 
We perform the spatial integral first 
\begin{eqnarray}
 \int_0^\infty d\sigma \frac{1}{ \cosh \sigma  - \cos p }
= 2\frac{ \tan^{-1} ( \cot \frac{p}{2} ) }{\sin p} 
= \frac{\pi - p}{ \sin p} .
\end{eqnarray}
Note that we have chosen the branch so that the resulting integral is positive for $p$ in the 
range $0$ to $2\pi \alpha$. 
The temporal integral is given by 
\begin{equation}
I( \alpha, 1) = \int_0^{2\pi \alpha}  F(p)  dp, \qquad \qquad F(p) = \frac{ ( 2\pi \alpha - p ) ( \pi - p) }{\sin p} .
\end{equation}
This integrand is obviously divergent at $p = 0$, 
we can regulate this integral by shifting the contour and considering 
\begin{equation}
I(\alpha, 1) = \lim_{\epsilon \rightarrow 0 } \int_0^{2\pi \alpha}
 \frac{1}{2}  (  F( p + i \epsilon ) + F( p - i \epsilon ) )  dp .
\end{equation}
Performing this integral and then evaluating the R\'{e}nyi divergence we obtain 
\footnote{We have also regulated the integral by placing a cut off $p =0$ and 
$I(\alpha, 1) = \lim_{\Lambda \rightarrow 0} \int_\Lambda^{2\pi \alpha} F(p) $
and obtain the same result as given in (\ref{delta1result}).  }
\begin{eqnarray}\label{delta1result}
D_{\alpha} ( \rho, \rho_\beta) &=&  
\mu^2\frac{\beta L}{\pi}\frac{1}{4(\alpha-1)}\left[ 
i \pi^3(2\alpha-1)+2 i \pi(2\alpha-1)\left(-4{\rm Li}_2(e^{2\mathrm{i}\pi\alpha})+
{\rm Li}_2(e^{4 i \pi\alpha})\right) \right. 
\nonumber \\
&&\hspace{3cm}\left. +16{\rm Li}_3(e^{2i\pi\alpha})-2{\rm Li}_3(e^{4{i}\pi\alpha})-14\zeta(3)  \right] .
\end{eqnarray}
 Note that  though the  result in (\ref{delta1result}) is not manifestly  real, it can be seen 
 real  by examining the numerical values of the function in  
 the figures \ref{closed1} and \ref{closed2}. 

%

\begin{figure}[t!]
  \centering
  \begin{subfigure}[b]{.49
 \linewidth}
    \includegraphics[height=.8\textwidth, width=1\textwidth]{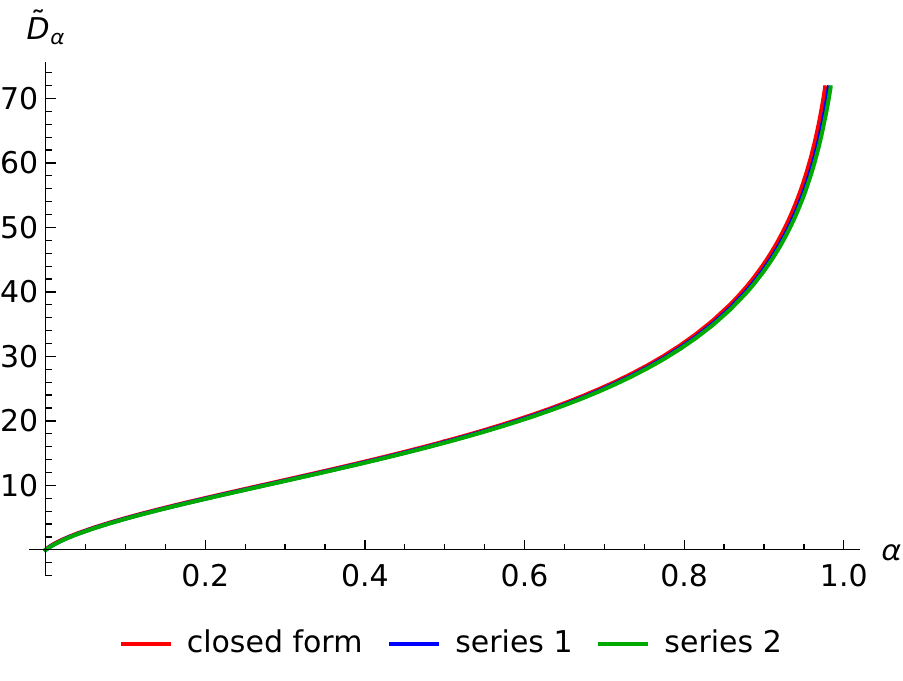}
    \caption{} \label{comparplot}
  \end{subfigure}
  \hspace{1pt}
  \begin{subfigure}[b]{.49\linewidth}
    \includegraphics[height=.8\textwidth, width=1\textwidth]{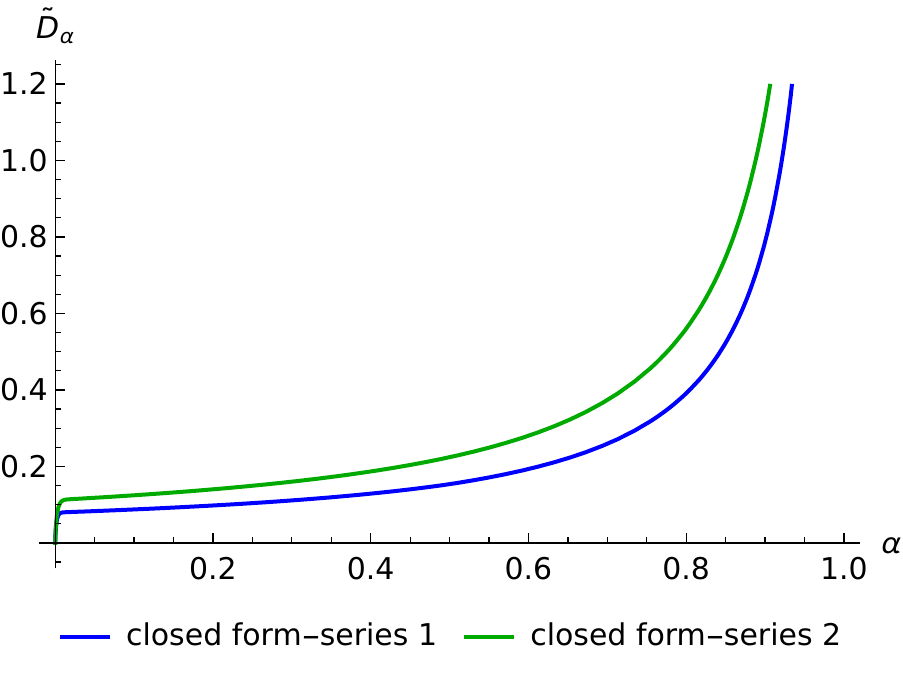}
  \caption{}    \label{efficiency} 
  \end{subfigure}
  \caption{(Left) Plots of $ \tilde{D}_{\alpha}=\frac{1}{\mu^2(\beta L/\pi)}D_{\alpha}(\rho_1||\rho_2)$ 
  vs $\alpha$ for $\Delta=1$. The closed form in  (\ref{delta1result}) 
  is shown as a continuous curve,  `series 1'
  constructed in this paper 
  and `series 2'  of \cite{Bernamonti:2018vmw} are as dashed curves of yellow and green respectively. 
  We have taken 1000 terms from each of the series. 
  (Right) Plots of $\tilde{D}_{\alpha}$ for  differences: (closed form  $-$  `series 1') and 
  (closed form $-$ `series 2').  It is seen that the series \eqref{sumformint} constructed  has smaller differences to the 
  closed form result. }
\end{figure}

We can use the closed form answer 
at $\Delta =1$   given in (\ref{delta1result}) 
to check the efficiency of the series obtained in 
 \cite[Appendix A, equation (A.19)]{Bernamonti:2018vmw} extrapolated to $\Delta =1$. 
 First we note the normalizations of the integral in (\ref{fundaint}) and  that in 
 \cite{Bernamonti:2018vmw} are related by $I(\alpha, \Delta)_{\rm ours} = 2^{\Delta - 3} I_{\rm theirs}$. 
 Let us call the  R\'{e}nyi divergence obtained by the series  for $\Delta =1$  in (\ref{sumformint}) as 
 `series 1' and that obtained in \cite{Bernamonti:2018vmw} after taking into account the 
 normalizations, `series 2'.
 In figure \ref{comparplot} we have plotted   the closed form result for the 
 R\'{e}nyi  divergences 
 from (\ref{delta1result}) 
 and `series 1' and `series 2' for 1000 terms in the series.
   In figure \ref{efficiency} we plot the differences between the 
 closed from result and `series 1' and `series 2' taking  1000 terms.
 As expected the closed form result  is always larger, however we can see the `series 1' obtained
is this paper has smaller differences  and  therefore converges faster  to the closed form answer.

As another consistency check, we consider the limit $\alpha \to 1$ of the integral $I(\alpha, \Delta)$ with 
$\Delta <1$ for covergence so that we can set $\epsilon =0$. 
The integral then becomes 
\begin{equation}\label{integrala}
I ( 1, \Delta) = 2^{\Delta-2}  \int_0^{2\pi} dp\int_0^\infty \frac{ 2 ( 2\pi -p)}{(\cosh \sigma - \cos p)^\Delta} .
\end{equation}
We can now compare  this  integral  done in  \cite[equation (13)]{Berenstein:2014cia}    for 
$d=2$ (also obtained for conformal perturbation theory)
\begin{eqnarray} \label{integralb}
C_{\Delta} &=& \int_{-\infty}^\infty d\tau  \int_0^\pi  d\theta  ~ {\rm Vol}  (S_0 )~
\frac{1}{ 2^{\Delta} ( \cosh\tau - \cos \theta)^\Delta }. 
\end{eqnarray}
This integral was evaluated with the result
\begin{eqnarray}
C_{\Delta} 
&=& \pi^{\frac{3}{2} } 2^{1-\Delta} 
\left[\frac{\Gamma(1-\Delta)\Gamma(\frac{\Delta}{2})}{\Gamma(1-\frac{\Delta}{2})^2\Gamma(\frac{1}{2}+\frac{\Delta}{2})}\right]. 
\end{eqnarray}
Now using the following identity for every integer $n$ 
\begin{equation}
\int_0^{2\pi} dp (2\pi -p) (\cos p)^n = 2\pi \int_0^\pi d\theta (\cos p )^n .
\end{equation}
 and comparing the integrals in  (\ref{integrala}) and (\ref{integralb}),  we expect 
\begin{equation}
I( 1, \Delta) = 4^{\Delta -1} \pi C_\Delta~.
\end{equation}
Let us now evaluate $I(1, \Delta)$.  
From the series representation  of the integral \eqref{sumformint} is easy to see that the result is
\begin{eqnarray}\label{integralc}
\hspace{-.7cm}I(1, \Delta) &=& 2^{2\Delta -2} \sum_{m=0}^\infty 
\frac{\Gamma\left(\frac{2m+\Delta}{2}\right)^2}{\Gamma(2m+1)\Gamma(\Delta)} \binom{2m}{m}\pi^2
=
 \frac{2^{(2\Delta-2)}\pi^2 \Gamma(1-\Delta)\Gamma(\frac{\Delta}{2})^2}{\Gamma(\frac{2-\Delta}{2})^2\Gamma(\Delta)}.
\end{eqnarray}
Evaluating the ratio  from (\ref{integralb}) and (\ref{integralc}) we indeed see
that  the ratio is precisely as expected and is given by 
\begin{equation}
\frac{I( 1, \Delta) }{C_\Delta} = 4^{\Delta -1}\pi~.
\end{equation}
We also see that, as expected  for  $\alpha=1$ the result in (\ref{integralc}) is $2^{\Delta -3}$ times the 
result evaluated using the series in \cite{Bernamonti:2018vmw}.

\section{Homogenous deformations by conserved currents}
\label{unifsour}

In this section we study R\'{e}nyi divergence when the source $\mu$ is spatially uniform and 
the operator ${\cal O}$ are conserved currents.  Thus the deformation we consider 
are 
\begin{equation}
S_{\rm{CFT} } ( \phi) \mapsto S_{\rm{CFT}}  (\phi) + \mu \int d^2 z  \left( 
{\cal O}(z)   + {\bar  {\cal O} } ( \bar z)  \right)
\end{equation}
In the subsequent subsections we 
 consider  ${\cal O }(z)$ to be spin-1, spin-2, spin-3 currents and show that the 
 R\'{e}nyi divergences for these deformations can be obtained in closed form to all 
 orders in conformal  perturbation theory. 
 Note that all these operators have dimensions $\Delta\geq 1$. 
 It was noted in \cite{Bernamonti:2018vmw}, that 
 for primaries with conformal dimensions $\Delta \geq 1$ and $h = \bar h$, there are
 divergences in conformal perturbation theory. 
 On the contrary, we will see that  in the case of holomorphic currents   the prescription developed in 
 \cite{Datta:2014ska,Datta:2014uxa,Datta:2014zpa}  to evaluate the integrals that occur in 
 conformal perturbation theory  leads to  finite answers for 
 R\'{e}nyi divergences. 

\subsection{$U(1)$ current }
\label{subsec:u1}
Let us consider the following deformation of the CFT action by a $U(1)$ current
\begin{equation}
S_{\rm{CFT} } ( \phi) \mapsto S_{\rm{CFT}}  (\phi) + \mu \int  d^2 z  \left( 
{j}(z)   + {\bar  j}  ( \bar z)  \right) .
\end{equation}
The $U(1)$ current admits the following OPE\footnote{
Note that the presence of $i^2 = -1$ in the  normalization of the OPE  is due to the 
fact that the chemical potential  for the global $U(1)$ charge 
in the Euclidean Lagrangian is purely imaginary. 
We have absorbed this factor of $i$ in the normalization of the currents. }
\begin{equation}
j( z) j(w)  \sim - \frac{\kappa}{ ( z- w) ^2},
\end{equation}
where $\kappa$ is the level of the current algebra. 
We focus on the holomorphic perturbation. The analysis for the anti-holomorphic perturbation 
proceeds identically and we can then add its contribution. 
The correlators of the anti-holomorphic   $U(1)$ current with the holomorphic current 
factorise except for contact-terms which involve delta functions. 
 In the prescription of doing the resulting integrals 
we ignore all  such contact-terms (see \cite{Datta:2014zpa} for details).   Therefore  the 
contribution from the anti-holomorphic sector can be taken into account by doubling
the contribution of the holomorphic sector. 

To proceed let us study the leading correction to the partition function:
  we evaluate 
\begin{equation}
{\rm Tr} ( \rho) = \cZ[\mu]  = \int [D\phi] \exp \left( - S[\phi] - \mu \int d^2 w  \,j (w)   \right) ~. 
\end{equation}
This path integral is evaluated on a cylinder of length $L$ with circumference $\beta$. 
$w$ is the co-ordinate on the cylinder which is related to the co-ordinate on the plane
by the map
$
z = e^{ { 2\pi w}/{\beta}} ~. 
$
Expanding  the partition function as a series in  $\mu$, 
we obtain
\begin{equation}\label{expanmu}
\frac{ \cZ[\mu]}{\cZ[0] } = 1  - \mu \int d^2 w  \langle  j( w)   \rangle  + 
\frac{\mu^2}{2} \int d^2 w_1 d^2 w_2 \langle j(w_1) j(w_2) \rangle + \cdots .
\end{equation}
Here the expectation values refer to correlators   evaluated on the cylinder. 
Since $j(w)$ is  a conformal primary, is expectation value on the cylinder vanishes. 
The two-point function of the currents  on the cylinder is given by  
\begin{equation}
 \langle j(w_1) j(w_2) \rangle  =   -\kappa \left(  \frac{\pi}{\beta} \right)^2
  \frac{1}{ \sinh^2 \frac{\pi}{\beta} ( w_1 - w_2)  }. 
  \end{equation} 
  To perform the integral we adopt the prescription developed in 
  \cite{Datta:2014ska,Datta:2014uxa,Datta:2014zpa}. 
  The spatial integrals are done  first followed by the temporal integrals. 
  Furthermore the last spatial integral is cut off by the length $L$  of cylinder and it gives 
  rise to the extensivity of the free energy. 
  Let us see this in detail for the  $\mu^2$ term in (\ref{expanmu})
\begin{align} \label{integralj}
 -\frac{1}{2} \mu^2 \kappa   \int_0^\beta d\tau_2  \int_{-L/2}^{L/2}   d \sigma_2 
   \int_0^\beta d\tau_1 
 \int_{-\infty}^\infty  d\sigma_1 &  \left(  \frac{\pi}{\beta} \right)^2
  \frac{1}{ \sinh^2 \frac{\pi}{\beta} (  \sigma_1 - \sigma_2 + i ( \tau_1 - \tau_2)  )  }
  \nonumber \\
  &  
  =  \pi  \kappa \mu^2 L\beta~.
  \end{align}
  Here we have used $w = \sigma + i \tau$ where $\sigma$ is the non-compact spatial 
  direction and $\tau$, the compact temporal direction. 
  Notice that in this prescription the integrals over the temporal directions are trivial. 
  This is because performing the spatial integral first picks out the conserved charge. 
  As we will see, 
  the fact that the temporal integrals  are trivial plays an important role for evaluating 
  when this deformation occurs as a Euclidean quench. 
  
  For the $U(1)$ current, all  $n$-point functions (on the cylinder) where $n$ is odd vanish, and 
  for even $n$ they factorise into   two-point functions. 
  Using this and the integral in (\ref{integralj}) we can exponentiate all the terms in the expansion 
  (\ref{expanmu}) to obtain 
  \begin{eqnarray}\label{part1} 
  \cZ[\mu]  &=& \cZ[{0}] \times \exp{  (  2 \pi \kappa  \mu^2  L \beta ) } . 
  \end{eqnarray}
  Here we have included the contributions from the anti-holomorphic sector. 
  The undeformed CFT partition function is easily evaluated  using the high temperature limit, 
  since the length of the cylinder is 
  large. Thus we have 
  \begin{equation}\label{part2}
  {\rm Tr} [\rho_\beta] = \cZ[0] =  \exp\left( \frac{\pi c L}{ 6\beta}  \right),
  \end{equation}
  where $c$ is the central charge of the CFT. 
  Combining (\ref{part1}) and (\ref{part2}), we  obtain at high temperatures
  \begin{equation} \label{part3}
  {\rm Tr} [\rho] = 
  \cZ[\mu]   = \exp{ \left( \frac{\pi c L}{ 6\beta} +  2\pi \kappa  \mu^2  L \beta \right) } . 
  \end{equation}
This result is also consistent with the modular transformation of the torus partition function in the Hamiltonian formalism. The grand-canonical partition function at low temperatures receives dominant contribution from the uncharged vacuum
\begin{align}\label{chZ}
\cZ(\tau,\nu)=\tr \left[ q^{L_0-c/24} \bq^{\bL_0-c/24} y^{J_0} \by^{\bar{J}_0} \right], \qquad \cZ_{\rm low}(\tau,\nu)\approx \exp\left(\frac{\pi c \beta}{6L}\right),
\end{align}
here, $q=e^{2\pi i \tau}$, $y=e^{2\pi i \nu}$ and we chose $\tau=i\beta/L$. Under modular transformations, the above partition function transforms as\footnote{The anti-holomorphic dependence is being suppressed for brevity.}
\begin{align}
\cZ\left(\frac{a\tau +b}{c\tau+d},\frac{\nu }{c\tau+d}\right)= \exp\left[\frac{ic\pi \kappa \nu^2}{c\tau +d }-\frac{ic\pi \kappa \bar\nu^2}{c\btau +d }\right]~\cZ(\tau,\nu).
\end{align}
Using the S-modular transformation, we can obtain the high temperature behaviour of the partition function 
\begin{align}
\cZ_{\rm high}(\tau,\nu) \approx \exp{ \left( \frac{\pi c L}{ 6\beta} +  {2\pi \kappa  \nu^2  L \over \beta} \right) }.
\end{align}
Upon identifying $\nu = \mu \beta$, this agrees precisely with the result \eqref{part3} obtained using the Lagrangian/path-integral formalism. This provides a verification of the integration prescription being used here. 
  
  Let us now turn to the Euclidean quench, the deformation is now given by 
  \begin{equation}
  S_{\rm{CFT} } ( \phi) \mapsto S_{\rm{CFT}}  (\phi) + \mu \int  d^2 z
  [ \theta( \tau) - \theta ( \tau - \alpha\beta) ]  \left( 
{j}(z)   + {\bar j}  ( \bar z)  \right) .
\end{equation}
Evaluating the path integral with this action results in  ${\rm Tr}  [\rho^{\alpha} \rho_\beta^{1-\alpha}] $. 
It is easy to carry out the integrations using the prescription just described. Since the temporal integrals are trivial,  performing the two temporal integrals in the case of the Euclidean quench  just results in  the replacement
$\beta^2 \mapsto  \alpha^2 \beta^2$. 
Therefore the result for the partition function of the quenched theory is given by 
\begin{eqnarray} \label{part4}
{\rm Tr} ( \rho^{\alpha} \rho_\beta^{1-\alpha} ) =  \exp{ \left( \frac{\pi c L}{ 6\beta} +   2\pi \kappa \alpha^2 \mu^2  L \beta \right) } .
\end{eqnarray}
Note that due to the triviality of the temporal integrals, the result for the 
Euclidean quench can be easily obtained from the knowledge of the deformed partition function 
(\ref{part3}); equivalently this causes a rescaling of the coupling $\mu \mapsto \alpha \mu$. 
We will see that this property is true for all uniform holomorphic deformations. In the Hamiltonian version this fact can be seen as follows. The partition function for the Euclidean quench setup is 
\begin{align}
{\rm Tr} ( \rho^{\alpha} \rho_\beta^{1-\alpha} )=\tr \left[ q^{\alpha(L_0-{c\over 24} + {\nu\over \tau}J_0)} \bq^{\alpha(\bL_0-{c\over 24}+{\bar\nu\over\btau}J_0)} q^{(1-\alpha)(L_0-{c\over 24}) } \bq^{(1-\alpha)(\bL_0-{c\over 24})}\right]. 
\end{align}
Now since $[L_0,J_0]=0$, the product of the operator exponentials can be trivially simplified to yield
\begin{align}
{\rm Tr} ( \rho^{\alpha} \rho_\beta^{1-\alpha} )=\tr \left[ q^{L_0-{c\over 24} + {\alpha\nu\over\tau}J_0} \bq^{\bL_0-{c\over 24}+{\alpha\bar\nu\over\btau}J_0} \right]=\tr \left[ q^{L_0-{c\over 24}} \bq^{\bL_0-{c\over 24}} y^{\alpha J_0} \by^{\alpha \bar{J}_0} \right]. 
\end{align}
Comparing this to \eqref{chZ}
this is a simple rescaling of the chemical potential, $\nu \mapsto \alpha\nu$, and this is also reflected in \eqref{part4}. This feature is universally true for all deformations that commute with the Hamiltonian. This aspect turns out to be very different when we deal with inhomogenous deformations or deformations by primary operators.

Now using (\ref{part2}), (\ref{part3}) and (\ref{part4}) in the definition of 
R\'{e}nyi divergence  (\ref{defrenyd}) we obtain 
\begin{equation} \label{renydivJ}
D_\alpha( \rho||\rho_\beta) = 2\pi  \alpha \mu^2  \kappa  L\beta.
\end{equation}
It is easy to see that this simple linear function in $\alpha$ satisfies all the properties  of R\'{e}nyi divergences
listed in Section \ref{sec:rd} (for $\kappa >0$).

\subsection{Stress tensor}

Next we consider the deformation by the stress tensor
\begin{equation}\label{deformstres}
S_{\rm{CFT} } ( \phi) \mapsto S_{\rm{CFT}}  (\phi) + \mu \int  d^2 z  \left( 
{T}(z)   + {\bar  T}  ( \bar z)  \right) . 
\end{equation}
As we have seen from the $U(1)$ deformation,  the first step is to study the 
deformed partition function. 
In \cite{Datta:2014zpa}, it was seen that using that such a deformation   
of a CFT defined on a torus 
leads to a shift in the temperature.  
This was seen till the quadratic order in perturbation theory where the resulting integrals
was carried out by the same prescription of performing the spatial integrals first and then 
the temporal integrals. 
Here, we will verify that the deformation given in  (\ref{deformstres}) on the cylinder 
indeed shifts the temperature to the fourth order in perturbation theory. 

Let us begin by considering the expansion of the partition function to $O(\mu^2)$. 
We first focus only on the  perturbation by the holomorphic  stress tensor $T(z)$. 
As mentioned for the $U(1)$ current, the correlators  between the holomorphic
stress tensor and its anti-holomorphic counterpart factorise with the exception of  
delta-function contact terms. 
 Our integration prescription ignores all 
contact terms and, as before, the anti-holomorphic contribution can be accounted for
by doubling the contribution of the holomorphic  sector. Till the second order in conformal perturbation theory, we have
\begin{eqnarray} \label{perttexp}
\log \frac{ \cZ [ \mu]  }{ \cZ[0]}  &=& - \mu \int d^2 w_1  \langle T (w_1)  \rangle  + \frac{\mu^2}{2}
\int d^2 w_1 d^2 w_2 \langle T(w_1 ) T(w_1)  \rangle  \\ \nonumber
&& - \frac{\mu^2}{2} \left(  \int d^2 w_1  \langle T (w_1)  \rangle \right)^2 + \cdots. 
\end{eqnarray}
To evaluate the correlators on the cylinder we use the  following relation 
satisfied by the stress tensor under conformal transformation 
  \begin{equation}\label{T1}
    T(w)=\Big(\frac{dz}{dw}\Big)^2 T(z)+\frac{c}{12}\{z,w\}
    \end{equation}
    and 
    \begin{equation}
    \{z,w\}=\frac{2z^{\prime\prime\prime}z^\prime-3z^{\prime\prime^2}}{2z^{\prime^2}}, 
    \end{equation}
    is the Schwarzian of the transformation. 
    We can take $z, w$  are the co-ordinates on the plane and the cylinder which are related
    by the map 
    \begin{equation}
    z = \exp \left( 2\pi \frac{w}{\beta} \right)~.
    \end{equation}
    Using this,  the 1-point and 2-point functions of the stress tensor on the cylinder 
    are given by 
    \begin{eqnarray} \label{1-1ptfns}
      \langle T(\omega)\rangle &=& \frac{c}{12}\Big(-\frac{2\pi^2}{\beta^2}\Big), \\ \nonumber
    \langle T(\omega_1)T(\omega_2)\rangle &=& 
    \frac{c}{2}\Big(\frac{\pi}{\beta}\Big)^4 \Big[\frac{1}{\sinh [\pi/\beta(\omega_2-\omega_1)]}\Big]^4+4\Big(\frac{\pi}{\beta}\Big)^4\Big(\frac{c}{12}\Big)^2~.
    \end{eqnarray}
    We now have to perform the integrals over the cylinder. 
    Performing the  integral for the 1-point function is trivial, it just involves multiplying the 
    1-point function with the volume of the cylinder $\beta L$. 
    The first non-trivial integral occurs at the quadratic order, it involves the 2-point function 
    \begin{eqnarray}
    \int_{0}^{\beta}d\tau_1 \int_{0}^{\beta} d\tau_2 \int_{-\frac{L}{2}}^{\frac{L}{2} }d\sigma_2  \int_{-\infty}^{\infty}d\sigma_1 \frac{c}{2}\Big(\frac{\pi}{\beta}\Big)^4 \Big[\frac{1}{\sinh [\pi/\beta(w_2-w_1)]}\Big]^4 =  \frac{2 c   L \pi^3}{ 3\beta}~.
    \end{eqnarray}
    where $w_i = \sigma_i  + i \tau_i$. We have used the same recipe to perform the integrals as in the $U(1)$ case. 
    Note that the disconnected term involving the square of the  one point function 
    cancels with the constant term which arises from the Schwarzian transformation 
    in (\ref{1-1ptfns}). This ensures that the free energy is extensive. 
    The result for the partition function to $O(\mu^2)$ is given by
    \begin{equation}
    \log   \frac{\cZ[\mu]}{ \cZ[0]}    = \frac{\pi c L }{6\beta} \left(   2\pi \mu + ( 2\pi \mu) ^2 + \cdots 
    \right) ~. 
    \end{equation}
    Here we have included the contribution from the anti-holomorphic sector. 
    
    We have performed the perturbative expansion to $O(\mu^4)$. The details of the 
    integrals involved are provided in Appendix  \ref{integrals1}. 
    The final result  to this order is given by 
    \begin{equation}\label{4thorder}
    \log ( \cZ[\mu] ) =  \frac{\pi c L }{6\beta}  \left( 1+ 2\pi\mu + ( 2\pi \mu)^2 + ( 2\pi  \mu)^3 + 
    ( 2\pi \mu )^4 + \cdots  \right) .
    \end{equation}
    In the above, the high temperature limit  of the undeformed CFT partition function has been used
    \begin{equation} \label{equ1}
   \log \cZ[0]  = \log {\rm Tr} ( \rho_\beta) = \frac{\pi c L}{6\beta}. 
    \end{equation}
    On examining the series in (\ref{4thorder})
    it is easy to note that the perturbed partition function can be obtained  by shifting the 
    temperature  
    \begin{equation}\label{beta-effective}
    \beta \rightarrow \beta' =  \beta  ( 1- 2\pi \mu). 
    \end{equation}
    Thus the partition function is given by 
    \begin{equation} \label{equ2}
    \log ( \cZ[\mu])  = \log ( {\rm Tr} \rho)  = \frac{\pi c L}{6 \beta ( 1- 2\pi  \mu) } .
    \end{equation}
    This is because the prescription of performing the spatial integrals first
    picks out the conserved charge. For the stress tensor deformation,  the 
    conserved charge  is the original  Hamiltonian (or energy) itself. 
    Therefore, the stress tensor deformation can be thought of as a shift in temperature. 
    This observation also indicates that the range of $\mu$  is 
    restricted to 
    \begin{equation}
    \mu < \frac{1}{2\pi}.
    \end{equation}
    This constraint enforces positivity of the rescaled effective temperature,  $\beta'$, from equation \eqref{beta-effective}. 
    
    Now we can address the Euclidean quench with the above ingredients;
    the deformation is given  by 
    \begin{equation}
     S_{\rm{CFT} } ( \phi) \mapsto S_{\rm{CFT}}  (\phi) + \mu \int  d^2 z
  [ \theta( \tau) - \theta (\tau -  \alpha\beta) ]  \left( 
{T}(z)   + {\bar T}  ( \bar z)  \right). 
    \end{equation}
    Going through the steps of expanding the path integral with this deformation 
    in powers of $\mu$ and 
    performing the resulting integrals we arrive at 
    \begin{eqnarray} \label{stressalp}
    \hspace{-.5cm}
    \log \left[  {\rm Tr} ( \rho^{\alpha} \rho_\beta^{1-\alpha} )  \right] 
    = 
    \frac{\pi c L }{6\beta}  \left( 1+ 2\pi \alpha\mu + ( 2\pi \alpha \mu)^2 + ( 2\pi  \alpha \mu)^3 + 
    ( 2\pi \alpha  \mu )^4 + \cdots  \right) . 
    \end{eqnarray}
    Note that the coupling $\mu$ is replaced by $\alpha \mu$ in the expansion.
    The reason for this is the same as that observed for the $U(1)$ deformation. 
    Performing the integrals over the spatial co-ordinates first renders the 
    integrations over the temporal co-ordinates trivial and therefore  each temporal 
    integral gets multiplied by a factor of $\alpha$ resulting in the expression
    in (\ref{stressalp}).    
    Again, the same reasoning holds to all orders in perturbation theory and therefore we 
    obtain 
    \begin{equation}  \label{equ3}
     \log \left[ {\rm Tr} ( \rho^{\alpha} \rho_\beta^{1-\alpha}  ) \right]
     = \frac{\pi c L}{6\beta ( 1- 2\pi \alpha \mu) }.
    \end{equation}
    
    We can now evaluate the R\'{e}nyi divergences, substituting (\ref{equ1}), (\ref{equ2}) and
    (\ref{equ3})  into the definition  (\ref{defrenyd}), we obtain
    	\begin{equation}\label{Da all order}
	D_\alpha(\rho||\rho_\beta)
	=\frac{\pi c L }{6 \beta}\frac{4 \pi^2 \alpha \mu^2}{(1-2 \pi \mu )(1-2 \pi \alpha \mu)}. 
    \end{equation}
    We can easily check that the above result for the R\'{e}nyi divergence satisfies 
    the first  4 properties listed in Section \ref{sec:intro}. 
    It is instructive to check the relation to the relative entropy  at 
   $\alpha = 1$.  In this limit the R\'{e}nyi divergence becomes
   \begin{equation}\label{Daa1}
	D_1(\rho||\rho_\beta)
	=\frac{\pi c L }{6 \beta}\frac{4 \pi^2 \mu^2}{(1-2 \pi \mu )^2}. 
    \end{equation}
    It can be verified that  the difference in free energies of the excited state
    $\rho$ and  the state $\rho_\beta$ using (\ref{D1free}) results precisely in (\ref{Daa1}). 
  This check not only confirms that the relation in (\ref{relalpha1}) 
  holds but also the fact that
  the undeformed Hamiltonian $H$ determines dynamics.

\subsection{Spin-3}

Our final example is the deformation by the spin-3 current, $W(z)$
\begin{equation}\label{deformspin3}
S_{\rm{CFT} } ( \phi) \mapsto S_{\rm{CFT}}  (\phi) + \mu \int  d^2 z  \left( 
{W}(z)   + {\bar  W}  ( \bar z)  \right) ~.
\end{equation} 
Such a deformation was studied in \cite{Datta:2014ska,Datta:2014uxa,Datta:2014zpa} 
to evaluate higher spin corrections to 
entanglement entropy. 
In \cite{Datta:2014ska}, the corrections to the partition function to $O(\mu^2)$ 
was also evaluated. 

We restrict our attention to the free boson CFT, but it will be clear from 
our analysis that the evaluation of R\'{e}nyi divergences can be extended 
any CFT which admits ${\cal W}_3$ within its chiral-algebra. 
Consider $N$ complex  free bosons $\phi_i$ which obey the OPE
\begin{equation}
\bar \phi_i (z, \bar z ) \phi^j (w, \bar w ) \sim - \delta^j_i \log|z-w|^2. 
\end{equation}
This theory admits a
spin-3 current  conserved current which is given by 
 \begin{equation}\label{defspin3cur}
 W(z)=\sqrt{\frac{5}{12\pi^2}}\sum_{i=1}^{\frac{c}{2}}  :(\partial^2\bar\phi_i(z)\partial\phi_i(z)-\partial\bar\phi_i(z)\partial^2\phi_i(z) ):~. 
\end{equation}
This current is  a conformal primary and its  leading OPE  is given by\footnote{Here we follow the normalisation of the spin-3 current used in \cite{Datta:2014ska}.}
\begin{equation}
W(z) W(0) \sim - \frac{ 5c}{ 6\pi^2 z^6} + \cdots,
\end{equation}
where $c= 2N$. 
Let us again focus on only the holomorphic correction to the partition function. 
To order $\mu^4$  we have the following expansion
\begin{align} \label{pertspin3}
\log\frac{\cZ[\mu]}{\cZ[0]} =\, & \frac{\mu^2}{2} \int d^2 w_1 d^2 w_2  
\langle W(w_1) W(w_2)  \rangle  \\ \nonumber
& + \frac{\mu^4}{ 4!}  
\left[  \int d^2 w_1 d^2 w_2  d^2 w_3 d^2 w_4 
\langle W( w_1) W( w_ 2 ) W( w_3) W( w_3) \rangle \right. \nn \\
&\qquad\quad \left. 
- 3  \left( \int d^2 w_1 d^2 w_2  
\langle W(w_1) W(w_2)  \rangle \right)^2 \right] + \cdots \nn . 
\end{align}
The correlators involving an odd number of insertions of $W$'s vanish. 
We follow the same prescription to evaluate the integrals as discussed 
for the $U(1)$ current and the spin-2 deformation. 
The correlators can be evaluated by 
Wick contractions; on the cylinder  they are given by\footnote{An expression for the  4-point function of the spin-3 currents 
in a theory with ${\cal W}_\infty[\lambda ]$ symmetry was derived in 
\cite{Long:2014oxa}, see equation (30).
The free boson theory lies at $\lambda=1$. 
We find our expression in (\ref{4ptfnw})  coincides  with that of \cite{Long:2014oxa}
only for large $c$. }
\begin{eqnarray}\label{4ptfnw}
\langle W(w_1) W(w_2) \rangle &=& -\frac{5c}{6\pi^2} \left( \frac{\pi}{\beta \sinh \frac{\pi}{\beta} 
( w_1 - w_2)   }  \right)^6 , \\ \nonumber
\langle W(w_1) W(w_2) W(w_3) W(w_4)  \rangle &=&
\frac{25 c^2}{ 36 \pi^4} 
 \left( \frac{\pi^2}{\beta^2 \sinh  \frac{\pi}{\beta} 
( w_1 - w_2) \sinh  \frac{\pi}{\beta} 
( w_3 - w_4)   }
 \right)^6  \\ \nonumber
&& 
\left[ 
1 + \tfrac{18}{c} \eta 
+ \tfrac{75}{c} \eta^2 + 
( 2 + \tfrac{54}{c} ) \eta^3 
+ 9 ( 1 + \tfrac{2}{c} ) \eta^4 
+ 6 \eta^5 
+ \eta^6 
\right]~.
\end{eqnarray}
where $\eta$ is the  related to the cross ratio by 
\begin{equation}
\eta = x + \frac{1}{x} - 2, \qquad
x = \frac{ \sinh \frac{\pi}{\beta} ( w_1 - w_3)  \sinh \frac{\pi}{\beta} ( w_2 - w_4) }{
\sinh \frac{\pi}{\beta} ( w_1 - w_4)  \sinh \frac{\pi}{\beta} ( w_2 - w_3)}. 
\end{equation}
We can perform the integrals in the expansion (\ref{pertspin3}) using the same 
prescription discussed earlier for the spin-1 and spin-2 deformations. 
The result for the integral which occurs at  the $\mu^2$ order is
\begin{equation}\label{spin32pint}
\int_0^\beta d\tau_2 \int_0^\beta d\tau_1 \int_{-\frac{L}{2}}^{\frac{L}{2} } d\sigma_2
\int_{-\infty}^\infty d\sigma_1 
\left( \frac{1}{\sinh \frac{\pi}{\beta} ( z_2 - z_1) } \right)^6 
= -\frac{16}{15} \frac{\beta^3 L}{ \pi}~. 
\end{equation}
Once again performing the spatial integrals first renders the temporal integrals trivial. 
The integral involving the 4-point function of the spin-3 current can also be peformed, 
this is done in the appendix \ref{integrals2}.  It can be seen that the 
contributions from the disconnected terms, that is the terms 
proportional to $c^2$ 
in the four point 
function (\ref{4ptfnw}) precisely cancel with the contributions from the 2-point function squared
contributions which occurs at order $\mu^4$. 
The result for the deformed partition function  is given by 
\begin{equation} \label{partw}
\log ( \cZ[\mu])= \log {\rm Tr } (\rho_\beta )= \frac{\pi c L}{ 6\beta} \left[
1 +  \frac{16}{3} \left(  \frac{\pi \mu}{\beta} \right)^2 + \frac{12800}{27} \left( \frac{\pi \mu }{\beta} \right) ^4 +
\cdots \right].
\end{equation}
The contributions from the anti-holomorphic sector have been included. 

Integrating  the spatial directions first while performing the integrals occurring in
the perturbative expansion ensures that the we have deformed the theory by 
the addition of chemical potential for the conserved spin-3 charge. 
Therefore  the perturbative expansion  (\ref{partw})  should coincide with the evaluation
of the partition function with the spin-3 chemical potential in the Hamiltonian formalism. 
This has been done first in  \cite{Kraus:2011ds},  furthermore in \cite{Beccaria:2013dua} 
an  expression 
for the partition function was obtained in closed from for all orders in the chemical potential. 
This is given by 
\begin{equation}
\log \cZ[\mu] =\frac{\pi c L}{6\beta}\left[\frac{3}{160 }\frac{\beta^2}{\pi^2\mu^2}+\frac{3}{4}\sqrt{\frac{3}{5}}\frac{\beta}{2 \pi  {i}\mu}-3\sqrt{2}\sqrt[4]{\frac{3}{5}}
\sqrt{\frac{\beta}{2 \pi {i}\mu}}
\zeta\Big(-\frac{1}{2},\frac{\beta}{16 \pi \mathrm{i} \mu }\sqrt{\frac{3}{5}}\Big)\right],
\end{equation}
where $\zeta$ refers to the Hurwitz zeta function. 
Expanding this perturbatively in $\mu$ results in 
\begin{eqnarray} \label{fullexpw}
\log \cZ[\mu] &=& \frac{\pi c L}{6\beta}\sum_{j=1}^{\infty}3(-1)^{\frac{3j-1}{2}}\frac{\prod_{n=0}^{j-1}(2n-1)}{ j!}4^j\left(\frac{3}{5}\right)^{\frac{1-j}{2}}\frac{  {B}_{j+1}}{j+1}\left(\frac{2\pi\mu}{\beta}\right)^{j-1},  \nonumber
\end{eqnarray}
where $B_{j+1}$ are the Bernoulli numbers.   (\ref{fullexpw}) 
precisely coincides with the expansion 
(\ref{partw}) to order $\mu^4$  that 
we obtained  by performing  the integrals in the perturbative expansion.

Let us now consider the Euclidean quench which is given by the deformation
\begin{equation}
     S_{\rm{CFT} } ( \phi) \mapsto S_{\rm{CFT}}  (\phi) + \mu \int  d^2 z
  [ \theta( \tau) - \theta ( \tau - \alpha\beta) ]  \left( 
{W}(z)   + {\bar W}  ( \bar z)  \right). 
    \end{equation}
    As we have discussed,  performing the spatial integrals first renders 
    the temporal integrals trivial. Therefore, for the  path integral for the 
    Euclidean quench can be obtained by replacing $\mu\rightarrow \alpha \mu$
    in the expansion (\ref{fullexpw}). 
    This is because each temporal integral yields $\alpha\beta$ instead of $\beta$. 
    Therefore we have 
    \begin{equation} \label{quenchw}
    \log [ {\rm Tr} ( \rho^{\alpha} \rho_\beta^{1-\alpha})  ]
    = \frac{\pi c L}{6\beta} \left[ 1 + 
    \frac{16}{3} \left(  \frac{\pi \alpha \mu}{\beta} \right)^2 + \frac{12800}{27} \left( \frac{\pi\alpha \mu }{\beta} \right) ^4 +  \frac{1126400}{9}\left(\frac{\pi\alpha\mu}{\beta}\right)^6 \cdots \right]. 
\end{equation}
We have also  evaluated  ${\rm Tr} (\rho^\alpha \rho_\beta^{1-\alpha})$ for the spin-3 deformation
using the Hamiltonian formulation for the free boson theory.
This is straightforward since the basis involving the mode numbers 
diagonalise both the Hamiltonian 
and the zero mode of the spin-3 current. We have verified to order $\mu^8$  that we indeed obtain
the expression in (\ref{quenchw}). 
Substituting these expansions in expression for 
the R\'{e}nyi divergence in (\ref{defrenyd})  we obtain
\begin{eqnarray} \label{renyidwdef}
D_{\alpha} ( \rho||\rho_\beta) &=& 
\frac{1}{\alpha -1} \left[ 
\log Z[\mu\alpha] - \alpha \log Z[\mu] - ( 1 -\alpha) \frac{\pi cL }{\beta} \right]
\\ \nonumber
&=& 
\frac{\pi c L}{6 \beta} \left[
\frac{16}{3} \left( \frac{\pi\mu}{\beta} \right)^2 \alpha + 
\frac{12800}{27} \left( \frac{\pi \mu}{\beta}\right)^4 \alpha( \alpha^2 + \alpha +1) 
+ \cdots \right]. 
\end{eqnarray}
Once again,  it can be verified that the above expression for R\'{e}nyi divergence 
satisfies all the properties listed in Section \ref{sec:intro}.

From the examples studied,  we conclude that 
R\'{e}nyi divergence between excited states obtained by deforming the CFT  by 
conserved currents and the thermal state can be evaluated once the 
partition function of the theory $\cZ[\mu]$  is known.  Here $\mu$ is the chemical potential for the 
charge corresponding to the conserved current. 
As we demonstrated by these examples, this is because in the perturbative expansion 
of the partition function integrating over the spatial directions 
ensures that we pick out the conserved charge. Therefore the temporal integrals are trivial. 
This results in ${\rm Tr} (\rho^{\alpha} \rho_\beta^{ 1-\alpha}) = \cZ[\alpha\mu]$ and we can 
then evaluate the  R\'{e}nyi divergence using (\ref{defrenyd}). 
Holographically, this implies that we can easily evaluate R\'{e}nyi divergences  between higher spin 
black holes in $\mathcal{W}_\infty[\lambda]$  and the BTZ black hole  using the partition functions
evaluated in \cite{Gutperle:2011kf,Gaberdiel:2012yb}.

\section{Inhomogeneous deformations } 
\label{inhodef}

The analysis in the previous section has shown that for excited states, in which the 
deformation commutes with the original Hamiltonian, the R\'{e}nyi  divergence 
can be obtained once one knows the partition function of the  deformed theory. 
The dependence of the  R\'{e}nyi divergence  on $\alpha$   for these class of 
deformations is quite simple.   
The deformation we considered was uniform along the spatial directions.
Therefore, while performing the spatial integrals first resulted in the conserved charge 
or the zero mode of the conserved current. 
For instance in the case of the $U(1)$ deformation,  we obtained the  conserved charge. 
It is this reason, that the deformation commuted with the Hamiltonian. The deformation by the primary operator, considered in Section \ref{analytdelta}, does not obey this property and therefore we get a non-trivial function of $\alpha$ for the \Re divergence. 
Another simple way to construct deformations that do not commute with the Hamiltonian 
is to consider inhomogenous deformations  along the spatial direction. 
Then performing the spatial integrals picks up non-zero modes of the conserved currents
which do not  commute with the Hamiltonian. 
A special class of inhomogenous deformations
sine-square deformation (SSD) has been of recent interest in the context of periodically driven Floquet CFTs \cite{ishibashi:ssd,wen:floquet,wen:ssd,vishwanath:ssd,chitra:ssd}.
In this section we evaluate the R\'{e}nyi divergences of  excited states 
constructed by deforming the CFT by 
SSDs, and its higher spin generalizations.

Before we begin the study of SSD deformations, we analyse the case of the simple harmonic oscillator deformed by a linear 
term in its potential.  The R\'{e}nyi divergence for this simple inhomogeneous  deformation 
is exactly calculable. 
We demonstrate that the  functional dependence of the R\'{e}nyi divergence  on $\alpha$ 
for this model is identical to that of the leading 
contribution to the R\'{e}nyi divergence
of SSDs.

\subsection{Deformed harmonic oscillator}
\label{subsec:sho}

Consider the simple harmonic oscillator with a  linear potential and a
time dependent forcing  with the Lagrangian
\begin{equation}
L = \frac{m}{2} \dot x^2  - \frac{1}{2} m \omega^2 x^2  + \mu f(t)  x~. 
\end{equation}
 The  Euclidean path integral of this system  is given by\footnote{We can obtain this by analytically continuing the result given in 
  \cite[Problem 3.11]{Feynman:1965:QMP}.}
\begin{eqnarray}
K( x_b, x_a; \tau_b, \tau_a) = \sqrt{ \frac{ m \omega}{ 2\pi   \sinh  \omega \beta  }}~
e^{ - S ( x_a, x_b; \tau_a, \tau_b ) },
\end{eqnarray}
where $\beta  = \tau_b - \tau_a$ 
and  the action is given by\footnote{We have set $\hbar =1$.}
\begin{eqnarray}
S  =  \frac{m\omega}{2\sinh \omega \beta  } 
&\Bigg[&(  x_a^2 + x_b^2 ) \cosh \omega \beta  - 2 x_ax_b 
 \\
& &- \frac{ 2\mu   }{m\omega} 
\int_{\tau_a}^{\tau_b}  f(\tau)   ( x_ b \sinh  \omega(  \tau -  \tau_a)   +
x_a \sinh  \omega( \tau_b - \tau)  ) \nn  \\
 & & -  \frac{2 \mu^2 }{m^2 \omega^2} 
 \int_{\tau_a}^{\tau_b}   d\tau f(\tau) 
 \sinh \omega (  \tau_b - \tau)  
 \left( \int_{\tau_a}^\tau  d\tau' f(\tau')  \sinh \omega ( \tau' - \tau_a)    \right ) \Bigg]\nn . 
 \end{eqnarray}
 For the case of the Euclidean quench
 \begin{equation}
 f(\tau) = \theta( \tau) - \theta ( \tau - \alpha\beta) ~. 
 \end{equation}
 We can then evaluate  ${\rm Tr} ( \rho^\alpha \rho_\beta^{ 1-\alpha} ) $ using this 
 path integral by setting $\tau_a = 0 , \tau_b = \beta, x_a = x_b = x$ and then 
 performing the integral over all  $x$.  This leads to the following 
 \begin{eqnarray}\label{shoreny1}
 {\rm Tr} ( \rho^\alpha \rho_\beta^{1-\alpha} ) 
 &=&  \int_{-\infty}^\infty dx\, K (x, x, \beta, 0 ) , \\ \nonumber
 &=&  \frac{1}{2 \sinh \frac{\beta\omega}{2}} 
 \exp \left[
 \frac{\mu^2 }{m \omega^3}   \left(  \frac{ \sinh \frac{\alpha\beta \omega}{2}  
 \sinh\frac{ ( \alpha -1)\beta\omega}{2}  }{ \sinh \frac{\beta\omega}{2} }   + \frac{ \alpha \beta \omega}{2} \right) \right].
 \end{eqnarray}
 From this result we can  obtain the undeformed  as well as the 
 deformed partition functions, which are given by 
 \begin{eqnarray}\label{shoreny2}
{\rm Tr }( \rho_\beta ) &=& \frac {1}{ 2\sinh \frac{\beta\omega}{2} }, \qquad
{\rm Tr} ( \rho) = \frac{1}{2 \sinh \frac{\omega\beta}{2} } \exp \left( \frac{\mu^2 \beta}{ 2 m \omega^2} \right) .
\end{eqnarray}
Substituting the  equations  (\ref{shoreny1})  and  \eqref{shoreny2} in to the definition of 
R\'{e}nyi divergence (\ref{defrenyd})  we obtain 
\begin{equation}\label{shoreny3}
D_\alpha ( \rho||\rho_\beta) = \frac{\mu^2}{m \omega^3  } 
\frac{ \sinh \frac{\alpha\beta\omega}{2}  \sinh \frac{ ( \alpha - 1) \alpha\beta\omega}{2} }
{( \alpha - 1)\sinh \frac{\beta\omega}{2} }~. 
\end{equation} 
It is easy to verify that this result satisfies all the properties given in Section \ref{sec:intro},
including the relation of the R\'{e}nyi divergence with the relative entropy given in 
(\ref{relalpha1}).

The above result can also be obtained using perturbation theory in the Hamiltonian formalism. The harmonic oscillator Hamiltonian  deformed by $x\,(\sim a_+ + a_-)$ is given by 
\begin{align}
H_g = \omega\left(a_+ a_-+\frac{1}{2}\right)+ g (a_+ + a_-). 
\end{align}
The relevant partition function that leads to the Renyi divergence is 
\begin{align}
\cZ = \tr \left[ q^{\alpha \left[\left(a_+ a_-+\frac{1}{2}\right)+ g (a_+ + a_-)\right]} q^{(1-\alpha) \left(a_+ a_-+\frac{1}{2}\right)}  \right], \qquad q=e^{-\beta\omega}.
\end{align}
The methods developed in the next section to tackle inhomogeneous deformations of CFTs can  be used to evaluate this partition function perturbatively in $g$. The details are provided in Appendix \ref{app:sho-ham}. The quadratic order result for the Renyi divergence matches with \eqref{shoreny3} above  obtained using the path-integral formalism. The coupling $g$ can be identified with its counterpart $\mu$ in the Lagrangian as
\begin{align}
g= \frac{\mu}{2\sqrt{m\omega^3}}. 
\end{align}

\subsection{Sine-square deformation}
The sine-square deformation (SSD) has been of recent interest in the context of periodically driven Floquet CFTs \cite{ishibashi:ssd,wen:floquet,wen:ssd,vishwanath:ssd,chitra:ssd}. This deformation forms a rare example which offers analytic tractability and exhibits an interesting phase diagram under Floquet driving. The SSD Hamiltonian is given by the following insertion of a sine-squared envelope function
\begin{align}
H_{\rm SSD} = 2\int_0^L dx~\sin^2 \left(\pi x \over L\right)~ T_{00}(x)~. 
\end{align}
Upon using the conformal transformation, $z=e^{\frac{2\pi x}{L}}$, and writing the stress-tensor in terms of the mode expansion, we get
\begin{align}
H_{\rm SSD}=\frac{2\pi}{L}\left[L_0+\bL_0 -\frac{1}{2}(L_1+L_{-1}) -\frac{1}{2}(\bL_{1}+\bL_{-1}) - \frac{c}{12}\right] ~. 
\end{align}
In what follows, we consider a slightly generalized version of the envelope function 
\begin{align}\label{Hg}
H_{g} &= \int_0^L dx~\left[1-g \cos \left(2\pi x \over L\right)\right]~ T_{00}(x) \,  , \nn \\	H_{g}&=\frac{2\pi}{L}\left[L_0+\bL_0 -\frac{g}{2}(L_1+L_{-1}) -\frac{g}{2}(\bL_{1}+\bL_{-1}) - \frac{c}{12}\right] ~. 
\end{align}
Here, $g$ parametrizes the deviation from a uniform profile.
The above Hamiltonian reduces to the sine-square deformation for $g\to 1$.
This form of the Hamiltonian is used in \cite{MacCormack:2018rwq} and is often referred to as the M\"obius Hamiltonian. Note that the Hamiltonian is built purely from the generators of the two $\sltwor$ sub-algebras.

The torus partition function of the deformed theory \eqref{Hg} can be computed exactly by utilizing the $\sltwor$ symmetry. The details of the calculation can be found in Appendix \ref{ssd-np}. The result is 
\begin{align}\label{deformed-Z}
\cZ_{g} (\tau, \bar \tau) &= \tr \left[ q^{L_0-\frac{g}{2}(L_{-1}+L_1)-c/24} \bar{q}^{\bar L_0-\frac{g}{2}(\bar L_{-1}+\bar L_1) -c/24}  \right],  \nn \\
&=  (q\bar{q})^{-(1-\sqrt{1-g^2}){c\over 24}} \cZ_{\rm CFT} (\tau\sqrt{1-g^2}, \bar \tau\sqrt{1-g^2}).
\end{align}
with $q=e^{2\pi i \tau}$. 
Apart from the power law pre-factor, the deformation \eqref{Hg} leads to an effective rescaling of the modular parameter $\tau\mapsto \tau\sqrt{1-g^2}$. For the rectangular torus, $\tau=i\beta /L$, and $0< g <1$ this implies an increase of temperature. For $g>1$ the deformation adds an angular chemical potential. Finally for $g=1$, the SSD limit, we have $\tau\mapsto 0$ which is the strict high temperature limit. It can therefore be seen that the deformation parameter $g$ also serves as a regulator. 

We record the perturbative corrections till the quadratic order
\begin{align}\label{deformed-Z-pert}
\cZ_{g} (\tau, \bar \tau)
&= Z_{\rm CFT} (\tau, \bar \tau) -\frac{g^2}{2} q^{ -c/24} \bar{q}^{  -c/24}(q(\log q)\pd_q + \bq(\log \bq)\pd_{\bq} ) f (\tau, \bar \tau) + \cdots~.
\end{align}
Here, $f (\tau, \bar \tau)  = (q\bq)^{c/24}\cZ_{\rm CFT}(\tau,\btau)$. The above correction will serve as a useful  check on our results for \Re divergences. 

\subsubsection*{\Re divergence}

The relevant partition function we need to calculate to obtain the \Re divergence is the following 
\begin{align}\label{rdZ}
\cZ_{g} (\tau, \bar \tau|\alpha) &= \tr \left[ q^{\alpha(L_0-\frac{g}{2}(L_{-1}+L_1)-{c\over 24})} \bar{q}^{\alpha(\bar L_0-\frac{g}{2}(\bar L_{-1}+\bar L_1) -{c\over 24})} q^{(1-\alpha)(L_0-{c\over 24})}
\bq^{(1-\alpha)(\bar{L}_0-{c\over 24})}
\right]. 
\end{align}
Note that unlike the holomorphic deformations, the deforming  operator does not commute with the Hamiltonian in this case. This makes the Taylor expansion in $g$ of the Boltzmann factors non-trivial. We also observe that this is the trace of the Euclidean version of the single-cycle Floquet operator for the M\"obius deformation. 

Although the torus partition function \eqref{deformed-Z} for the M\"obius deformation alone can be obtained non-perturbatively in $g$, it is rather difficult to evaluate \eqref{rdZ} which has the insertion of an additional Boltzmann factor. We proceed perturbatively.\footnote{So strictly speaking, we are studying a perturbative version of the SSD or the M\"obius deformation in this work.} 
Let us consider the holomorphic pieces of the deformed and undeformed Hamiltonian (apart from the Casimir energy shift) and the corresponding partition function
\begin{align}
H_g '= X - g Z~, \qquad H_0' = X~, \qquad \cZ_{\rm hol}=\tr\left[e^{-\alpha b H'_g} e^{-(1-\alpha)b H'_0}\right] ~, 
\end{align}
where $Z=(L_{1}+L_{-1})/2$.
The anti-holomorphic piece can be straightforwardly added later on.
The derivative of the operator exponential is given by the Duhamel's formula
\begin{align}\label{int-rep}
\frac{\pd}{\pd g} e^{-\alpha b H'_g} = \alpha b    \int_{0}^{1} e^{-\alpha b (1-s) H'_g} \frac{\pd H'_g}{\pd g} e^{-\alpha b s H'_g}~,
\end{align}
with $b=-2\pi i \tau=-\log q$. 
The derivative of the relevant partition function for \Re divergence is then
\begin{align}\label{1-tr}
\frac{\pd \cZ_{\rm hol}}{\pd g}  = \alpha b\, \tr\left[e^{-(1-\alpha) b H'_0}\int_0^1ds ~e^{-\alpha b (1-s) H'_g} Z e^{-\alpha b s H'_g} \right]~.
\end{align}
For $g=0$ the above RHS is proportional to $\tr[Z e^{-bH'_0}]$. Since $Z$ isn't made of zero-modes, its expectation value in any state (primary or descendant) will vanish. Therefore, there is no correction at the linear order in perturbation theory. 

We need the second derivative for the next-order correction. Before doing that, let's calculate the $g$-derivative of the integrand appearing in \eqref{1-tr}. Using \eqref{int-rep} again to act on the Boltzmann factors, we obtain
\begin{align}
\cA(g)=&\frac{\pd}{\pd g}\left(e^{-\alpha b  (1-s) H'_g} Z e^{-\alpha b  s H'_g} \right)\nn \\ =& \left[ \alpha b  (1-s)\int_0^1 du~ e^{\alpha b (1-s)(u-1)H'_g} Z e^{-\alpha b (1-s)H'_g}\right]  Z e^{-\alpha b  s H'_g}\nn \\
&+ e^{-\alpha b (1-s)H'_g} Z \left[\alpha b  s \int_0^1 dv~e^{\alpha b  s (v-1)H'_g} Z e^{-\alpha b  s v H'_g}\right]~.
\end{align}
The second derivative of the trace at $g=0$ can now be evaluated
\begin{align}
&\frac{\pd^2 \cZ_{\rm hol}}{\pd g^2}  \bigg|_{g=0} = \alpha b   \int_0^1 ds~\tr\left[e^{-(1-\alpha) b   H'_0} \cA(0)\right]\nn \\
=~& (\alpha b  )^2 \int_0^1 ds~ (1-s) \int_0^1 du ~\tr \left[ e^{-(1-\alpha) b   H'_0} e^{\alpha b  (1-s)(u-1)H'_0} Z e^{-\alpha b  (1-s)H'_0} Z e^{-\alpha b   s H'_0} \right] \nn \\
&+  (\alpha b  )^2  \int_0 ^1 ds ~s \int_0^1 dv ~
\tr \left[ e^{-(1-\alpha) b   H'_0} e^{-\alpha b  (1-s)H'_0} Z e^{\alpha b   s(v-1)H'_0} Z e^{-\alpha b   v s H'_0} \right].
\end{align}
We use the cyclicity of trace to rearrange the operators within the traces and change integration variables of the in the second term above: $t=1-s$ and $v=1-u$. It can then be seen both terms above are equal to each other and the second derivative simplifies to be 
\begin{align}\label{2nd-derivative}
&\frac{\pd^2 \cZ_{\rm hol}}{\pd g^2}  \bigg|_{g=0} 
=~ 2 (\alpha b  )^2 \int_0^1 ds ~(1-s) \int_0^1 du~ \tr \left[ e^{- b  (1-\alpha u (1-s)) H'_0} Z e^{-\alpha b   (1-s) u H'_0} Z \right].
\end{align}
Next, we use the identity \eqref{LL-shuffle} derived in the Appendix \ref{ham-alg}. For $(yw)=q$ we have
\begin{align}
\tr[w^{X} Z y^{X} Z] = \frac{y+w}{4} \tr [L_{1}L_{-1}q^{L_0}].
\end{align}
Hence, the second order correction is
\begin{align}\label{renyi-div-ord2x}
\frac{\pd^2 \cZ_{\rm hol}}{\pd g^2} \bigg|_{g=0}  =& ~2 (\alpha b)^2 \left(\frac{1}{4} \int_0^1 ds~(1-s)\int_0^1 du \left(q^{\alpha u (1-s)}+q^{1-\alpha u (1-s)}\right)\right) \tr \left[L_{1}L_{-1}q^{L_0}\right] \nn \\
=& -\frac{1}{2} \left(4 q^{1/2} \sin (\pi  (\alpha -1) \tau ) \sin (\pi  \alpha  \tau )+\alpha(1-q)\log q\right) \tr \left[L_{1}L_{-1}q^{L_0}\right].
\end{align}
We now restore the anti-holomorphic part and incorporate the Casimir energy shifts in the Hamiltonians. The corrected partition function is
\begin{align}\label{rdZ2}
&\cZ_{g} (\tau, \bar \tau|\alpha) =  \cZ_{\rm CFT} (\tau, \bar \tau) \nn \\
&- \frac{g^2}{4}  \left(4 q^{1/2} \sin (\pi  (\alpha -1) \tau ) \sin (\pi  \alpha  \tau )+\alpha(1-q)\log q\right) ~\tr \left[L_{1}L_{-1}q^{L_0-c/24}\bq^{\bL_0-c/24}\right]\nn \\
&- \frac{g^2}{4}  \left(4 \bq^{1/2} \sin (\pi  (\alpha -1) \btau ) \sin (\pi  \alpha  \btau )+\alpha(1-\bq)\log \bq\right) ~\tr \left[\bL_{1}\bL_{-1}q^{L_0-c/24}\bq^{\bL_0-c/24}\right]\nn \\
& + O(g^3).
\end{align}
Using the expressions for the traces from Appendix \ref{app:traces} (see also \cite{Maloney:2018hdg}), we have the result 
\begin{align}\label{rdZ3}
&\cZ_{g} (\tau, \bar \tau|\alpha) =~  \cZ_{\rm CFT} (\tau, \bar \tau) \nn \\
&- \frac{g^2}{2}  {(q\bq)^{-c/24}} \left(4 q^{1/2} \sin (\pi  (\alpha -1) \tau ) \sin (\pi  \alpha  \tau )+\alpha(1-q)\log q\right) \frac{q}{(1-q)}\pd_q f(\tau,\btau)\nn \\
& - \frac{g^2}{2}{(q\bq)^{-c/24}} \left(4 \bq^{1/2} \sin (\pi  (\alpha -1) \btau ) \sin (\pi  \alpha  \btau )+\alpha(1-\bq)\log \bq\right)\frac{\bq}{(1-\bq)}\pd_\bq f(\tau,\btau)\nn \\& + O(g^3).
\end{align}
We can check whether this result is consistent with the $\alpha\to 0,1$ limits. For $\alpha\to 0$ we have the undeformed CFT throughout the temporal cycle and the above correction vanishes as expected. On the other hand, for $\alpha\to 1$, we have the M\"obius deformation turned on at all times and we precisely recover the result \eqref{deformed-Z-pert}. Choosing a rectangular torus, $\tau=i\beta/L$, the high temperature version of \eqref{rdZ3} is
\begin{align}
\cZ_{g}(\beta,\alpha) \approx \cZ_{\rm CFT}(\beta)\bigg[1+ {g^2}  \left(4   \sinh (\pi  (\alpha -1) \tfrac{\beta}{L} ) \sinh (\pi  \alpha  \tfrac{\beta}{L} )-4\pi^2 \alpha \tfrac{\beta^2}{L^2}\right)\frac{c L^3}{24\pi  \beta^3}  + O(g^3) \bigg].
\end{align} 
With these ingredients in place, the R\'{e}nyi divergence can be found from its definition
\begin{align}\label{rd-Z-def}
D_\alpha(\rho_{gD}||\rho_\beta) = \frac{1}{\alpha-1} \log \frac{\tr[\rho_{g}^\alpha\rho_\beta^{1-\alpha}]}{\tr[\rho_{g}]^\alpha\,\tr[\rho_\beta]^{1-\alpha}} =\frac{1}{\alpha-1} \log \frac{Z_{g}(\beta,\alpha)}{Z_{g}(\beta,1)^{\alpha}Z_{\rm CFT}(\beta)^{1-\alpha}} ~. 
\end{align}
In the leading order in small $g$, we have
\begin{align} \label{renydssd1}
D_\alpha(\rho_{\rm SSD}||\rho_\beta) 
&= 
-{2} {g^2}q^{3/2} ~{\sin (\pi  (\alpha -1) \tau ) \,\sin (\pi  \alpha  \tau )\over (\alpha-1)(1-q)} \pd_q  {\log} f(\tau,\btau) +\text{anti-holomorphic}  + O(g^4) ,
\end{align}
where, once again, $f (\tau, \bar \tau)  = (q\bq)^{c/24}\cZ_{\rm CFT}(\tau,\btau)$. 
We note that the dependence of $\alpha$ at this order is exactly the same as that of the deformed harmonic oscillator \eqref{shoreny3}. 
A plot of the Renyi divergence with $\alpha$ is shown below. 
The relative entropy is given by the $\alpha\to 1$ limit
\begin{align}
S(\rho_{\rm SSD}||\rho_\beta)=
-{2} {g^2}~  q^{3/2} ~{(\pi \tau)\sin (\pi    \tau ) \over 1-q} \pd_q  {\log}f(\tau,\btau)  +\text{anti-holomorphic}+ O(g^4) ~.
\end{align}

\begin{figure}[!h]
	\centering
	\includegraphics[width=1\linewidth]{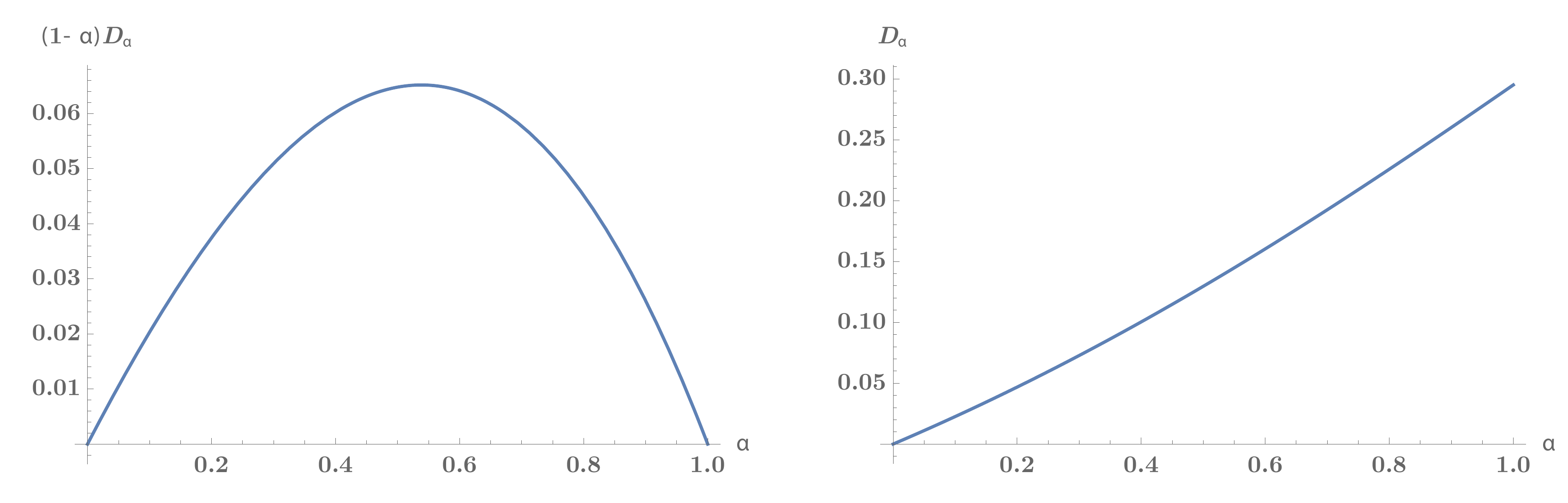}
		\caption{Renyi divergence \eqref{renydssd1} for the sine-squared deformation  with $L = 1, c = 30, \beta = 0.1, g=0.04$.}
	\label{fig:ssd-renyi-divs-1}
\end{figure}

\subsection{Higher spin generalisation}
We can also consider inhomogneous deformation by higher spin currents 
\begin{align}
\delta H_{s} = \int_0^L dx~\left[\frac{g}{2^{k}}   \left(e^{-{2\pi x \over L}(s-2+k)}+e^{-{2\pi x \over L}(s-2-k)}\right)\right]~ V^{(s)}_{00...}(x)
\end{align}
Here, $V^{(s)}$ denotes a higher spin current with spin $s$. 
In the previous section the case with $s=2$ and $k=1$ was considered.  Recall that, in terms of modes 
\begin{align}
V^{(s)}(z) = \sum_{n=-\infty}^\infty \frac{V^{(s)}_n}{z^{n+s}}~. 
\end{align}
The deformation is then
\begin{align}
\delta H_{s} = \frac{2\pi}{L} \frac{g}{2^{k}}\int_C \frac{dz}{2\pi i} \left[z^{s-1+k}+ {z^{s-1-k}}\right]V^{(s)}(z) = \frac{2\pi}{L}  \frac{g}{2^{k}}  [V^{(s)}_{-k}+V^{(s)}_{k}]~. 
\end{align}
The deformation by the spin-3 current turns out to be simpler than other ones.  The torus partition function can be exactly found just like the M\"obius deformation, it leads to effective rescaling of the modular parameter. For the deformation by $W_{-2}+W_2$ we have
\begin{align}\label{w3-pf}
\cZ_{g} (\tau, \bar \tau) &= \tr \left[ q^{L_0+\frac{g}{4}(W_{-2}+W_2)-c/24} \bar{q}^{\bar L_0+\frac{g}{4}(\bar W_{-2}+\bar W_2) -c/24}  \right] \nn \\
&=  (q\bar{q})^{-(1-\sqrt{1+4g^2}){c\over 24}} \cZ_{\rm CFT} (\tau\sqrt{1+4g^2}, \bar \tau\sqrt{1+4g^2})~.
\end{align}
The details of this calculation can be found in Appendix \ref{w3-np}.

\subsubsection*{Renyi divergence}

Once again we are interested in the trace with two  Boltzmann factor insertions
\begin{align}\label{rdZ-hs}
\cZ_{g} (\tau, \bar \tau|\alpha) &= \tr \left[ q^{\alpha(L_0+\frac{g}{2^k}(\Vs{-k}+\Vs{k})-{c\over 24})} \bar{q}^{\alpha(\bar L_0+\frac{g}{2^k}(\bVs{-k}+\bVs{k}) -{c\over 24})} q^{(1-\alpha)(L_0-{c\over 24})}
\bq^{(1-\alpha)(\bar{L}_0-{c\over 24})}
\right]. 
\end{align}
We restrict $k$ to $1<|k|<s$ so that the corrections are determined by the global hs$[\lambda]$ algebra. 
For calculating the corrections perturbatively, the steps till \eqref{2nd-derivative} are the same -- we just need to replace the deforming operator appropriately. The second order correction is (writing just the holomorphic piece for the time being and $b=-\log q$ as before)
\begin{align}\label{2nd-derivative-WW}
\frac{\pd^2 \cZ_{\rm hol} }{\pd g^2}\bigg|_{g=0} 
=~& \frac{(\alpha b)^2}{2^{2k-1} } \int_0^1 ds ~(1-s) \int_0^1 du~ \tr \left[ e^{-b(1-\alpha u (1-s)) L_0} \Vs{+} e^{-\alpha b (1-s) u L_0}\Vs{+} \right] ,\nn \\
\end{align}
where, $\Vs{+}=\Vs{k}+\Vs{-k}$. 
The higher spin currents can be shuffled using the following identity
\begin{align}\label{WW-shuffle-x}
\tr \left[(\Vs{k}+\Vs{-k})y^{L_0}(\Vs{k}+\Vs{-k})w^{L_0}\right] 
&= (y^k+w^k)\, \tr \left[\Vs{k} \Vs{-k}q^{L_0}\right].
\end{align}
This is proved in the Appendix  \ref{ham-alg}. The trace appearing above can be calculated in the same manner as the Virasoro case -- see equation (\ref{vv-thermal}). The final result for the second order correction is 
\begin{align}\label{corr2}
\frac{\pd^2 \cZ_g }{\pd g^2}\bigg|_{g=0}
=& - \tfrac{2^{1-2k}q^{k}}{k} \tfrac{ 4q^{k/2} \sin (\pi  (\alpha -1) k \tau ) \sin (\pi  \alpha  k \tau ) +k \alpha(1-q^k)\log q}{1-q^{k}}~{(q\bq)^{-c/24}}C^2_{ss}(k,\lambda)~\pd_q  f(\tau,\btau) + \cdots\nn \\
&+ \text{anti-holomorphic}~.
\end{align}
Here, $C_{NN}^2(k,\lambda)$ is the following structure constant appearing in the hs$[\lambda]$ algebra (see e.g.~\cite[Appendix A]{Gaberdiel:2011wb})
\begin{align}\label{ww-alg-0}
[V^{(s)}_{-k} , V^{(s)}_{k}] = C_{ss}^2(k,\lambda) L_0 + \text{zero-modes of even spins}.
\end{align}
Equation \eqref{corr2} reduces to the M\"obius deformation for $s=2$ and $k=1$ given in equation \eqref{renyi-div-ord2x}. It also reduces to perturbatively expanded $\mathcal{W}_3$ result \eqref{w3-pf} for $\alpha=1$, $\lambda=-3$, $s=3$ and $k=2$. The `$\cdots$' in \eqref{corr2} involve expectation values of even-spin currents. These vanish in the high temperature limit if we choose to work in the primary basis. 

The partition function \eqref{rdZ-hs} is then 
\begin{align}
&\cZ_{g} (\tau, \bar \tau|\alpha) = ~\cZ_{\rm CFT}(\tau, \bar \tau) \nn \\
&- g^2 \left[\tfrac{2^{-2k}q^{k}}{ k} \tfrac{ 4q^{k/2} \sin (\pi  (\alpha -1) k \tau ) \sin (\pi  \alpha  k \tau ) +k \alpha(1-q^k)\log q}{1-q^{k}}~{(q\bq)^{-c/24}}C^2_{ss}(k,\lambda)~\pd_q  f(\tau,\btau) +\text{anti-hol} \right]\nn \\
&+O(g^3)~.
\end{align}
Here we have not written terms arising from even-spin currents. As before, the Renyi divergence can be found straightforwardly from the above deformed partition function. The leading order result is
\begin{align} \label{ssdarbk}
D_\alpha(\rho_g||\rho_\beta) \approx- g^2 \bigg[&\frac{2^{-2k}q^{k}}{ k} \frac{ 4q^{k/2} \sin (\pi  (\alpha -1) k \tau ) \sin (\pi  \alpha  k \tau ) }{(1-q^{k})(1-\alpha)}~C^2_{ss}(k,\lambda)~\pd_q  {\log} f(\tau,\btau)\nn \\ & 
+\text{anti-holomorphic} \bigg].
\end{align}
We note that this result takes a universal form for all higher-spin inhomogeneous deformations considered here. The dependence on the spin of the deforming current appears only through the structure constant $C_{ss}^2$. 

We can also consider the deformation of the CFT Hamiltonian by the operator $J_{1}+J_{-1}$ built from the modes of $U(1)$ current. The \Re divergence can be analogously computed and we have the leading result
\begin{align}
D_\alpha(\rho_g||\rho_\beta) \approx g^2 \bigg[&\frac{ q^{3/2} \sin (\pi  (\alpha -1) \tau ) \sin (\pi  \alpha  \tau ) }{(1-q)(1-\alpha)}~\kappa
+\text{anti-holomorphic} \bigg],
\end{align}
where, $\kappa$ is the level of the current algebra. 

The $\alpha$ dependence in the leading correction for the inhomogeneous deformations considered here takes exactly the same form as the deformed harmonic oscillator \eqref{shoreny3}. The frequency $\omega$ of the harmonic oscillator gets replaced  by the mode number $k$. Note that the $\alpha$ dependence is fixed by the same integral which appears in \eqref{renyi-div-ord2x}, \eqref{2nd-derivative-WW} and \eqref{sho-int}. The integrand is, in turn, determined by the following commutators with the undeformed Hamiltonians 
\begin{align}
[H_{\rm CFT},L_k] = -k L_k, \qquad [H_{\rm CFT},V^{(s)}_{k}] = -kV^{(s)}_{k}, \qquad [H_{\rm SHO},a_{\pm}] = \pm \omega a_\pm ,
\end{align}
where, we set the length of the spatial circle $L=2\pi$ and $\hbar=1$.
In particular, these relations facilitate  rearrangement of operators within the traces and Boltzmann factors (see Appendix \ref{ham-alg}). The similar form of these algebraic relations is the reason behind the universality at the leading order.

\section{Generalized  second laws of thermodynamics}
\label{gslt}

One of the aims of   \cite{Bernamonti:2018vmw}  to 
study R\'{e}nyi divergences in conformal field theories
was the possibility that they provide additional constraints on thermodynamical evolution
for out of equilibrium systems besides the ordinary second law. 
These constraints are a natural extrapolation of the  simple thermodynamical criterion  that 
a transition from the state $\rho_1$ to $\rho_2$ is allowed provided the free energy 
$F(\rho_1) > F(\rho_2)$.  Note that we are  considering open systems, and each of these 
density matrices are in contact with its respective heat  baths. 
We have seen  in equation (\ref{relalpha1}), 
that one of the properties of   the R\'{e}nyi divergence 
$D_{\alpha} ( \rho||\rho_\beta)$ is that
at $\alpha =1$, it reduces to 
  the differences between the  free energies of the  density matrices
 $\rho$ and $\rho_\beta$. 
 Using this property we can arrive at the following  equivalent statement  of the second law. 
 Consider the difference 
 \begin{equation}
 \delta D_1(\rho_1||\rho_2) \equiv  D_1(\rho_1||\rho_\beta) - D_1( \rho_2||\rho_\beta) .
 \end{equation}
 Then if  $ \delta D_1(\rho_1||\rho_2)  >0$,  the  occurrence of the state $\rho_2 $ 
 in a transition from  $\rho_1$ to $\rho_\beta$ is allowed by the 
 conventional second law. 
The generalized  2nd law is the statement   that state $\rho_2$ is allowed 
in a transition  from  $\rho_1$ to $\rho_\beta$  only if 
\cite{Brand_o_2015,Bernamonti:2018vmw}
\begin{equation}
 \delta D_\alpha(\rho_1||\rho_2)   \equiv 
 D_\alpha(\rho_1||\rho_\beta) - D_\alpha( \rho_2||\rho_\beta)>0 , \qquad
 {\mbox{for all }}  \alpha . 
\end{equation}
A heuristic understanding of this generalization can be obtained by considering 
the microcanonical ensemble. 
This implies $\rho_\beta   \rightarrow \rho^*$ where $\rho^*$ is the equal a-priori probability 
distribution. The above statement  is  then equivalent to maximizing the R\'{e}nyi entropy
during a transition.

\begin{figure}[!h]
  \centering
  \begin{subfigure}[b]{0.45\linewidth}
    \includegraphics[height=.6\textwidth, width=1\textwidth]{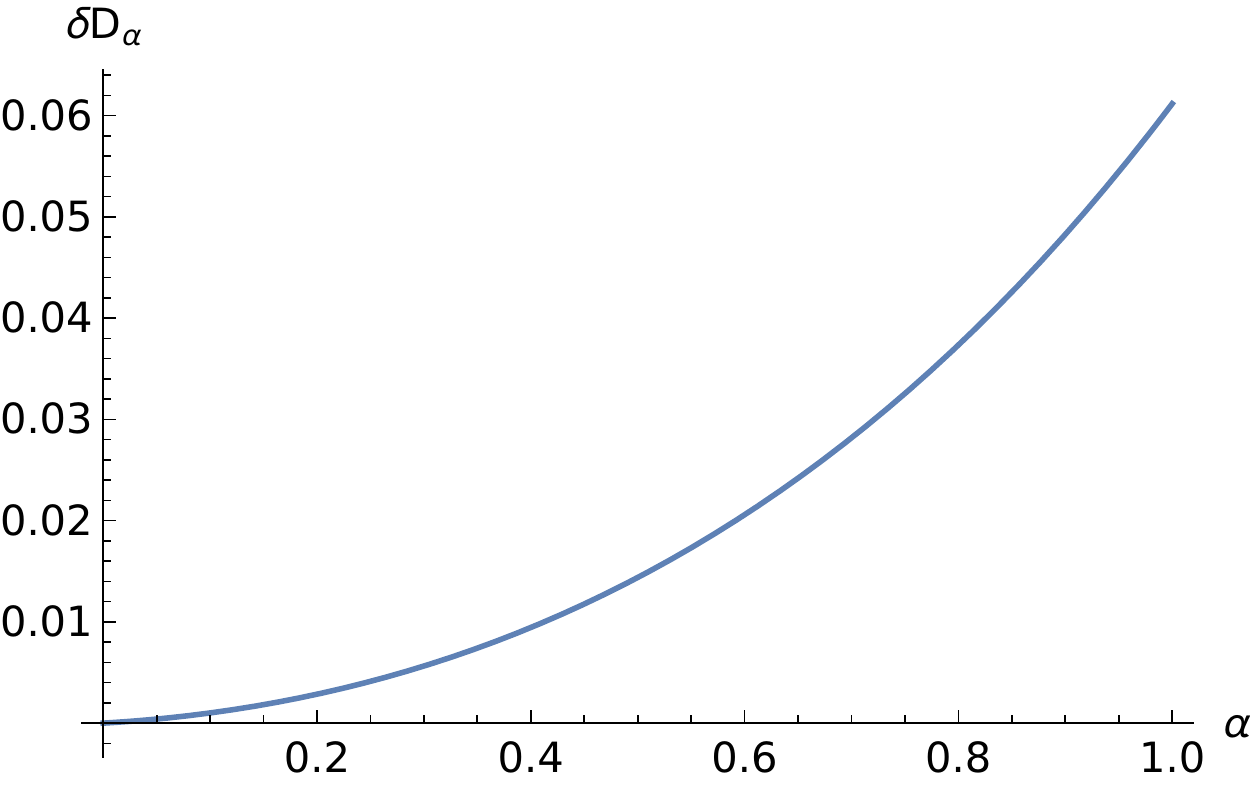}
    \caption{} \label{figsample1}
  \end{subfigure}
  \hspace{2pt}
  \begin{subfigure}[b]{0.45\linewidth}
    \includegraphics[height=.6\textwidth, width=1\textwidth]{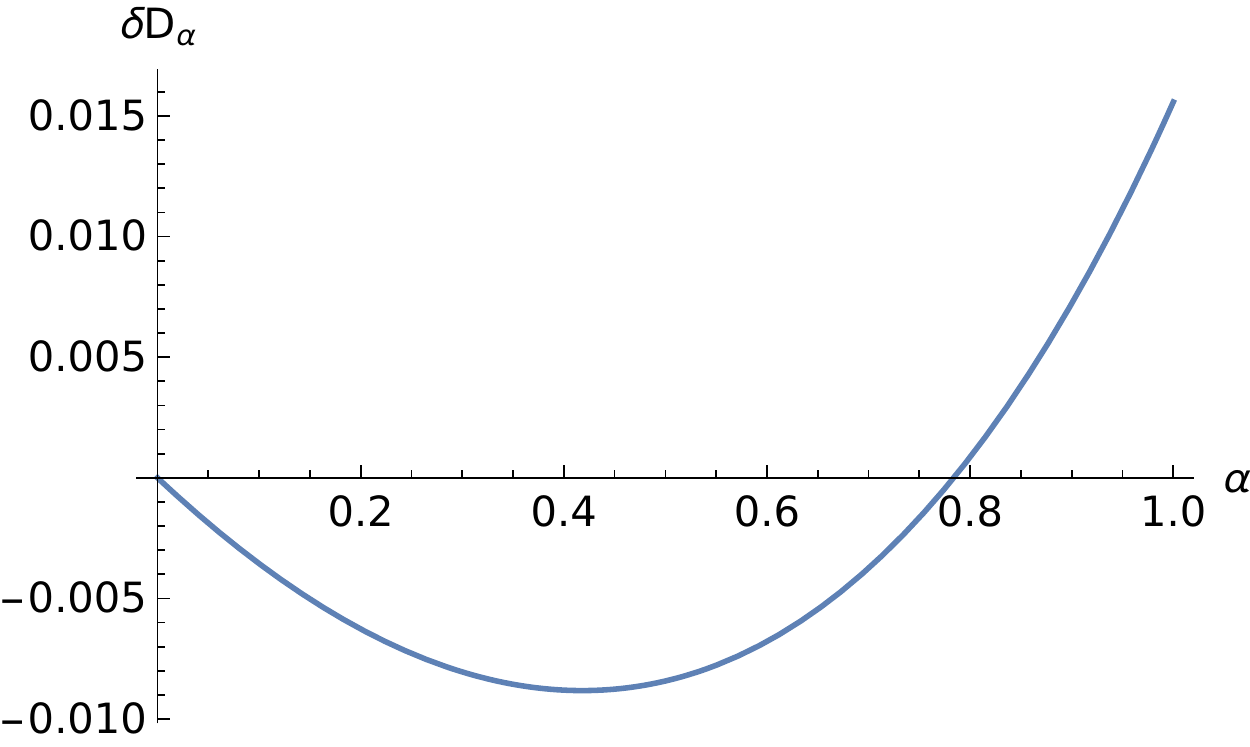}
    \caption{} \label{figsample2}
  \end{subfigure}
  \caption{(Left) A sample plot  of $\delta D_{\alpha}(\rho_1||\rho_2) $
   showing the case where there are no 
  additional constraints from the generalized second law for the transition from 
  $\rho_1\rightarrow \rho_2$. (Right) A sample plot of $\delta D_{\alpha}(\rho_1||\rho_2) $
  showing that the generalized 
  second law disallows the transition from $\rho_1\rightarrow \rho_2$ even though
  it is allowed by the conventional second law.  } 
\end{figure}

Thus, the  generalized second laws in principle represent more constraints that the 
traditional second law. 
However, it need not be always the case. 
Consider a plot of of $\delta D_{\alpha}( \rho_1||\rho_2) $  as shown in the figure \ref{figsample1}. 
It is clear from the shape of the graph that the transition from $\rho_1$ to $\rho_2$ 
which is allowed by the traditional second law at $\alpha=1$ continues to 
be allowed for all $0<\alpha<1$.  In this situation, the generalized second law 
does not provide any further constraint than that by the conventional second law.
However if the shape of the graph is as shown in figure \ref{figsample2}, then the 
generalized state that the transition $\rho_1$ to $\rho_2$ which is allowed by the 
conventional second law at $\alpha=1$ is forbidden  by the generalized second law. 
It is important that this discussion is for open systems  in contact with 
the heat bath and not for closed system where we also need to ensure conservation of 
energy.

In this section we will use the results for R\'{e}nyi divergences of sections 
and obtain situations of the kind shown in figure \ref{figsample2}. 
We will see in all the case we study there are situations in which 
the generalised second laws do indeed 
represent additional constraints. 
This study extends the observations of \cite{Bernamonti:2018vmw}  which were 
made for excitations
$\rho_1, \rho_2$ corresponding to deformations by conformal primaries of 
weights $\Delta_1, \Delta_2$ respectively.

\subsection{Higher temperature to  spin-1 charged states}
 
\begin{figure}[h!]
  \centering
  \begin{subfigure}[b]{.49
 \linewidth}
    \includegraphics[height=.8\textwidth, width=1\textwidth]{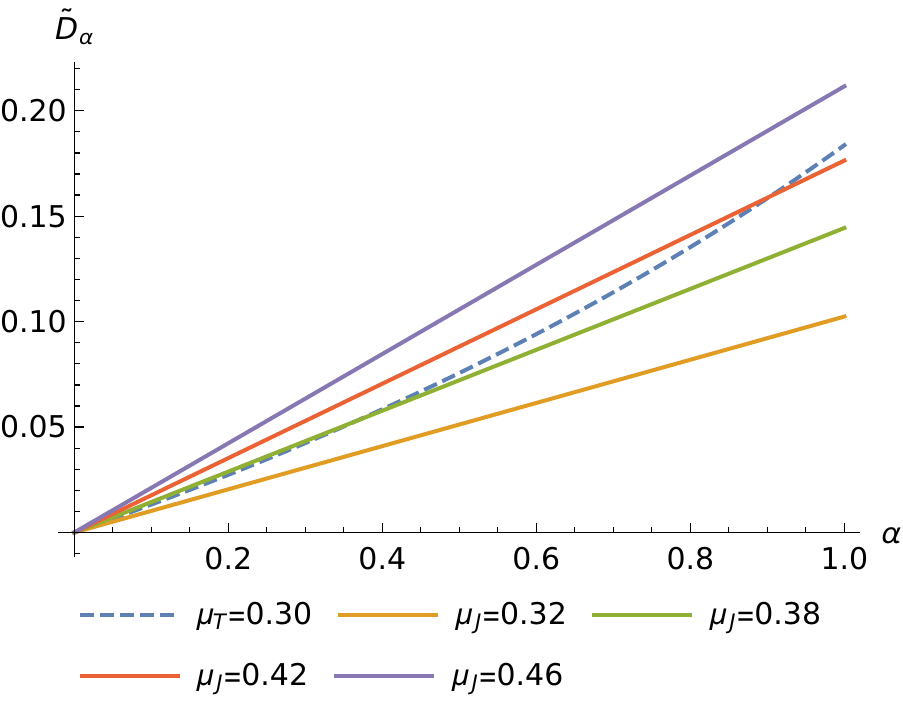}
    \caption{} \label{figTJtransition1}
    \label{TandJ}
  \end{subfigure}
  \hspace{1pt}
  \begin{subfigure}[b]{.49\linewidth}
    \includegraphics[height=.8\textwidth, width=1\textwidth]{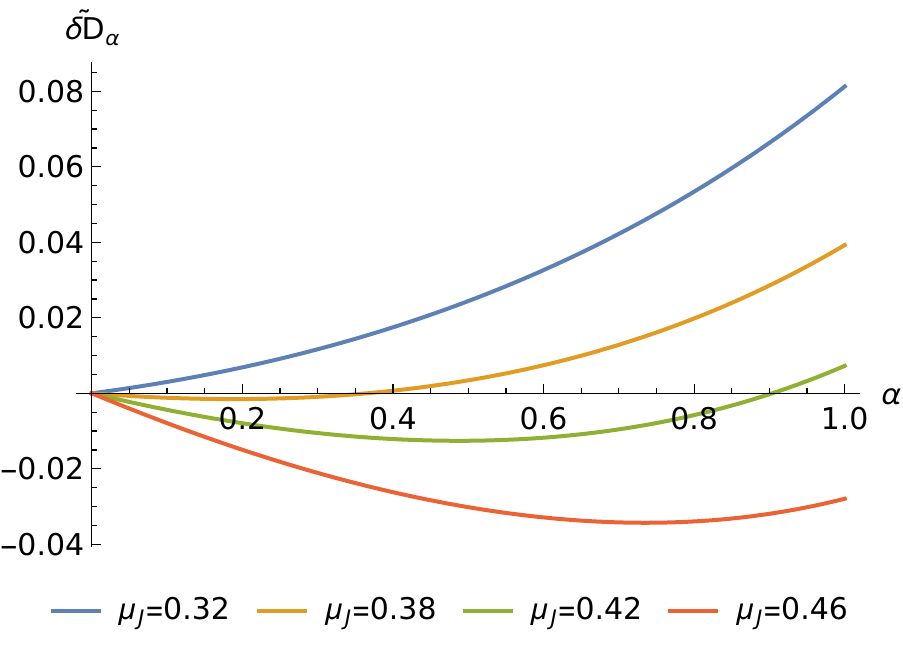}
  \caption{}    \label{figTJtransition2} 
  \end{subfigure}
  \caption{ (Left)  $ \tilde{D}_{\alpha}=\frac{ 6\beta }{\pi cL} D_{\alpha} $ vs $\alpha$ for fixed $\mu_T =0.3$.  Values of $\mu_J$ decrease  on the curves from top to bottom.  (Right) plot 
  of $ \delta\tilde{D}_{\alpha}=\frac{6\beta}{\pi cL } \delta D_\alpha ( \rho_T||\rho_J) $ vs $\alpha$ for fixed $\mu_T=0.3$. Note the  transition 
  represented by the curves 
   $\mu_J = 0.42, 0.38$ are allowed by the conventional first law but disallowed by the generalised second laws of thermodynamics. } 
\end{figure}

Let us consider the case of of the R\'{e}nyi divergence of excited state corresponding to 
the  stress 
temperature deformation, $\rho_T$ and that of the $U(1)$ current deformation $\rho_J$. 
From (\ref{renydivJ}) and (\ref{Da all order}), we write these results as 
\begin{eqnarray}
D_{\alpha, J}( \rho_J||\rho_\beta) & =& \frac{\pi c L}{6 \beta}    \mu_J^2 \alpha  , \\ \nonumber
D_{\alpha, T}(\rho_T||\rho_\beta)  &=& \frac{\pi c L}{6 \beta} \left( 
 \frac{\mu_{T}^2\alpha}{(1-\mu_T)(1-\alpha\mu_T)} \right) .
\end{eqnarray}
Here the $U(1)$ deformation has been redefined as 
\begin{eqnarray}
2\pi \kappa \mu^2 \beta^2  \rightarrow \mu^2_J~,
\end{eqnarray}
and the stress tensor deformation is defined as 
\begin{eqnarray}
2\pi \mu \rightarrow \mu_T~. 
\end{eqnarray}
With these definitions both $\mu_J, \mu_T$ are dimensionless, $\rho_\beta$ is the reference 
thermal state. 
 Note that the excited state corresponding to the stress tensor 
deformation can be thought of a thermal state at temperature
\begin{equation}\label{newtemp}
T'  = \frac{T}{1- \mu_T}~. 
\end{equation}
Let us define the   difference of the  R\'{e}nyi divergences which is given by 
\begin{equation}
\delta D_{\alpha}  ( \rho_T||\rho_J) = \frac{\pi c L}{6 \beta} \left( 
 \frac{\mu_{T}^2\alpha}{(1-\mu_T)(1-\alpha\mu_T)} -  \mu_J^2 \alpha \right)~. 
 \end{equation}
 In the figure  \ref{figTJtransition1} we have the R\'{e}nyi divergences  for 
 $\mu_T = 0.3$ along with various values of $\mu_J$. 
 It can be seen that $\mu_J =0.42, 0.38$, that the excited state  $\rho_J$ is allowed by 
 the traditional second law but disallowed by the generalized second law. 
  This is also clear from the plot of $\delta D_{\alpha} ( \rho_T||\rho_J)$ in 
  figure  \ref{figTJtransition2}.  Note that  for  $\mu_J = 0.46$, the charged state is prevented 
  to 
  occur as an intermediate state even by the second law. 
  
  Since we have analytical expressions for the R\'{e}nyi divergences, it is easy to find the 
  domains for which the intermediate state $\rho_J$ is allowed to occur by the 
  generalized second law for transitions from higher temperature to lower temperature. 
  Therefore  form (\ref{newtemp}) we have $0 <\mu_T <1$ and without loss of generality we can chose
  $\mu_J$ to be positive. 
  \begin{itemize}
  \item
  For small enough chemical potentials given by 
  \begin{equation}
  \mu_J < \frac{\mu_T}{\sqrt{ 1- \mu_T} } = \frac{ T'-T}{\sqrt{TT'} },  \qquad
  D_{\alpha, J} ( \rho_J||\rho_\beta) < D_{\alpha, T}( \rho_T ||\rho_\beta), 
  \end{equation}
  we see that the charged excited state $\rho_J$ is allowed by the both the  conventional 
  second law and the generalised second law. 
  \item
  For chemical potentials in the range 
  \begin{equation}
   \frac{\mu_T}{\sqrt{ 1- \mu_T} } <\mu_J < \frac{\mu_T}{1-\mu_T} = \frac{T'-T}{T}, 
  \end{equation}
  the excited state $\rho_J$ is allowed by the conventional second law, but disallowed by the
  generalised second law.  Note that for this range we have
  $\delta D_{1} ( \rho_T||\rho_J) >0$. 
  \item
  For large chemical potentials in the range
 \begin{equation}
  \frac{\mu_T}{1-\mu_T} < \mu_J , \qquad D_{\alpha, T}( \rho_T ||\rho_\beta) < D_{\alpha, J} ( \rho_J||\rho_\beta) , 
  \end{equation}
the transition intermediate state $\rho_J$ does not occur in the transition 
from $\rho_T$ to $\rho_\beta$.  
  \end{itemize}
  
\subsection{Transitions between two  temperatures}

We  have seen that given  the stress tensor deformation $\mu_1$, the 
density matrix of the excited state corresponds to that of temperature
\begin{equation}
T_1 = \frac{T}{ 1- \mu_1}.
\end{equation}
The R\'{e}nyi divergence is given by 
\begin{equation}
D_{\alpha}( \rho_1||\rho_\beta) = \frac{\pi cL }{6 \beta} \left( \frac{ \mu_1^2 \alpha}{ (1-\mu_1) ( 1- \mu_1\alpha )}  \right).
\end{equation}

\subsubsection*{Transitions from higher temperature to lower temperature}

\begin{figure}[h!]
  \centering
  \begin{subfigure}[b]{.49
 \linewidth}
    \includegraphics[height=.8\textwidth, width=1\textwidth]{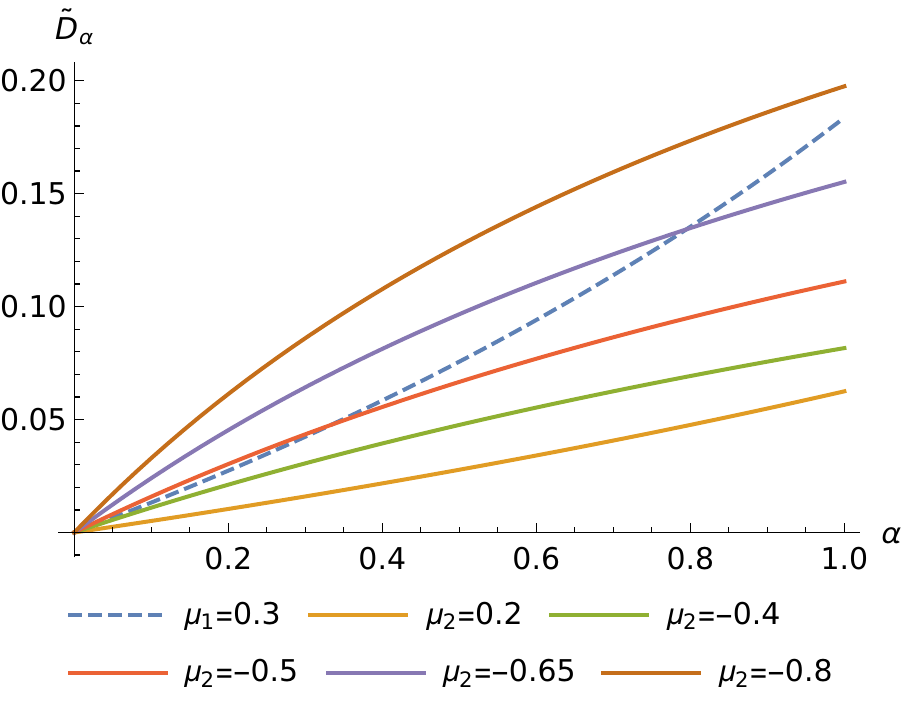}
    \caption{} \label{figTTtransition1}
  \end{subfigure}
  \hspace{1pt}
  \begin{subfigure}[b]{.49
 \linewidth}
    \includegraphics[height=.8\textwidth, width=1\textwidth]{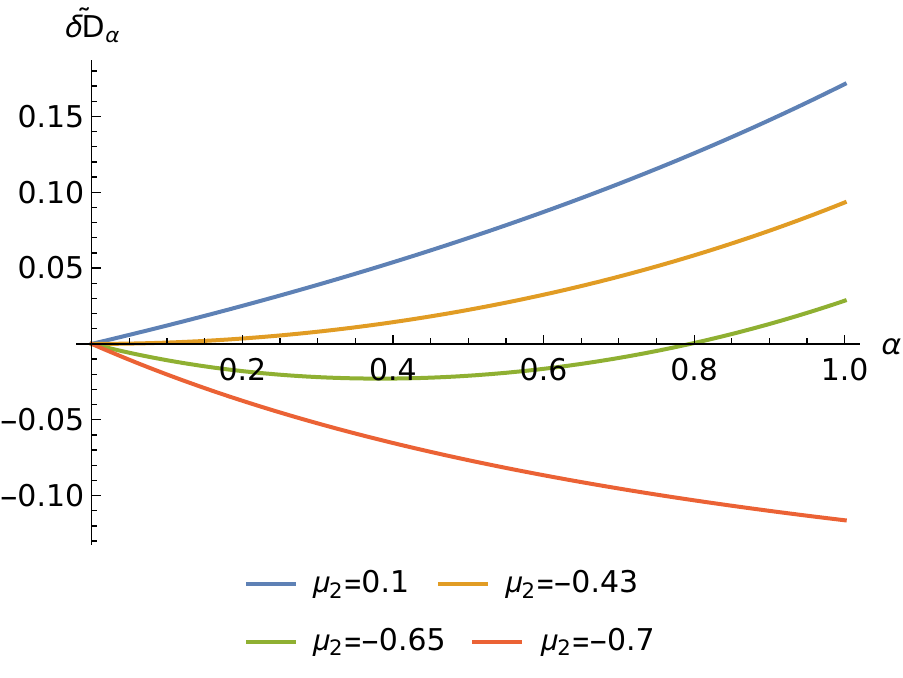}
    \caption{} \label{figTTtransition2}
  \end{subfigure}
	\caption{ (Left) 
	 $\tilde{D}_{\alpha}=\frac{6\beta}{\pi cL}  D_{\alpha}$ vs $\alpha$ for fixed $\mu_1=0.3$,   we see the the generalised second law prevents
	 transitions to  states with  $\mu_2$ in the range  $\frac{\mu_1}{2\mu_1-1}=-0.75 < \mu_2 <\frac{\mu_1}{\mu_1-1}=-0.428$. 
	 The conventional second law disallows states with  $\mu_2 < -0.75$.  (Right)  
	 $ \delta \tilde{D}_{\alpha}=\frac{6\beta}{\pi cL} \delta D_{\alpha}(\rho_1||\rho_2)$  vs $\alpha$ for fixed $\mu_1=0.3$ and different $\mu_2$.  }
\end{figure}

Let the state $\rho_1$  be at higher temperature,   $0< \mu_1  <1$ . Now  consider another  state $\rho_2$  corresponding to the 
stress tensor deformation $\mu_2$.  We can now ask in the transition from $\rho_1$ to  $\rho_\beta$ is it possible that the 
state $\rho_2$ be allowed by the traditional second law, but dis-allowed by the generalised second law. 
In figure  \ref{figTTtransition1}  we have plotted various values of $\mu_2$  for a fixed value of $\mu_1 = 0.3$. 
We find that  for $ -0.75 < \mu_2 < -0.428$, that is a band of lower temperatures than $T$, it is indeed the case 
that there are situations where  the traditional first law  at $\alpha=1$ allows the state $\rho_2$ to occur 
in the transition from $\rho_1 \rightarrow \rho_\beta$, however the generalised second law prohibits this process. 
This is also confirmed in figure \ref{figTTtransition2} which plots the difference in R\'{e}nyi  divergences  defined by 
\begin{equation}
\delta_\alpha D_\alpha( \rho_1||\rho_2) = \frac{\pi cL }{6 \beta}
\left( \frac{ \mu_1^2 \alpha}{ 1-\mu_1) ( 1- \mu_1\alpha ) } -  \frac{ \mu_2^2 \alpha}{ (1-\mu_2) ( 1- \mu_2\alpha ) }  \right) ~.
\end{equation}

Since we have analytical and simple  algebraic expressions for the R\'{e}nyi divergences,  by studying various inequalities 
we can make the following general statements for the transition form $( \rho_1, \mu_1) $ at higher temperature  to $\rho_\beta$. 
\begin{itemize}
\item In the range $\frac{1}{2} \leq \mu_1 < 1$,  we see that for $\mu_2<\mu_1$ and for  $0<\alpha<1$ we have 
\begin{equation}
D_\alpha ( \rho_1|| \rho_2)   >0   \quad {\rm for} \; \; \frac{1}{2} <\mu_1 < 1  \quad{\rm and }\; \; \mu_2 <\mu_1~.
\end{equation}
This of course implies that there are no situations in which the generalised second law implies more constraints than the 
conventional second law. 
\item
In the range $0<  \mu_1 < \frac{1}{2}$, 
for states with 
\begin{equation}
D_\alpha ( \rho_1|| \rho_2)   >0   \quad {\rm for} \;\; 0  <\mu_1 < \frac{1}{2}  \quad {\rm and}\;\;   - \frac{\mu_1}{ 1- \mu_1} <\mu_2 < \mu_1 ~.
\end{equation}
Again this implies that there are no situations in which the generalised second law implies more constraints than the 
conventional second law. 
\item
But in the range 
\begin{equation}
0  <\mu_1 < \frac{1}{2}  \quad {\rm and}\;\;    - \frac{\mu_1} { 1- 2\mu_1}  <\mu_2 <  - \frac{\mu_1}{ 1- \mu_1}, 
\end{equation}
we  see that the generalised second law prevents the occurrence of the state $( \rho_2 , \mu_2)$ in the transition 
from $(\rho_1, \mu_1)$ to $\rho_\beta$ which is allowed by the traditional second law. 
\item
Finally in the range
\begin{equation}
0  <\mu_1 < \frac{1}{2}  \quad {\rm and}\; \;    -\infty   <\mu_2 <  - \frac{\mu_1}{ 1-2 \mu_1}. 
\end{equation}
The traditional second law itself prevents the occurrence of the state $( \rho_2 , \mu_2)$ . 
\end{itemize}
In summary we conclude that in a transition from as state of higher temperature $(\rho_1,\mu_1)$ to 
$\rho_\beta$, 
the states which are allowed to occur  by the generalised   second law satisfies the condition, 
\begin{equation}
- \frac{\mu_1}{ 1- \mu_1} < \mu_2 <\mu_1. 
\end{equation}
It is interesting to note that  $T_1= \frac{T}{ 1- \mu_1}$  the lowest  temperature $T_2^* = - \frac{\mu_1}{ 1- \mu_1}  $ allowed satisfies the following property
\begin{equation}
T_1 T_2^* = T . 
\end{equation}
That is $T$ is the geometric mean of $T_1$ and $T_2^*$. 

\subsubsection*{Transitions from lower temperature to higher temperature}

In the transition from the  excited state  $\rho_1$ with  $-\infty <\mu_1< 0$ to   $\rho_\beta$ we find that there are no 
situations in which the generalized second law prevents a transition which is allowed  by the traditional 
second law. 
The higher temperature states $(\rho_2, \mu_2) $  which are allowed as intermediate states in this transition 
satisfy the  condition 
\begin{equation}
 -\infty <\mu_1< 0  \quad {\rm and } \;\;\; \mu_1 < \mu_2 < \frac{-\mu_1 }{ 1- 2\mu_1}. 
\end{equation}
It is again interesting to note that  $T_1= \frac{T}{ 1- \mu_1}$ and highest  temperature  $ T_2^*  = \frac{-\mu_1 }{ 1- 2\mu_1}$ 
which is allowed satisfies the property
\begin{equation}
{T_1 + T_2^*} =   2T. 
\end{equation}
Here $T$ is the arithmetic mean of $T_1$ and $T_2^*$.

\subsection{Spin-3 charged state to higher temperature}

Let us consider  transitions from the excited state corresponding to the spin-3 deformation $(\rho_W, \mu_W) $  to 
the state $(\rho_T, \mu_T) $ which is  the excited state corresponding to the stress tensor deformation. 
From (\ref{renyidwdef}) and (\ref{Da all order}) the  R\'{e}nyi divergences  for these states are given by  
\begin{eqnarray}
D_{\alpha} ( \rho_W||\rho_\beta) &=& 
\frac{\pi c L}{6 \beta} \left[
\frac{16}{3} \left( \frac{\pi\mu_W}{\beta} \right)^2 \alpha + 
\frac{12800}{27} \left( \frac{\pi \mu_W}{\beta}\right)^4 \alpha( \alpha^2 + \alpha +1) 
+ \cdots \right],  \nonumber \\
D_{\alpha, T}(\rho_T||\rho_\beta)  &=& \frac{\pi c L}{6 \beta} \left( 
 \frac{\mu_{T}^2\alpha}{(1-\mu_T)(1-\alpha\mu_T)} \right) 
\end{eqnarray}
Let us also define 
\begin{equation}
\delta D_\alpha ( \rho_W|| \rho_T) =  D_{\alpha} ( \rho_W||\rho_\beta)  - D_{\alpha, T}(\rho_T||\rho_\beta) ~.
\end{equation}

\begin{figure}[h]
	\begin{center}
		\includegraphics[width=.6\linewidth, height=.45\textwidth]{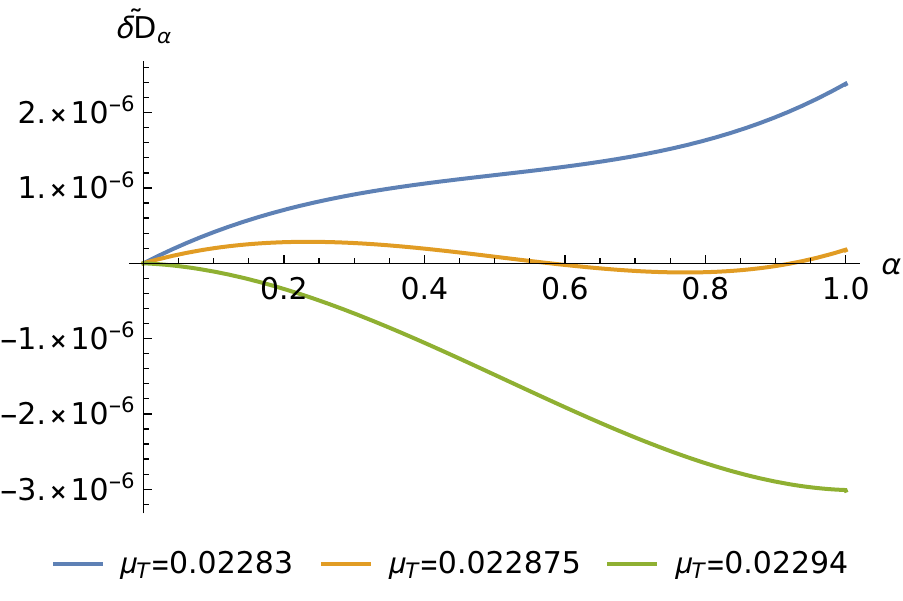}
	\end{center}
	\caption{ $\tilde{D}_{\alpha}=\left( \frac{6\beta}{\pi c L}\right) \delta D_\alpha ( \rho_W|| \rho_T)$ 
	 vs $\alpha$ for fixed $\mu_W=0.02$.  The transition represented by the curve for $\mu_T =0.022875$ is 
	 allowed by the conventional second law but disallowed by the 
	 generalized second laws of thermodynamics. } \label{WT1}
\end{figure}

We can study if there are additional constraints from the generalised second law for the occurrence of the 
state $(\rho_T, \mu_T)$ in the transition from $(\rho_W||\rho_\beta)$ to $\rho_\beta$. 
The series representing $D_{\alpha} ( \rho_W||\rho_\beta) $ is an asymptotic series for small $\mu_W$. 
Therefore to trust our results we need to take the values of $\mu_W$ and the number of terms in the
expansion with care. 
In  figure \ref{WT1} we have plotted $\delta D_\alpha ( \rho_W|| \rho_T)$ for a fixed $\mu_W = 0.002$ for three 
values of $\mu_T$. 
We have kept to order $\mu_W^{20}$ terms in the expansion of $D_{\alpha} ( \rho_W||\rho_\beta) $, and have verified that 
the asymptotic expansion gives the same  curves  by keeping  terms  to order $\mu_W^{10}$  or  to order   $\mu_W^{50}$. 
Thus the asymptotic expansion gives stable values for  $\delta D_\alpha ( \rho_W|| \rho_T)$ for  $\mu_W = 0.002$. 
Note that the curve for $\mu_T =0.022875$ represents the situation for which the state
$\rho_T$ is allowed for the transition from $\rho_W$ to $\rho_\beta$  by the  conventional second law of thermodynamics
but is disallowed by the generalized second laws. 
We have  also seen, that 
there are such additional constraints from the 
generalized second laws for the occurrence of $\rho_{W}$ in transitions from the state  at  higher temperature 
to the charged spin-3 state $\rho_W$.

\subsection{Transitions involving SSD states}

\begin{figure}[h]
	\begin{center}
	  \begin{subfigure}[b]{.49\linewidth}
		\includegraphics[height=.8\textwidth, width=1\textwidth]{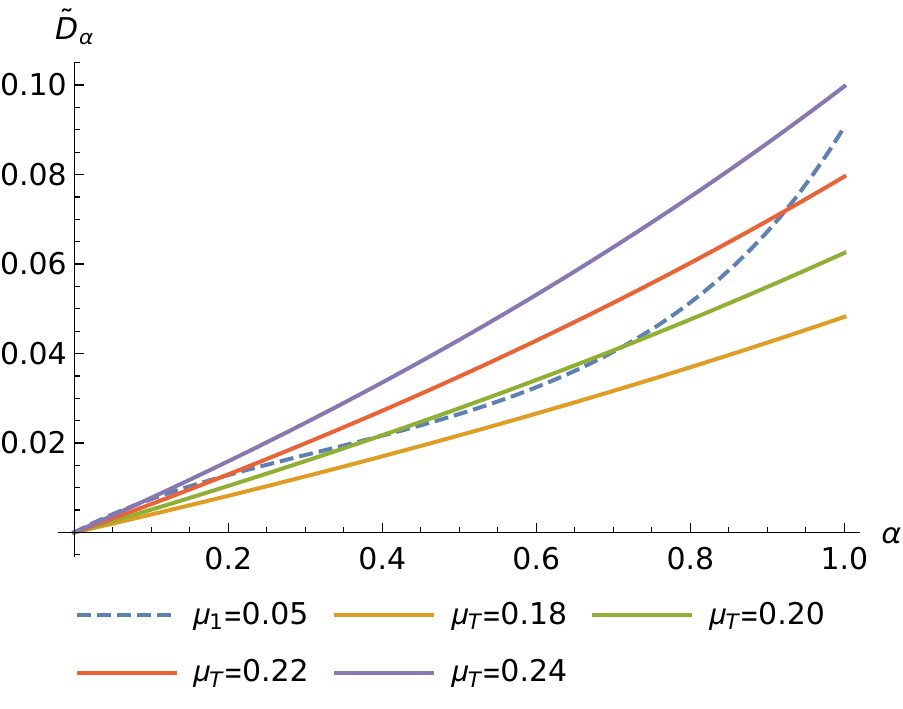}
		\caption{} \label{figtssd1}
		\end{subfigure}
		\hspace{1pt}
		\begin{subfigure}[b]{.49 \linewidth}
\includegraphics[height=.8\textwidth, width=1\textwidth]{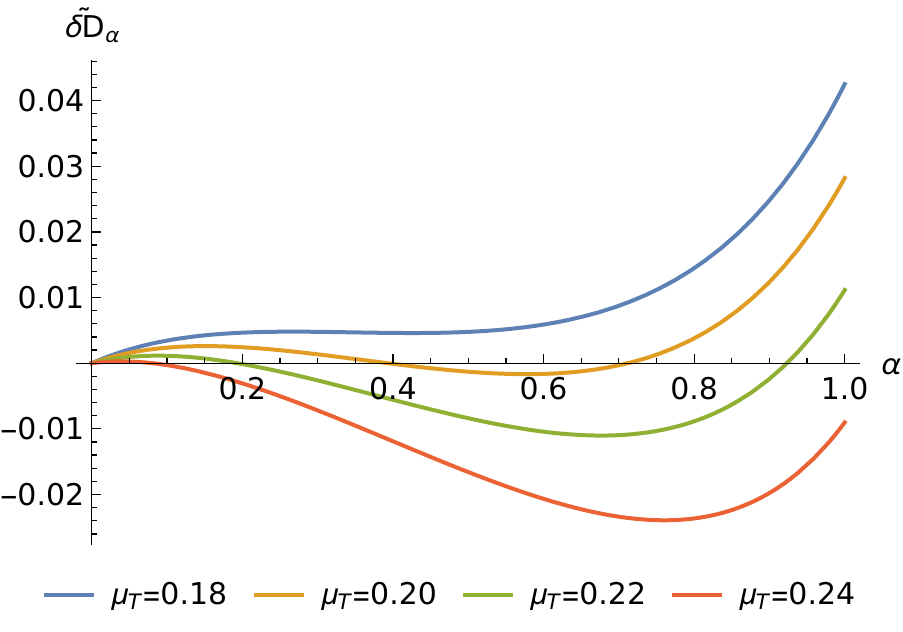}
		\caption{}\label{figtssd2}
		\end{subfigure}
		\end{center}
	\caption{(Right)  $\tilde{D}_{\alpha}=\frac{6\beta}{\pi cL} D_\alpha$ vs $\alpha$ for different values of $\mu_T$ and  fixed $\mu_{1}=0.05$.  
	(Left) $\delta\tilde{D}_{\alpha}=\delta D_{\alpha} ( \rho_{k=1}|| \rho_T)$ for a fixed $\mu_{1}=0.05$. 
	We have set $\beta/L =1$
	The transition represented by the curve $\mu_T  = 0.22$
	 is allowed by the conventional second law but 
	 disallowed by the generalized second laws. }
\end{figure}

The R\'{e}nyi divergences  of the excited state  $(\rho_{k=1} $  corresponding to the SSD deformation $g ( L_1 + L_{-1})$ in the 
Hamiltonian 
to the quadratic order in perturbation theory on the 
torus  is given by 
\begin{eqnarray}
D_{\alpha}( \rho_{k=1}||\rho_\beta) 
&=&  \mu_1^2 \frac{\pi c L}{6 \beta} \frac{ 
 \sinh \frac{ \pi\alpha\beta}{L} \sinh \frac{\pi ( \alpha-1) \beta}{L} }{ \alpha -1} . 
\end{eqnarray}
Here we have taken  a rectangular torus by setting 
  $\tau = \frac{i\beta}{L}$ and absorbed the prefactors $2 g^2 q \partial_q \log f(\tau, \bar \tau)$
  into the coupling  $\mu_{1}$.  This is possible since we work at a fixed $\tau$. 
We can now consider transitions between the SSD deformed state $(\rho_{k=1}, \mu_1)$  and  the excited 
state corresponding to the stress tensor deformation $(\rho_T, \mu_T)$ with R\'{e}nyi divergence
\begin{eqnarray}
D_{\alpha}( \rho_T||\rho_\beta) &=& \frac{\pi c L}{6\beta}  \left( 
 \frac{\mu_{T}^2\alpha}{(1-\mu_T)(1-\alpha) } \right) .
 \end{eqnarray}
 Let us define  the difference
 \begin{equation}
 \delta D_{\alpha} ( \rho_{k=1}|| \rho_T) = D_{\alpha}( \rho_{k=1}||\rho_\beta) - D_{\alpha}( \rho_T||\rho_\beta) .
 \end{equation}
 In figure  \ref{figtssd1} we have plotted the R\'{e}nyi divergences of the excited  states corresponding to the $(\rho_T, \mu_T)$ 
 for different values of $\mu_T$ along with  the R\'{e}nyi divergence of the SSD state $(\rho_T, \mu_{1})$ at $\mu_1 = 0.05$. 
 We see that   for  $\mu_T =0.22$, the transition from the SSD excited state is allowed by the conventional second law, 
 but disallowed by the generalised second laws. This can also be seen in figure \ref{figtssd2} which plots the difference 
  $\delta D_{\alpha} ( \rho_{k=1}|| \rho_T)$. 
  We have chosen value of  $\beta/L =1$ and $\mu_1$ so that it 
  \begin{equation}
  \mu_1^2 \ll \frac{1}{\pi\sinh \pi }  
  \end{equation}
  This ensures that  the R\'{e}nyi divergence at $\alpha =1$, its largest value  
is small so that perturbation theory can be trusted. 

Finally let us consider transitions from the SSD state $(\rho_{k=2}, \mu_2)$  with R\'{e}nyi divergence
given by. Here the SSD state is obtained by adding the deformation $g( V_2^{(s)} + V_{-2}^{(s)})$. 
\begin{equation}
D_{\alpha} ( \rho_{k=2}|| \rho_\beta) =  \mu_2^2 \frac{\pi c L}{6 \beta} \frac{ 
 \sinh \frac{ 2 \pi\alpha\beta}{L} \sinh \frac{2 \pi ( \alpha-1) \beta}{L} }{ \alpha -1}
\end{equation}
To obtain this expression we have taken  $\tau = \frac{i \beta}{L}$ in  (\ref{ssdarbk}) 
and absorbed all the resulting prefactors into $\mu_2^2$. 
We can also consider the difference 
\begin{eqnarray}
\delta D_{\alpha} ( \rho_{k=2}|| \rho_{k=1}) = 
\mu_1^2 \frac{\pi cL}{c\beta}   \left( 
\frac{  \gamma
 \sinh \frac{ 2 \pi\alpha\beta}{L} \sinh \frac{2 \pi ( \alpha-1) \beta}{L} }{ \alpha -1}
 -
 \frac{ \sinh \frac{  \pi\alpha\beta}{L} \sinh \frac{\pi ( \alpha-1) \beta}{L} }{ \alpha -1}
 \right) , \nonumber \\
\end{eqnarray}
where the ratio  $\gamma = ( \frac{\mu_2}{\mu_1})^2$. 

\begin{figure}[h!]
	\begin{center}\begin{subfigure}[b]{.49\linewidth}
		\includegraphics[height=.8\textwidth, width=1\textwidth]{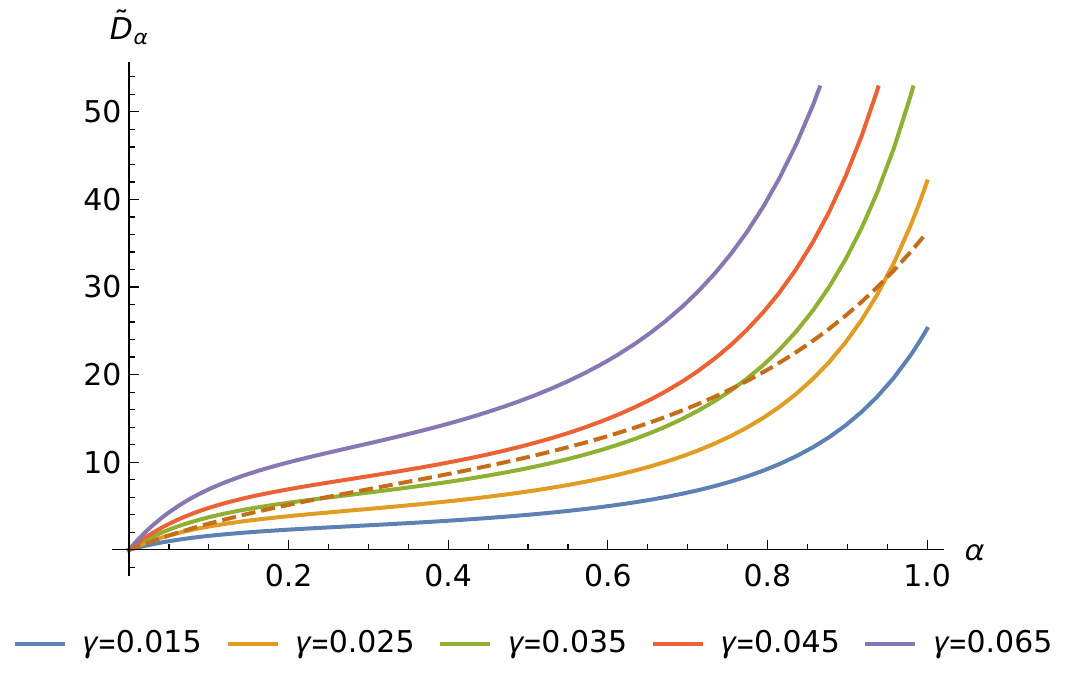}
		\caption{} \label{figtssd12}
		\end{subfigure}
		\hspace{1pt}
		\begin{subfigure}[b]{.49 \linewidth}
		\includegraphics[height=.8\textwidth, width=1\textwidth]{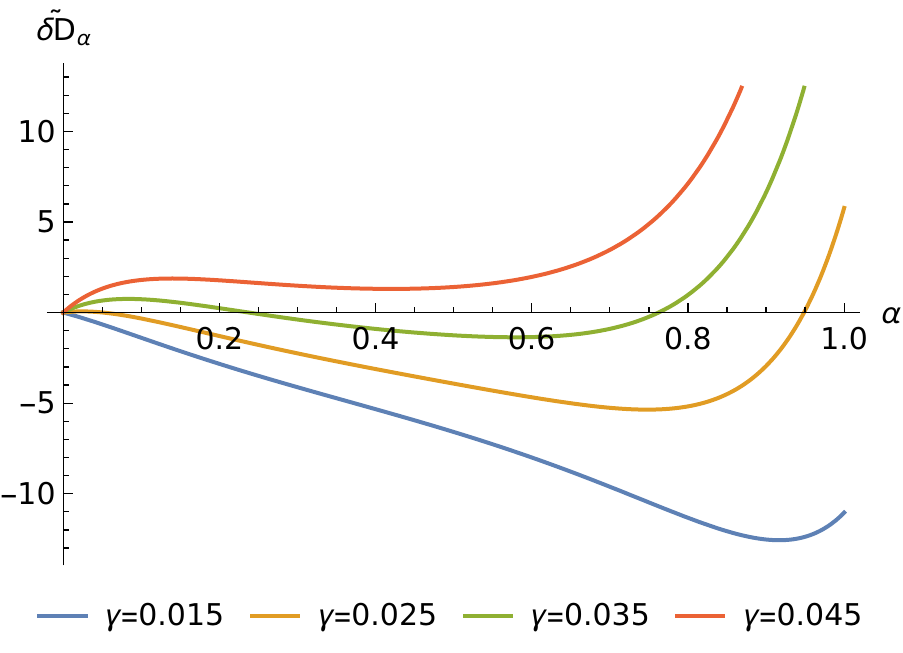}
		\caption{}\label{figtssd22}
		\end{subfigure}
		\end{center}
	\caption{ (Left) 
	$ \tilde{D}_{\alpha}=\left( \frac{6\beta}{\pi c L\mu_1^2}\right)D_{\alpha}$ with 
	 $\frac{\beta}{L} =1$.The dashed curve represents 
	$\frac{ 6\beta}{\pi c L \mu_1^2}D_{\alpha}(\rho_{k=1}||\rho_\beta)$. The 
	coloured curves correspond to 
	$\frac{ 6\beta}{\pi c L \mu_1^2} D_{\alpha}(\rho_{k=1}||\rho_\beta)$  for different values of 
	$\gamma$.  The curves $\gamma=0.035, 0.025 $ represent a transition from  the SSD $k=2$
	excited state to the $k=1$ state which is allowed by the second law but prohibited 
	by the generalised second law. (Right)  $\delta\tilde{D}_{\alpha}=\left( \frac{6\beta}{\pi c L\mu_1^2}\right)\delta D_{\alpha} ( \rho_2|| \rho_1)$  for 
	various values of $\gamma$.  }
\end{figure}

In figure \ref{figtssd12}, we  have plotted the R\'{e}nyi divergences of excited state 
$(\rho_{k=2}, \mu_2)$,   
$(\rho_{k=1} , \mu_1)$   for various values of the ratio $\gamma$. 
The R\'{e}nyi divergences are normalized by dividing $\frac{\pi cL \mu_1^2}{6\beta}$. 
The dashed  curve represents the SSD deformation for $k=1$, while the solid
curves are that for $k=2$. 
The curves corresponding to $\gamma = 0.035, 0.025$ are transitions
  from the $k=2$ SSD excited state
to the $k=1$ SSD state which is allowed by the conventional second law, but  prohibited by 
generalised second laws.  We have set $\frac{\beta}{L} =1$. 
Figure \ref{figtssd22}, show the same by plotting $\delta D_{\alpha} ( \rho_{k=2}|| \rho_{k=1})$ 
for various 
values of $\gamma$. 

\section{Conclusions}
\label{conclusion}

In this paper we evaluated the R\'{e}nyi divergences between the  thermal density matrix and 
excited states
created by deforming  the Hamiltonian by scalar primaries, conserved currents and SSD deformations
in 2d CFTs.  One of our aims was to obtain  tractable examples so that the properties of R\'{e}nyi divergences
can be studied further.  
Our study  has shown that the  $\alpha$ dependence of 
leading correction to the R\'{e}nyi divergences is universal as long as the  deforming operators 
 together with the Hamiltonian satisfy the Heisenberg algebra. 
Indeed as far as we are aware, the   $\alpha$ dependence of R\'{e}nyi divergences
have been obtained only numerically  as in \cite{Bernamonti:2018vmw}. 
Systems such as the ones  considered in this paper  with 
simple  $\alpha$ dependence  for which R\'{e}nyi divergences can be obtained exactly 
have not been studied. 
We used these simple examples to show that the generalized second laws 
of thermodynamics put forward in \cite{Brand_o_2015,Bernamonti:2018vmw}  do place
more constraints on the evolution of non-equilibrium systems compared to the traditional second law. 
The methods developed in this paper -- conformal perturbation theory for holomorphic 
deformations and Hamiltonian perturbation theory -- can be used for other systems.

It is worthwhile to consider \Re divergences beyond the ones considered in this work and \cite{Bernamonti:2018vmw}. Specifically, an interesting future problem is to study the \Re divergence for the $T\bar{T}$ deformation \cite{Cavaglia:2016oda,Smirnov:2016lqw}. This entails the calculation of 
$$\cZ_\alpha (\lambda) = \tr[q^{\alpha H_{T\bar T} (\lambda)} q^{(1-\alpha)H_{\rm CFT}} ]~,$$
where, $\lambda$ is the $T\bar T$ coupling. 
Analysing this quantity would shed light on the differences in the nature of eigenstates of the deformed and undeformed theories. 
Note that although the spectrum of $H_{T\bar T}$ and $H_{\rm CFT}$ are related by a one-to-one map, the Hamiltonians themselves do not commute. This renders the calculation of $\cZ_\alpha(\lambda)$ non-trivial.
One way to proceed is to derive a flow equation for the above quantity along the lines of \cite{Cardy:ttb} or potentially bootstrapping using modular properties \cite{Aharony:2018bad,Datta:2018thy}. Moreover, it would also be fruitful to check whether the integration prescription used in this work or variants discussed in \cite{Datta:2014zpa} can be employed to treat this deformation perturbatively. 

Finally,  the universality in the $\alpha$ dependence of  R\'{e}nyi divergences for 
SSD deformations encourages us that it is worthwhile to engineer holographic setups
 for such quenches. 
The SL$(2, \mathbb{R})$ Chern-Simons theory  in 3-dimensions  and its higher spin generalization
provide  natural frameworks to think of these deformations. This will not only teach us more 
 about  holography in these systems, but also lead to  constructions with Euclidean time dependences in the bulk. 
 Such  aspects of holography have not been fully explored. It will be  interesting to see
 whether such holographic setups together with the generalised second laws
  will have lessons for the processes of black hole formation and evaporation. 
  
 \acknowledgments

We thank Pawel Caputa, Diptarka Das, Michael Gutperle, Per Kraus and Tomas Prochazka for fruitful discussions. SD thanks International Centre for Theoretical Sciences (ICTS) for support and hospitality during the program --  Thermalization, Many body localization and Hydrodynamics (ICTS/hydrodynamics2019/11).


\appendix

\section{Details  for spin-2 and spin-3 deformations}
\label{appendA}

In this appendix we provide the details of the correlations functions and the integrals 
for evaluating the partitions functions for spin-2 and spin-3 deformations. 
For these deformations we  carry out the perturbative expansion to order $\mu^4$. 

\subsection{Spin-2 deformation to  the quartic order } \label{integrals1}

To carry out the the perturbative expansion of the partition function in (\ref{perttexp}) to order
$\mu^4$ we would require the $3$ point function and the $4$ point function of the 
stress tensor on the cylinder as well as the integrals on the cylinder. 

We begin with the $3$-point function of the stress tensor on the plane, 
which is given by 
\begin{align}
& \langle T(\omega_1) T(\omega_2) T(\omega_3) \rangle  =  
\left( \frac{ \pi}{\beta}  \right)^6 \frac{c}{ 
\sinh^2\frac{\pi}{\beta} ( w_1 - w_2) 
\sinh^2\frac{\pi}{\beta} ( w_2 - w_3) 
\sinh^2 \frac{\pi}{\beta} ( w_1 - w_3) }   \nonumber \\
& \qquad\qquad -\frac{c^2}{12} \left(  \frac{\pi}{\beta} \right)^6 
\left( \frac{1}{\sinh^4 \frac{\pi }{\beta} ( w_1 - w_2) } 
+ \frac{1}{\sinh^4 \frac{\pi }{\beta} ( w_2 - w_3) } 
+ \frac{1}{\sinh^4 \frac{\pi }{\beta} ( w_1 - w_3) }  \right)   \nonumber \\
& \qquad\qquad-  \frac{c^3  \pi^6}{216 \beta^6} ~. 
\end{align}
We now integrate the $3$-point function on the cylinder using the prescription 
discussed in section \ref{unifsour}. 
This results in 
\begin{eqnarray}  \label{T3integrations2}
\int d^2 \omega_1 \int d^2 \omega_2  \int d^2 \omega_3 
\langle T(\omega_1) T(\omega_2) T(\omega_2) \rangle  
=  -\frac{4 c L \pi^4}{\beta}-\frac{c^2 L^2 \pi^5}{3 \beta^2}-\frac{c^3 L^3 \pi^6}{216 \beta^3} .
\nonumber \\
\end{eqnarray}

The $4$-point function of the stress tensor on the cylinder can be obtained 
using the conformal Ward identity. It  is given by 
\begin{eqnarray}
& & \langle T(\omega_1) T(\omega_2) T(\omega_3) T(\omega_4)  \rangle 
=
\langle T(\omega_1) T(\omega_2) T(\omega_3) T(\omega_4)  \rangle^{(0)} =
\\ \nonumber
& & - \frac{c^2}{6} \left( \frac{\pi }{\beta} \right)^8  \left( 
\frac{1}{ 
\sinh^2\frac{\pi}{\beta} ( w_1 - w_2) 
\sinh^2\frac{\pi}{\beta} ( w_2 - w_3) 
\sinh^2 \frac{\pi}{\beta} ( w_1 - w_3) }
+ 3 \; {\rm permutations}  \right) 
\\ \nonumber
&  &\qquad \qquad \qquad\qquad \qquad + \frac{c^3}{72} \left( \frac{\pi }{\beta} \right)^8  
\left( \frac{1}{ \sinh^2 \frac{\pi}{\beta} ( w_1 - w_2) } + 5 \; {\rm permutations} 
\right) 
\\ \nonumber
& & \qquad\qquad\qquad\qquad\qquad +\frac{c^4}{1296}  \left( \frac{\pi }{\beta} \right)^8  .
\end{eqnarray}
where
\begin{eqnarray}
& & \langle T(\omega_1) T(\omega_2) T(\omega_3) T(\omega_4)  \rangle^{(0)}
=\frac{c^2}{4} \left( \frac{\pi}{\beta} \right)^8
\left( \frac{1}{\sinh^4 \frac{\pi}{\beta}  w_{12}  \sinh^4 \frac{\pi}{\beta}  w_{34} }
 + ( 2\leftrightarrow 3) + ( 1\leftrightarrow 3)  \right. 
 \nonumber \\ \nonumber
 & & \frac{8}{c} \frac{1}{ \sinh^2\frac{\pi}{\beta} w_{12} 
 \sinh^2\frac{\pi}{\beta} w_{34} \sinh^2\frac{\pi}{\beta}w_{23} \sinh^2\frac{\pi}{\beta}  w_{14} }
 +\frac{8}{c} \frac{1}{ \sinh^2\frac{\pi}{\beta} w_{13} 
 \sinh^2\frac{\pi}{\beta} w_{24} \sinh^2\frac{\pi}{\beta}  w_{23} \sinh^2\frac{\pi}{\beta}  w_{14} }
 \\ 
 & &\left.  - \frac{8}{c} \frac{1}{ \sinh\frac{\pi}{\beta} w_{12} 
\sinh\frac{\pi}{\beta} w_{34}  \sinh\frac{\pi}{\beta} w_{13}
\sinh\frac{\pi}{\beta} w_{24} 
\sinh^2\frac{\pi}{\beta} w_{23} 
\sinh^2\frac{\pi}{\beta}  w_{14} }
\right) .
\end{eqnarray}
Here $w_{ij} = w_i  - w_j$. 
The integrals on the cylinder are given by 
\begin{eqnarray}
& &  \left( \frac{\pi}{\beta} \right)^8 \int 
\frac{1}{\sinh^4 \frac{\pi}{\beta}  w_{12}  \sinh^4 \frac{\pi}{\beta}  w_{34} }
= \frac{16}{9} \frac{\pi^6 L^2}{\beta^2}  ,\\ \nonumber
& & \left( \frac{\pi}{\beta} \right)^8 \int 
\frac{1}{ \sinh^2\frac{\pi}{\beta} w_{12} 
 \sinh^2\frac{\pi}{\beta} w_{34} \sinh^2\frac{\pi}{\beta}w_{23} \sinh^2\frac{\pi}{\beta}  w_{14} }
 = \frac{32}{45} ( 15 - \pi^2) \frac{\pi^5 L}{\beta} , \\ \nonumber
& &   \left( \frac{\pi}{\beta} \right)^8 \int 
   \frac{1}{ \sinh\frac{\pi}{\beta} w_{12} 
\sinh\frac{\pi}{\beta} w_{34}  \sinh\frac{\pi}{\beta} w_{13}
\sinh\frac{\pi}{\beta} w_{24} 
\sinh^2\frac{\pi}{\beta} w_{23} 
\sinh^2\frac{\pi}{\beta}  w_{14} }  \\ \nonumber 
&&\qquad \qquad \qquad \qquad \qquad \qquad \qquad \qquad 
\qquad \qquad \qquad \qquad  =\left( \frac{16}{3}  - \frac{64}{45}\pi^2\right)  \frac{\pi^5 L}{\beta} .
\end{eqnarray}
Here $\int = \int \prod_{i=1}^{4} d^2 w_i$ and the integrals over the cylinder are done using the 
prescription in section \ref{unifsour}. 
We can now use these results in the expansion of the partition function $ Z[\mu]$ to the 
quartic order in $\mu$ which 
leads to (\ref{4thorder}). 

\subsection{Spin-3 deformation to  the quartic order } \label{integrals2}

To obtain the partition function for the spin-3 deformation to the quartic order in $\mu$
we require the  $4$-point function of the spin-3 currents. 
For the free field realization of the spin-3 current  in terms of $\frac{c}{2}$ complex bosons 
given in (\ref{defspin3cur})  we can evaluate the $4$-point function by Wick contractions. 
This results in 
\begin{eqnarray}\label{4ptfnwa}
\langle W(w_1) W(w_2) W(w_3) W(w_4)  \rangle &=&
\frac{25 c^2}{ 36 \pi^4} 
 \left( \frac{\pi^2}{\beta^2 \sinh  \frac{\pi}{\beta} 
( w_1 - w_2) \sinh  \frac{\pi}{\beta} 
( w_3 - w_4)   }
 \right)^6  \\ \nonumber
&& 
\left[ 
1 + \tfrac{18}{c} \eta 
+ \tfrac{75}{c} \eta^2 + 
( 2 + \tfrac{54}{c} ) \eta^3 
+ 9 ( 1 + \tfrac{2}{c} ) \eta^4 
+ 6 \eta^5 
+ \eta^6 
\right].
\end{eqnarray}
where $\eta$ is the  related to the cross ratio by 
\begin{equation}
\eta = x + \frac{1}{x} - 2, \qquad
x = \frac{ \sinh \frac{\pi}{\beta} ( w_1 - w_3)  \sinh \frac{\pi}{\beta} ( w_2 - w_4) }{
\sinh \frac{\pi}{\beta} ( w_1 - w_4)  \sinh \frac{\pi}{\beta} ( w_2 - w_3)}. 
\end{equation}
Let us label each of the integrals that occur in on integrating the $4$-point function 
in (\ref{4ptfnwa}) as 
\begin{eqnarray}
&& \int \langle W(w_1) W(w_2) W(w_3) W(w_4)  \rangle
=  \\ \nonumber
&& \qquad\qquad\qquad \frac{25 c^2}{36\pi^4}  
 \left( I_0  + \frac{18}{c} I_1 + \frac{75}{2c} I_2 +  ( 2 + \frac{54}{c} ) I_3 
+ 9 ( 1 + \frac{2}{c} ) I_4 +  6 I_5 +  I_6 \right) .
\end{eqnarray}
Here each of the integrals are defined according to the respective order the corresponding term occurs in the $4$ point function. 
For example 
\begin{eqnarray}
I_1 &=& \left(\frac{\pi}{\beta}\right)^{12} \int \frac{1}{\sinh^4\frac{\pi}{\beta}w_{12}\sinh^4\frac{\pi}{\beta}w_{34}\sinh\frac{\pi}{\beta}w_{13}\sinh\frac{\pi}{\beta}w_{14}\sinh\frac{\pi}{\beta}w_{23}\sinh\frac{\pi}{\beta}w_{24}} ~,
\nonumber  \\
&=&  \Big(\frac{224}{81}+\frac{256}{945}\pi^2\Big)\frac{ 2\pi^9 L}{\beta^5}~.
\end{eqnarray}
The list of all the integrals are given below
	\begin{align*}
	 I_0&=\frac{256}{225}\frac{\pi^{10} L^2}{\beta^6},  & I_2&=\frac{256}{9}\frac{\pi^{10} L^2}{\beta^6}, \\ I_3&=\Big(\frac{64}{9}+\frac{256}{315}\pi^2\Big)\frac{2\pi^{9} L}{\beta^5}, 
	&I_4&=\Big(\frac{2176}{81}+\frac{512}{189}\pi^2\Big) \frac{2\pi^9 L}{\beta^5} ,    \\I_5&= \Big(-192+\frac{1024}{105}\pi^2\Big) \frac{2\pi^9L}{\beta^5},       & I_6&=\Big(\frac{3136}{9}+\frac{256 \pi L }{225 \beta}-\frac{11264}{315}\pi^2\Big) \frac{2\pi^9}{\beta^6}  .
	\end{align*}
	Substituting  these results  along with the integrals of the $2$-point function of the 
	spin-3 currents in (\ref{spin32pint}) 
	in the expansion of the partition given in (\ref{pertspin3}) we obtain
	(\ref{partw}).

\section{Torus partition functions}
\label{appendB}

\subsection{Sine-squared deformation}
\label{ssd-np}
The $g$-parametrized Hamiltonian for the sine-squared deformation is 
\begin{align}\label{hamiltonian}
H=\frac{2\pi}{L}\left[L_0+\bL_0 -\frac{g}{2}(L_1+L_{-1}) -\frac{g}{2}(\bL_{1}+\bL_{-1}) - \frac{c}{12}\right]~. 
\end{align}
The partition function corresponding to the above Hamiltonian can be found by using the $\sltwor$ symmetry\footnote{We thank Per Kraus for suggesting this method.}. We define the following linear combinations of the generators
\begin{align}
X = L_0\ ,  \qquad Y= \frac{L_{-1}-L_{1}}{2} \ ,\qquad  Z= \frac{L_{-1}+L_1}{2}~. 
\end{align}
It can then be seen that
\begin{align}
[X,Y] = Z, \qquad [X,Z] = Y,  \qquad [Z,Y]=X. 
\end{align}
Let us now consider the adjoint action
\begin{align}
M_\lambda X M_{-\lambda}=e^{\lambda Y} X e^{-\lambda Y} &= X + \lambda[Y,X] + \frac{\lambda^2}{2!} [Y,[Y,X]] + \frac{\lambda^3}{3!} [Y,[Y,[Y,X]]] +\cdots\nn \\
&= X - \lambda Z + \frac{\lambda^2}{2!} X - \frac{\lambda^3}{3!} Z +\cdots\nn \\
&= (\cosh \lambda) X - (\sinh \lambda) Z~. 
\end{align}
Another way to obtain the same result is by using the 2-dimensional representation of $\sltwor$.
We re-write the above as 
\begin{align}
W\equiv X - (\tanh \lambda) Z  = M_\lambda \left(X\over \cosh \lambda\right) M_{-\lambda}
\end{align}
This brings us to the Boltzmann factor. Since $M_\lambda M_{-\lambda}=1$ we have
\begin{align}\label{boltz-adjx}
e^{2\pi i \tau  W}= e^{2\pi i \tau M_\lambda \left(X\over \cosh \lambda\right) M_{-\lambda}} =  M_{\lambda}e^{2\pi i \frac{\tau}{\cosh\lambda} X}  M_{-\lambda}. 
\end{align}
Therefore, using cyclicity of the trace
\begin{align}\label{tracex}
\tr \left[ q^{W} \bar{q}^{\bar W}  \right] = \tr \left[ y^{W} \bar{y}^{\bar W}  \right] \qquad \text{where, } y=e^{2\pi i \tau \over \cosh(\lambda)}~. 
\end{align}
If we define the coupling as 
$
g = \tanh(\lambda),
$
then the modular parameter gets rescaled as 
\begin{align}
\tau \mapsto \frac{\tau}{\cosh(\lambda)} = \tau\sqrt{1-g^2}~. 
\end{align}
To summarize, the torus partition functions are
\begin{align}\label{deformed-Zxx}
\cZ_{\rm CFT} (\tau, \bar \tau) &= \tr \left[ q^{L_0-c/24} \bar{q}^{\bar L_0 -c/24}  \right]=q^{ -c/24} \bar{q}^{  -c/24} f(\tau,\bar \tau )   
\nn \\
\cZ_{g} (\tau, \bar \tau) &= \tr \left[ q^{L_0+\frac{g}{2}(L_{-1}+L_1)-c/24} \bar{q}^{\bar L_0+\frac{g}{2}(\bar L_{-1}+\bar L_1) -c/24}  \right] \nn \\&= q^{ -c/24} \bar{q}^{  -c/24} f (\tau\sqrt{1-g^2}, \bar \tau\sqrt{1-g^2})~. 
\end{align}

\subsection{$\mathcal{W}_3$ generalization}
\label{w3-np}
The above procedure can be generalized for the $\mathcal{W}_3$ case. The deformed Hamiltionian is 
\begin{align}
H=\frac{2\pi}{L}\left[L_0 +\frac{g}{4}(W_{-2}+W_2)\right]~. 
\end{align}
Using the same methods of the previous subsection, we have
\begin{align}
X = L_0\ ,  \qquad Y_3= \frac{W_{-2}-W_{2}}{4} \ ,\qquad  Z_3= \frac{W_{-2}+W_{2}}{4}~. 
\end{align}
The SL(3,R) algebra implies that
\begin{align}
[X,Y_3] =  2Z_3, \qquad [X,Z_3] = 2Y_3,  \qquad [Z_3,Y_3]=X/2. 
\end{align}
Let us now consider the action
\begin{align}
M_\lambda X M_{-\lambda}=e^{\lambda Y_3} X e^{-\lambda Y_3} &= X + \lambda[Y_3,X] + \frac{\lambda^2}{2!} [Y_3,[Y_3,X]] + \frac{\lambda^3}{3!} [Y_3,[Y_3,[Y_3,X]]] +\cdots\nn \\
&= X -2 \lambda Z_3 - \frac{\lambda^2}{2!} X + \frac{\lambda^3}{3!} 2Z_3 + \frac{\lambda^4}{4!} X - \frac{\lambda^5}{5!} 2Z_3   -++-\cdots\nn \\
&= \left(1 - \frac{\lambda^2}{2!}+ \frac{\lambda^4}{4!} - \cdots \right) X -\left( \lambda -  \frac{\lambda^3}{3!} + \frac{\lambda^5}{5!} +\cdots \right)  2Z_3~ \nn \\
&= (\cos \lambda) X - 2 (\sin \lambda) Z_3.
\end{align}
Hence,
\begin{align}
M_\lambda {X\over \cos\lambda} M_{-\lambda}=
&=  X - 2 (\tan \lambda) Z_3.
\end{align}
If $g=2\tan\lambda$, then $\cos \lambda=(1+4g^2)^{-1/2}$. So
\begin{align}
(1+4g^2)^{1/2}M_\lambda {X} M_{-\lambda}
&=  X - g Z_3.
\end{align}
This leads to a rescaling of the modular parameter 
$
\tau \mapsto \tau (1+4g^2)^{1/2}
$.
That is
\begin{align}\label{deformed-Z3}
\cZ_{\rm CFT} (\tau, \bar \tau) &= \tr \left[ q^{L_0-c/24} \bar{q}^{\bar L_0 -c/24}  \right]=q^{ -c/24} \bar{q}^{  -c/24} f(\tau,\bar \tau )~,   
\\
\cZ_{g} (\tau, \bar \tau) &= \tr \left[ q^{L_0+\frac{g}{2}(L_{-1}+L_1)-c/24} \bar{q}^{\bar L_0+\frac{g}{2}(\bar L_{-1}+\bar L_1) -c/24}  \right]~, \nn \\
&= q^{ -c/24} \bar{q}^{  -c/24} f (\tau\sqrt{1+4g^2}, \bar \tau\sqrt{1+4g^2})~.\nn
\end{align}
The simplifications seen above, however, does not generalize to deformations of this kind by currents with spin-4 or greater. The reason is that for 
\begin{align}
X = L_0\ ,  \qquad Y_s= \frac{W^{(s)}_{-k}-W^{(s)}_{k}}{2^{s-1}} \ ,\qquad  Z_s= \frac{W^{(s)}_{-k}+W^{(s)}_{k}}{2^{s-1}}~. 
\end{align}
we have
\begin{align}
[X,Y_s] =  (s-1)Z_s, \qquad [X,Z_s] = (s-1)Y_s,  \qquad [Z_s,Y_s]\sim X + \text{even-spin-currents}. 
\end{align}
However, we can still proceed using perturbation theory. 

\section{Details on Hamiltonian perturbation theory}
\label{ham-alg}

\subsection{Some trace identities}
\label{app:traces}
We derive a few identities which are used for doing perturbation theory for the inhomogeneous deformations of CFTs. 

For $(yw)=q$ we have
\begin{align}\label{LL-shuffle}
&\tr \left[(L_1+L_{-1})y^{L_0}(L_1+L_{-1})w^{L_0}\right] = \tr \left[L_1y^{L_0}L_{-1}w^{L_0}\right] +\tr  \left[L_{-1}y^{L_0}L_{1}w^{L_0}\right] \nn \\
&= y \,\tr \left[L_1 L_{-1}y^{L_0}w^{L_0}\right] + w \,\tr \left[y^{L_0}L_1L_{-1} w^{L_0}\right] = (y+w)\, \tr \left[L_1 L_{-1}(yw)^{L_0}\right] \nn \\
&= (y+w)\, \tr \left[L_1 L_{-1}q^{L_0}\right]
\end{align}
We have used cyclicity of the trace and the fact that moving a $L_1$ to the right of $u^{L_0}$ produces an extra factor of $u$.
The higher spin generalization of this is similar. We use the commutator $[L_0,V^{(s)}_{n}] = -nV^{(s)}_{n}$ to get
\begin{align}\label{WW-shuffle}
&\tr \left[(\Vs{k}+\Vs{-k})y^{L_0}(\Vs{k}+\Vs{-k})w^{L_0}\right] = \tr \left[\Vs{k}y^{L_0}\Vs{-k}w^{L_0}\right] +\tr  \left[\Vs{-k}y^{L_0}\Vs{k}w^{L_0}\right] \nn \\
&= y^k \,\tr \left[\Vs{k}\Vs{-k}y^{L_0}w^{L_0}\right] + w^k \,\tr \left[y^{L_0}\Vs{k}\Vs{-k} w^{L_0}\right]  
= (y^k+w^k)\, \tr \left[\Vs{k} \Vs{-k}(yw)^{L_0}\right] \nn \\
&= (y^k+w^k)\, \tr \left[\Vs{k} \Vs{-k}q^{L_0}\right].
\end{align}

We also need the following expectation value on the torus. These have been worked out in 
\cite{Maloney:2018hdg} 
\begin{align}
Q&=\tr \left[  L_{-1}L_1 q^{L_0}\bar{q}^{\bar L_0} \right] = \tr \left[  L_{1} q^{L_0}\bar{q}^{\bar L_0}  L_{-1}\right]  
= q\tr \left[ L_1 L_{-1} q^{L_0}\bar{q}^{\bar L_0}\right] \nn \\
&= q \tr \left[(2L_0 + L_{-1}L_1) q^{L_0}\bar{q}^{\bar L_0}\right] 
= 2q \tr \left[L_0 q^{L_0}\bar{q}^{\bar L_0}\right] +q Q~.
\end{align}
Solving for $Q$
\begin{align}
\tr \left[  L_{-1}L_1 q^{L_0}\bar{q}^{\bar L_0} \right] = \frac{2q  }{1-q }  \,\tr \left[L_0 q^{L_0}\bar{q}^{\bar L_0}\right] = \frac{2q^2}{1-q}  \pd_{q} f
\end{align}
and 
\begin{align}\label{QQ}
\tr \left[  L_{1}L_{-1} q^{L_0} \bar{q}^{\bar L_0}\right] = q^{-1}Q=\frac{2q}{1-q}   \pd_{q} f~.
\end{align}
The analogous  ingredient for the higher spin case is the following quantity
\begin{align}
R=\tr \left[ V^{(s)}_{k}V^{(s)}_{-k}q^{L_0}    \right] 
\end{align}
To do this, we recall the commutator
\begin{align}
[L_0,V^{(s)}_{n}] = -nV^{(s)}_{n}
\end{align}
Therefore
\begin{align}
V^{(s)}_{m}q^{L_0} =q^{-m}q^{L_0} V^{(s)}_{m}
\end{align}
So
\begin{align}
R&=q^{k}\tr \left[ V^{(s)}_{k}q^{L_0}  V^{(s)}_{-k}  \right] = q^{k}\tr \left[ V^{(s)}_{-k}  V^{(s)}_{k}q^{L_0}  \right]  \\
&= q^{k}\tr \left[ [V^{(s)}_{-k} , V^{(s)}_{k}]q^{L_0}  \right] + q^{k}\tr \left[ V^{(s)}_{k}V^{(s)}_{-k}q^{L_0}    \right]  
= q^{k}\tr \left[ [V^{(s)}_{-k} , V^{(s)}_{k}]q^{L_0}  \right] + q^{k} R. \nn 
\end{align}
The hs$[\lambda]$ algebra gives the following commutator for the spin-$N$ current
\begin{align}\label{ww-alg}
[V^{(s)}_{-k} , V^{(s)}_{k}] = C_{ss}^2(k) L_0 + \text{zero-modes of even spins},
\end{align}
The structure constant $C_{NN}^2(k)$ can be obtained case-by-case [Gaberdiel-Hartman appendix].
Hence
\begin{align}\label{vv-thermal}
R= \frac{q^{k}}{q^{k}-1}\tr \left[ [V^{(s)}_{-k} , V^{(s)}_{k}]q^{L_0}  \right]  
&=  \frac{q^{k}}{q^{k}-1}C^2_{ss}(k)\tr \left[ L_0q^{L_0}  \right] + \cdots\nn \\&
=  \frac{q^{k}}{q^{k}-1}C^2_{ss}(k)~\pd_q \tr \left[ q^{L_0}  \right] + \cdots. 
\end{align}
So, this can be calculated from the derivative of the partition function as before -- cf.~\eqref{QQ}. Note that the `$\cdots$' above denote 1-point functions of higher-even-spin currents. If we choose to work in the primary basis of the $\mathcal{W}_\infty[\lambda]$ algebra and these vanish in the high temperature limit.

The analogous result for $U(1)$ currents is
\begin{align}
\tr \left[J_{k}J_{-k}q^{L_0}  \right]  = \frac{q^k}{1-q^k} \tr  \left[[J_{k},J_{-k}]q^{L_0}  \right] = \frac{q^k}{1-q^k} ~\kappa k ~\tr\left[q^{L_0}  \right] .
\end{align}
where we used $[J_m,J_n]=\kappa m\delta_{m+n,0}$.

Finally, we need the following identity for the deformation of the harmonic oscillator by a linear ramp\footnote{There are some differences of signs in contrast to the CFT cases\begin{align}
	y^{-a_+ a_-} a_\pm y^{a_+ a_-} = a_\pm y^{ { \mp 1}}, \qquad 
	y^{-L_0} L_{\pm 1} y^{L_0} = L_{\pm 1} y^{ { \pm 1}}. \nn 
	\end{align}
	This is because of the commutation relation $[a_-,a_+]=1$ and
$
	[L_0,L_{\pm 1}] = {\mp } L_{\pm 1} ,\, [a_+a_-,a_{\pm 1}] = {\pm } a_{\pm 1}
$. That is, $L_{-1}$ and $a_+$ are raising operators while $L_{+1}$ and $a_-$ are lowering operators.   } 
\begin{align}\label{aa-shuffle}
&\tr \left[(a_+ + a_ -)y^{a_+ a_-}(a_++a_-)w^{a_+ a_-}\right] = \tr \left[a_+y^{a_+ a_-}a_-w^{a_+ a_-}\right] +\tr  \left[a_-y^{a_+ a_-}a_+w^{a_+ a_-}\right] \nn \\
&= y^{-1} \,\tr \left[a_+ a_-y^{a_+ a_-}w^{a_+ a_-}\right] + w^{-1} \,\tr \left[y^{a_+ a_-}a_+a_- w^{a_+ a_-}\right]
= q^{-1}(y+w)\, \tr \left[a_+ a_-q^{a_+ a_-}\right].
\end{align}

\subsection{Deformed harmonic oscillator}
\label{app:sho-ham}
In this subsection, we consider the deformed harmonic oscillator with an additional linear potential ($\hbar=1$)
\begin{align}
H_g = \omega\left(a_+ a_-+\frac{1}{2}\right)+ g (a_+ + a_-). 
\end{align}
We shall evaluate the \Re divergence in the Hamiltonian formalism. This serves as a consistency check with the path integral calculation in Section \ref{subsec:sho} and also makes the agreement of the $\alpha$ dependence with the inhomogeneous CFT deformations more transparent. 
The partition function corresponding to the Euclidean quench setup is 
\begin{align}
\cZ_\alpha = \tr \left[ q^{\alpha \left[\left(a_+ a_-+\frac{1}{2}\right)+ g (a_+ + a_-)\right]} q^{(1-\alpha) \left(a_+ a_-+\frac{1}{2}\right)}  \right], \qquad q=e^{-\beta\omega}.
\end{align}
We evaluate the above partition function perturbatively in $g$ till the quadratic order. The object above can be rewritten as 
\begin{align}
\cZ_\alpha &= e^{-\beta \omega/2}~\tr \left[  y^{  a_+a_-  + g(a_+ + a_-) } w^{    a_+a_-  } \right],\qquad y= e^{-\alpha\beta\omega}, w= e^{-(1-\alpha)\beta\omega}. 
\end{align}
As before the first order correction vanishes as the expectation values, $\vev{a_\pm}=0$.   The second derivative is 
\begin{align}\label{2nd-derivative-sho}
&\frac{\pd^2 \cZ_\alpha}{\pd g^2}  \bigg|_{g=0} 
=~ 2 (\alpha b  )^2 e^{-\beta \omega/2}\int_0^1 ds ~(1-s) \int_0^1 du~ \tr \left[ e^{- b  (1-\alpha u (1-s)) a_+a_-}Xe^{-\alpha b   (1-s) u a_+a_-} X \right].
\end{align}
with $b=\beta\omega$ and $X=a_++a_-$. Using the relation \eqref{aa-shuffle} we get
\begin{align}\label{sho-int}
\frac{\pd^2 \cZ_\alpha}{\pd g^2}  \bigg|_{g=0} 
=&~ 2 (\alpha b  )^2 q^{-1}\left[\int_0^1 ds ~(1-s) \int_0^1 du~ (q^{  (1-\alpha u (1-s)) }+ q^{\alpha   (1-s) u }) \right] \nn \\
& ~\qquad\times \tr \left[a_+a_-e^{-b (a_+a_-+1/2)}\right].
\end{align}
This is the same integral encountered in CFT deformations. The result is
\begin{align}\label{d2}
\frac{\pd^2 \cZ_\alpha}{\pd g^2}  \bigg|_{g=0} 
=&- 2  q^{-1}\left[4 q^{1/2} \sinh (\pi  (\alpha -1) \tfrac{\beta \omega}{2\pi} ) \sinh (\pi  \alpha  \tfrac{\beta \omega}{2\pi} )+\alpha(1-q)\log q \right]\nn \\
& ~\qquad\times \tr \left[a_+a_-e^{-b (a_+a_-+1/2)}\right]. 
\end{align}
Using the expression for \Re divergence, equation \eqref{rd-Z-def}, we obtain
\begin{align}
D_\alpha(\rho||\rho_\beta)\approx 4g^2 \frac{ q^{-1/2} \sinh (   (\alpha -1) {\beta \omega}/{2} ) \sinh (  \alpha   {\beta \omega}/{2 } )}{(1-\alpha)} ~\pd_b \log \tr \left[e^{-b a_+a_-}\right]. 
\end{align}
The part containing the trace above is 
\begin{align}
\pd_b \log \tr \left[e^{-b a_+a_-}\right] =\pd_b \log \frac{1}{1-e^{-b}}= \frac{1}{1-e^{b}}.
\end{align}
So, the \Re divergence is
\begin{align}\label{sho-final}
D_\alpha(\rho||\rho_\beta)\approx - \, 4g^2 \,  \frac{ \sinh (   (\alpha -1) {\beta \omega}/{2} ) \sinh (  \alpha   {\beta \omega}/{2 } )}{(1-\alpha)\sinh(\beta\omega/2)}  ~. 
\end{align}
This agrees with the result \eqref{shoreny3} obtained using the path integral formalism. From the path integral calculation, we have seen that there are no higher order corrections to the \Re divergence (or the logarithm of the deformed partition function). This implies that the correction \eqref{d2} exponentiates, analogous to the $U(1)$ current deformations considered in Section \ref{subsec:u1}.
 
\bibliographystyle{JHEP}
\bibliography{references}        
 \end{document}